\documentclass[prd,nofootinbib,preprint,superscriptaddress]{revtex4}
 \pdfoutput=1
\usepackage[usenames,dvipsnames]{color}  
\usepackage{graphicx}
\usepackage{setspace}
\usepackage{subfigure}
\usepackage{amsmath}
\usepackage{amssymb}
\usepackage[colorlinks=true,citecolor=blue,urlcolor=blue, pdfborder={0 0 0}]{hyperref}
\usepackage[normalem]{ulem}
\usepackage{xcolor}
\usepackage{multirow}

\makeatletter
\def\p@subsection{}
\makeatother

\definecolor{darkred}{rgb}{0.6,0,0}

\definecolor{linkcolor}{rgb}{0,0,0.5}

\def\gsim{\raise0.3ex\hbox{$\;>$\kern-0.75em\raise-1.1ex\hbox{$\sim\;$}}}
\def\lsim{\raise0.3ex\hbox{$\;<$\kern-0.75em\raise-1.1ex\hbox{$\sim\;$}}}

\def\beqn#1{\begin{equation}\label{#1}}
\def\eeqn{\end{equation}}

\def\beqa#1{\begin{eqnarray}\label{#1}}
\def\eeqa{\end{eqnarray}}

\def\Z2{$\mathcal{Z_2}$}

 \newcommand{\mg}[1]{{\color{magenta}  #1}}
\newcommand {\ignore}[1]{}

\def\321{$\mathrm{SU(3) \otimes SU(2) \otimes U(1)}$ }

\def\bl#1{\textcolor{blue}{#1}}

\def\mg#1{\textcolor{magenta}{#1}}

\newcommand{\AddrIACS}{ School of Physical Sciences, Indian Association for Cultivation of Science,\\
2A $\&$ 2B Raja S C Mullick Road, Kolkata 700032, India}
 
 \newcommand{\AddrIISERB}{Department of Physics,
 Indian Institute of Science Education and Research - Bhopal \\
 Bhopal Bypass Road, Bhauri, Bhopal, India}

\begin{document}

\title{ Hubble Tension and Cosmological Imprints of  $U(1)_X$ Gauge Symmetry: $U(1)_{B_3-3 L_i}$ as a case study}
\author{Dilip Kumar Ghosh}\email{dilipghoshjal@gmail.com}
\affiliation{\AddrIACS}
\author{Purusottam Ghosh}\email{pghoshiitg@gmail.com}
\affiliation{\AddrIACS}
\author{Sk Jeesun}\email{skjeesun48@gmail.com}
\affiliation{\AddrIACS}
\author{Rahul Srivastava}\email{rahul@iiserb.ac.in}
\affiliation{\AddrIISERB}

\begin{abstract} 
The current upper limit on $N_{\rm eff}$ at the time of CMB by Planck 2018 can place stringent constraints in the parameter space of  BSM paradigms where their additional interactions may affect neutrino decoupling.
Motivated by this fact in this paper we explore the consequences of light gauge boson ($Z'$) emerging from local $U(1)_X$ symmetry in $N_{\rm eff}$ at the time of CMB.
First, we analyze the generic $U(1)_X$ models with arbitrary charge assignments for the SM fermions
 and show that, in the context of $N_{\rm eff}$  the generic $U(1)_X$ gauged models can be broadly classified into two categories, depending on the charge assignments of first generation leptons.
We then perform a detailed analysis with two specific  $U(1)_X$ models:  $U(1)_{B_3-3L_e}$ and $U(1)_{B_3-3L_\mu}$ and explore the contribution in $N_{\rm eff}$ due to the presence of $Z'$ realized in those models.
For comparison, we also showcase the constraints from low energy experiments like: Borexino, Xenon 1T, neutrino trident, etc. 
We show that in a specific parameter space, particularly in the low mass region of $Z'$, 
the bound from $N_{\rm eff}$ (Planck 2018) is more stringent than the experimental constraints. Additionally, a part of the regions of the same parameter space may also relax the $H_0$ tension.
\end{abstract}

\maketitle


\section{INTRODUCTION}
\label{sec:intro}
The cosmological parameter $N_{\rm eff}$, associated with the number of relativistic degrees of freedom, is crucial in describing the dynamics of the thermal history of the early universe. 
At very high temperature of the universe, the photon (along with electrons) and neutrino bath were coupled, whereas at low temperature the interaction rate drops below the Hubble expansion rate, and the two baths decouple \cite{Dodelson:2003ft}.
$N_{\rm eff}$ is parameterised in terms of the ratio of energy densities of neutrino and photon bath.
Within the Standard Model (SM) particle contents, the two aforementioned baths were coupled through weak interactions at high temperature and as temperature drops ($T\sim 2$ MeV) they decouple.
Assuming such scenario the predicted value of $N_{\rm eff}^{\rm SM}$ turns out to be $3.046$ \cite{Mangano:2005cc,Grohs:2015tfy}.
This value of $N_{\rm eff}$ deviates from the number of neutrinos ($3$) in the SM 
particle content is due to various non-trivial effects like non-instantaneous neutrino decoupling, finite temperature QED corrections and flavour oscillations of neutrinos \cite{Mangano:2005cc,Grohs:2015tfy}.
However, in observational cosmology also, 
measurements of the cosmic microwave background (CMB), baryon acoustic oscillations (BAO), and 
other cosmological probes provide constraints on $N_{\rm eff} $.
The current Planck 2018 data has precise measurement  of $N_{\rm eff}$ at the time of CMB with  $95\%$ confidence level, $N_{\rm eff} = 2.99^{+0.34}_{-0.33}$ \cite{Planck:2018vyg}. 
Thus the current upper limit from Planck 2018 data shows that there can be additional contribution (apart from SM predicted value) to $N_{\rm eff}$, indicating the scope for new physics. 

It is evident that $N_{\rm eff}$ will change in the presence of any beyond standard model (BSM) particles with sufficient interactions with either of the photon or neutrino bath at temperatures relevant for neutrino decoupling \cite{Escudero:2018mvt,Escudero:2019gzq,Esseili:2023ldf,Ganguly:2022ujt} or in presence of any extra radiation \cite{Abazajian:2019oqj,Poulin:2018cxd,Ghosh:2022fws,Ghosh:2023ocl}.
Thus the upper limit on $N_{\rm eff}$ from CMB can be used to constrain such BSM paradigms dealing with any extra energy injection.  
Several studies have been performed to explore the imprints of BSM models in $N_{\rm eff}$ like models with early dark energy \cite{Poulin:2018cxd,Poulin:2018dzj}, relativistic decaying dark matter \cite{Bringmann:2018jpr} and non-standard neutrino interactions (NSI) \cite{Escudero:2018mvt,Escudero:2019gzq,Li:2023puz,Biswas:2022fga,Berbig:2020wve,Fabbrichesi:2020wbt}. 
From the perspective of neutrino physics, the last one is very interesting since the non-standard interactions of light neutrinos may alter the late-time dynamics between photon and neutrino baths, contributing to $N_{\rm eff}$ and the same NSI interactions can be probed from ground based neutrino experiments as well \cite{Shutt:2002rg,Majumdar:2021vdw}.

On the other hand, the anomaly-free $U(1)_X$ gauge extended BSM models are well-motivated from several aspects like non zero neutrino masses \cite{Ma:2015raa}, flavor anomalies \cite{Bonilla:2017lsq} etc. 
The anomaly condition allows the introduction of right-handed neutrinos in the theory and thus it can also explain non-zero neutrino masses via the Type-I seesaw mechanism \cite{Ma:2015raa}. 
These scenarios naturally involve a gauge boson (${Z^\prime}$) that originates from the $U(1)_X$ abelian gauge symmetry and has neutral current interactions with neutrinos and electrons which may have some nontrivial role in neutrino decoupling and hence in deciding $N_{\rm eff}$ \cite{Escudero:2019gzq,Dutta:2020jsy,Esseili:2023ldf,Li:2023puz}.
The detectability of gauge boson throughout the mass scale ($M_{Z^\prime}$) also motivates such scenarios. For TeV-scale $Z'$, constraints arise from collider experiments \cite{CMS:2016cfx,ATLAS:2019erb,Das:2016zue,Accomando:2017qcs}, whereas in the sub-GeV mass region, low energy scattering experiments (neutrino electron scattering \cite{Coloma:2022umy}, neutrino-nucleus scattering \cite{COHERENT:2021xmm} etc.) are relevant to constrain the parameter space. However, both types of direct searches become less sensitive in the mass $M_{Z^\prime} \lesssim \mathcal{O}$(MeV) and lower coupling ($g_X \lesssim 10^{-5}$) region. In that case, the CMB observation on $N_{\rm eff}$ plays a crucial role and can impose severe constraints on the $M_{Z^\prime} - g_X$ plane, which stands as the primary focus of our study.

In cosmology, there exists some discrepancy between the values of expansion rate $H_0$ obtained from CMB and local measurement \cite{DiValentino:2021izs,Riess:2018byc, Riess:2016jrr, Riess:2021jrx, Planck:2018vyg,Vagnozzi:2019ezj,Vagnozzi:2023nrq}.
Using direct observations of celestial body distances and velocities, the SHOES collaboration calculated $H_0 = 73.04 \pm 1.04 ~{\rm Km~s^{-1}~Mpc^{-1}}$ \cite{Riess:2021jrx}.
However, the CMB measurements like Planck 2015 TT data predict $H_0=68.0^{+2.6}_{-3.0}~{\rm Km~s}^{-1}\rm{Mpc}^{-1}$ at $1\sigma$  \cite{Planck:2015fie}
and from recent Planck 2018 collaboration, it turns out to be $H_0 = 67.36 \pm 0.54 {~\rm Km~s}^{-1}~\rm{Mpc}^{-1}$ \cite{Planck:2018vyg}, where both the collaboration analyzes the CMB data under the assumption
of $\Lambda$CDM cosmology. 
So there is a disagreement between the local and CMB measurements roughly at the level of $4 \sigma-6\sigma$ \cite{DiValentino:2021izs}. 
Although such discrepancy may arise from systematic error in measurements \cite{Efstathiou:2013via,Freedman:2017yms}, it also provides a hint for BSM scenarios affecting the dynamics of the early universe.  
One possible way out to relax this so called ``$H_0$ tension" involves increasing $N_{\rm eff}$ (in 
the approximate range $\sim3.2-3.5$) \cite{Bernal:2016gxb,Escudero:2019gzq,Araki:2021xdk}\footnote{However, it is essential to acknowledge that increasing $N_{\rm eff}$ also provokes the tension related to the other parameter of $\Lambda$CDM model, $\sigma_8$ \cite{Planck:2018vyg,Hildebrandt:2018yau}.}.
{It is also worth highlighting that,
though BSM radiation with $N_{\rm eff}\sim3.2-3.5$ ameliorate the $H_0$ tension to $3.6\sigma$, it can not resolve the tension completely as revealed by recent studies \cite{Bernal:2016gxb,deJesus:2022pux}}.
As discussed earlier $U(1)_X$ local gauge extension leads to new gauge boson and new interaction with SM leptons which might affect neutrino decoupling and hence contribute to $N_{\rm eff}$.
Thus $U(1)_X$ gauge extension can be one possible resolution to the $H_0$ tension also.
Two kinds of $U(1)_X$ scenarios have been considered so far in the literature to address the Hubble tension problem and/or the excess in $N_{\rm eff}$ at CMB: $U(1)_{\mu-\tau}$ \cite{Escudero:2019gzq} and more recently with $U(1)_{B-L}$ \cite{Esseili:2023ldf}. However, there exist several other well motivated $U(1)_X$ models (different anomaly free combinations of $B$ and $L$ numbers \cite{Coloma:2022umy,AtzoriCorona:2022moj}) whose consequences in $N_{\rm eff}$ have not been explored till date. Depending on the value of $L$, each model exhibits distinct signatures in $N_{\rm eff}$, demanding individual treatment \footnote{{Note that, the value of $N_{\rm eff}$ does not depend on the $B$ number as will be discussed in Sec.\ref{sec:neff}.}}. Therefore an analysis of $N_{\rm eff}$ in the context of light gauge boson in generic $U(1)_X$ model is required, which is the main objective of this work.

In this work, we begin with a generic $U(1)_X$ extension and explore the aforementioned cosmological phenomena more comprehensively. 
For simplicity, we assume the light gauge boson was in a thermal bath, and the right-handed neutrinos are heavy enough ($> \mathcal{O}(10^2)$ MeV) that they hardly play any role in neutrino decoupling. 
We study the dynamics of neutrino decoupling in the presence of the light gauge boson ($Z'$) in a generic $U(1)_X$ scenario with different $U(1)_X$ charges assigned to SM fermions. 
For some specific values of such charges, the generic $U(1)_X$ models can be  
interpreted as popular $U(1)_X$ models as we will discuss in the later part.
We evaluate $N_{\rm eff}$ by solving a set of coupled Boltzmann equations that describe the evolution of light particles i.e. electron, SM neutrino, and light gauge boson ( $M_{Z'}\sim \mathcal{O}(10)$ MeV). 
Our analysis shows that depending on the $U(1)_X$ charge assignment of the first generation lepton,
the generic $U(1)_X$ gauged models can be classified in the context of $N_{\rm eff}$: $Z^\prime$ having tree-level coupling with $e^\pm$ and $Z^\prime$ having induced coupling with $e^\pm$. This classification leads to two different types of $N_{\rm eff}$ characteristics depending on the $U(1)_X$ charges of first generation leptons. 
Thus we identify the region of parameter space that can be excluded by Planck 2018 observations and indicate the constraints from $N_{\rm eff}$.
We also illustrate the parameter space in which Hubble tension can be relaxed where the value of $N_{\rm eff}$ falls within the range of $3.2$ to $3.5$ \cite{Bernal:2016gxb}.
Although $N_{\rm eff}$ for a generic MeV scale particle has been discussed in ref.\cite{Sabti:2019mhn}, it rely on model independent formalism in contrast to our case.
We also analyze the cosmological observations for specific gauged $U(1)_X$ models, $U(1)_{B_3-3L_i}$ ($i=e,\mu,\tau$).
Our analysis show that, despite the resemblance with $B-L$ or $L_\mu-L_\tau$ model, the bound from $N_{\rm eff}$ significantly differs in ${B_3-3L_i}$ models due to the different $U(1)_X$ charges of first generation leptons.
In context of ground based experiments, these extensions encounter fewer constraints in comparison to others, as these are exclusively linked only to the third generation of quarks and one generation of leptons.
Our results show that the upper limit on $N_{\rm eff}$ from Planck 2018 \cite{Planck:2018vyg} can put stringent bounds on the parameter space of light $Z'$ in the mass region $\lesssim \mathcal{O}(10^2)$ MeV where the other experimental
bounds are comparatively relaxed.

The paper is organized as follows. In sec.\ref{sec:model}, we begin with a brief model-independent discussion on the generic $U(1)_X$ model. Sec.\ref{sec:neff} is dedicated to a comprehensive analysis and discussion of the dynamics governing neutrino decoupling in terms of the cosmological parameter $N_{\rm eff}$ in the presence of the light $Z^\prime$ originated from the generic $U(1)_X$. Subsequently, in sec.\ref{sec:class}, we present the numerical results in terms of $N_{\rm eff}$ for this generic $U(1)_X$ model. In sec.\ref{sec:examples}, we discuss the analysis presented earlier, which applies to the specific $U(1)_X$ model, $U(1)_{B_3-3L_i} (i=e,\mu,\tau)$. We present a brief discussion on $H_0$ tension in sec.\ref{sec:h0}. Finally, in sec.\ref{sec:conclusions}, we summarise and conclude with the outcomes of our analysis. In Appendices \ref{sec:apxA} to \ref{sec:apxC}, we provide various technical details for the calculation of $N_{\rm eff}$.

\section{Model Independent Discussion: Effective $Z'$ Models}
\label{sec:model}


In this section, we look at effective light $Z'$ models. We take a model-independent but minimalist approach and  only assume that the $Z'$ originates from the breaking of a new local $U(1)_X$ abelian gauge symmetry under which the SM particles are charged as shown in Table \ref{tab:u1x}.
The anomaly cancellation conditions for a typical $U(1)_X$ local gauge symmetry requires the presence of additional chiral fermions 
beyond the SM fermions\footnote{Note that for one generation of SM fermions, the only $U(1)$ symmetry which is anomaly free is the SM hypercharge $U(1)_Y$ symmetry. 
For the full three generations of SM fermions, it is possible to have additional $U(1)_X$ symmetries which can be made anomaly free without adding any new chiral fermions such as $U(1)_{q_i -q_j}$ ;
$i,j = 1,2,3$ with $q_i, q_j$ denoting charges of a given generation of SM quarks but such symmetries have other phenomenological problems and are not of interest to the current study.}. 
To keep the minimal scenario we assume that the only BSM fermions needed for anomaly cancellation are the (three) right handed neutrinos $\nu_R$ which are taken to be charged under the $U(1)_X$ symmetry. 
With the addition of $\nu_R$, several different type of gauged $U(1)_X$ models including the popular $U(1)_X$ models 
can be constructed\footnote{Some models like $L_i-L_j$ do not need $\nu_R$ for anomaly cancellation \cite{Coloma:2022umy,Banerjee:2021laz}.}. We will discuss some of the models later in Sec.~\ref{sec:examples}.

For this purpose within this section, we will not specify the nature of $U(1)_X$ symmetry nor the charges of SM particles and $\nu_R$ under it and will treat them as free parameters and proceed with a generic discussion. 
We will also not go into details of anomaly cancellation constraints. All these things will be clarified in the following sections during 
our discussions on some well motivated $U(1)_X$ models.
 \begin{table}[h]
\begin{center}
\begin{tabular}{| c || c | c |}
  \hline
\hspace{0.1cm} Fields   \hspace{0.1cm}   & \hspace{0.1cm} $SU(3)_c \times SU(2)_L \times U(1)_Y$   \hspace{0.1cm}       & \hspace{0.1cm} $U(1)_{{X}}$    \hspace{0.1cm}   \\
\hline \hline
$Q_i$       & $(3, 2, \frac{1}{3})$                           &  $\mathbb{X}_{Q_i}$     \\   
$u_i$       & $(3, 1, \frac{4}{3})$                           &  $\mathbb{X}_{u_i}$      \\ 
$d_i$       & $(3, 1, -\frac{2}{3})$                          &  $\mathbb{X}_{d_i}$      \\
$L_i$       & $(1, 2, -1)$                                    &  $\mathbb{X}_{L_i}$      \\
$\ell_i$       & $(1, 1, -2)$                                    &  $\mathbb{X}_{\ell_i}$         \\
$\nu_{R_i}$ & $(1,1,0)$                                       &  $\mathbb{X}_{\nu_i}$     \\
\hline \hline
$\Phi$         & $(1, 2, 1)$                                     &  $\mathbb{X}_\Phi$ \\
$\sigma$    & $(1, 1, 0)$                                     &  $\mathbb{X}_\sigma$       \\         
\hline 
  \end{tabular}
\end{center}
\caption{Particle content and $U(1)_X$ gauge charge assignments of Standard Model and new particles. In this section, we will take $\mathbb{X}_\Phi = 0$ as we are interested in a light $Z'$ gauge boson. To simplify our notation later we denote $\mathbb{X}_{L_i} = \mathbb{X}_{\ell_i}=X_i~(i\equiv 1-3,~{\rm signify}~ e,\mu,\tau)$, which are relevant for our discussion on $N_{\rm eff}$ (see the text for detail).}
  \label{tab:u1x}
\end{table}
%
In Table \ref{tab:u1x}, in addition to $\nu_{R}$  (typically required for anomaly cancellation)  we also add an SM singlet scalar $\sigma$ carrying $U(1)_X$ charge whose vacuum expectation value (VEV) will break the $U(1)_X$ symmetry. 
Note that apart from the minimal particle content of  Table ~\ref{tab:u1x}, most of the $U(1)_X$ models available in literature may also contain additional BSM particles\footnote{Such new particles are typically much heavier than MeV scale and will not change our analysis \cite{Bonilla:2017lsq}.}. These additional particles are model 
dependent and we refrain from adding them to proceed with a minimal setup.

Now some general model independent simplifications and conclusions can be immediately drawn for the charges of the particles under  $U(1)_X$ symmetry listed in the aforementioned Table~\ref{tab:u1x}.
\begin{enumerate}

\item  \textbf{Light $Z'$:}  In the upcoming sections, we delve into an effective resolution of the
well-known Hubble parameter tension \cite{Bernal:2016gxb} by increasing $N_{\rm eff}$ considering a light $Z^\prime$ with mass 
$M_{Z'} \sim \mathcal{O}(\rm MeV)$ mass range\cite{Escudero:2019gzq}. To avoid any fine-tuning in keeping the $M_{Z^\prime} $ very light 
compared to the SM $Z$ gauge boson mass i.e. $M^2_{Z^\prime} \ll M^2_{Z}$, we consider the corresponding $U(1)_X$ 
charge of the SM Higgs doublet $\mathbb{X}_\Phi = 0$, so that $M^2_{Z^\prime} $ does not receive any contribution 
from the SM vev.

\item \textbf{Mass generation for quarks and charged leptons:} For the choice of $\mathbb{X}_\Phi = 0$, to generate the SM quark and charged lepton masses, we must take $\mathbb{X}_{Q_i} = \mathbb{X}_{u_i} = \mathbb{X}_{d_i}$ and $\mathbb{X}_{L_i} = \mathbb{X}_{\ell_i}$ such that the standard Yukawa term $y^u_{ij} \bar{Q}_i\tilde \Phi u_j$ involving only SM fields can be written in the canonical form as shown in eq.(\ref{eq:yuk}). 
It is important to note that although within each generation, the $U(1)_X$ charges of quark doublet should be the same as that of the up and down quark singlets, the charges  may differ when comparing across different generations i.e. 
$\mathbb{X}_{Q_1} \neq \mathbb{X}_{Q_2} \neq \mathbb{X}_{Q_3}$. 
The same applies to charged leptons. In fact, in later sections, we will indeed consider flavour dependent $U(1)_X$ symmetries.
To simplify our notation throughout the remaining part of this paper we will denote $\mathbb{X}_{L_i} = \mathbb{X}_{\ell_i}=X_i$.

\item {\bf Quark mixing:} As a follow up point note that if the charges of all three generations of quarks are unequal i.e. if we take  $\mathbb{X}_{Q_1} \neq \mathbb{X}_{Q_2} \neq \mathbb{X}_{Q_3}$ then the Yukawa matrices $Y^u$ and $Y^d$ and hence the resulting mass matrices will only have diagonal entries and we will not be able to generate CKM mixing. Thus, the charges of some (but not all) generations should match with each other in order to allow the generation of quark mixing. The same is true for charged lepton mass matrices. Again we will elaborate it further with specific examples in the coming sections.

\item {\bf BSM fermions:} The $U(1)_X$ charge of right handed neutrinos is typically fixed by anomaly cancellation conditions as we will discuss in a later section with specific examples. 
As mentioned earlier, for the sake of simplicity, we will not consider $U(1)_X$ symmetries involving chiral fermions beyond the fermion content of Table~\ref{tab:u1x}. 
 
 \end{enumerate}

Based on the aforementioned assumptions one can write the Yukawa and scalar potential for the general $U(1)_X$ model as:
 \begin{eqnarray}
  \mathcal{L}_{Yuk} \, = \, y^u_{ij} \bar{Q}_i \tilde{\Phi} u_j \, + \, y^d_{ij} \bar{Q}_i \Phi d_j
  \, + \, y^e_{ij} \bar{L}_i \Phi \ell_j \, + \, \mathcal{L}_{\nu} + h.c.
  \label{eq:yuk}
 \end{eqnarray}
and
\begin{eqnarray}
 V(\Phi,\sigma) \, = \, \mu^2_\Phi \Phi^\dagger \Phi \, + \, \mu^2_\sigma \sigma^\dagger \sigma \, + \, \lambda_{\Phi} \Phi^\dagger \Phi \Phi^\dagger \Phi \, + \, \lambda_{\sigma} \sigma^\dagger \sigma \sigma^\dagger \sigma \, + \, \lambda_{\Phi\sigma} \Phi^\dagger \Phi \sigma^\dagger \sigma~~.
 \label{eq:pot}
\end{eqnarray}
 
 It is worth highlighting a couple of salient features of the Yukawa terms and the scalar potential of this scenario.

\begin{enumerate}

\item  In eq.\eqref{eq:yuk}, $\mathcal{L}_{\nu}$ refers to the terms needed for light active neutrino mass generation. These terms depend on the details of the $U(1)_X$ symmetry, the charges of leptons under it as well as the nature of neutrinos (Dirac or Majorana) and the mechanism involved for mass generation. Furthermore, this typically requires the presence of additional scalars or fermions or both, beyond the particle content listed in Tab.~\ref{tab:u1x}. 
Note that even in the massless limit and absence of any other interactions, the $\nu_R$ is still interacting with the rest of the particles through its $U(1)_X$ gauge interactions and
depending on the charges and strength of the $U(1)_X$ gauge coupling ($g_X$), they can be in thermal equilibrium with the rest of the plasma at a given epoch in the evolution phase of the early universe.

\item Since we want a very light $Z'$, therefore the vev of $\sigma$ field $\langle \sigma \rangle = v_\sigma$
 which is responsible for $Z^\prime$ mass generation should be small. Furthermore, after SSB, if we want the real physical scalar $\sigma_R$ to be light as well, we should have $\lambda_{\Phi \sigma} \ll 1$. The condition $\lambda_{\Phi \sigma} \ll 1$ will also imply that the 125 GeV scalar ($h$) is primarily composed of the real part of $\Phi^0$ (a neutral component of $\Phi$) and hence the LHC constraints on it can be trivially satisfied. 
 
 \end{enumerate}
 
 {Throughout this work we also assumed that the tree level gauge kinetic mixing between $U(1)_Y$ and $U(1)_X$ symmetries is negligible. Hence for a specific $U(1)_X$ model, we have only two BSM free parameters, the gauge coupling ($g_X$) and mass ($M_{Z'}$) of the $Z'$ gauge boson.}
With these general assumptions, one can examine the potential range of a $Z'$ parameter that may account for the observed excess of $N_{\rm eff}$ in the Planck 2018 data, leaving the charge of the leptons and the gauge coupling as free parameters. 

%

\section{$\nu_L$ decoupling in presence of light $Z'$ and ${ N_{\rm eff}}$}
\label{sec:neff}
It is a well known fact that
one of the most precisely measured quantities from cosmology is the effective neutrino degrees of freedom ($N_{\rm eff}$), which may get altered by the presence of light BSM particles.
This change in $N_{\rm eff}$ from the SM value can in principle provide one of the solutions to relax the Hubble tension \cite{Bernal:2016gxb}.
 In this work, we focus on a scenario with light $Z'$ interacting with SM neutrinos ($\nu_i~,(i\equiv e,\mu,\tau)$) that may lead to change in $N_{\rm{eff}}$.
After having a brief description of generic features of light $Z'$ models emerging from $U(1)_{X}$ symmetries in the previous section, we now move to the most crucial part of our paper which is the cosmological implications of those scenarios.
Before delving into the analysis of $N_{\rm eff}$ in the presence of such light $Z'$, we would like to mention some key aspects of $N_{\rm eff}$ within the SM framework.

In a standard cosmological scenario at temperature $T\sim 20$ MeV \footnote{ By the temperature of the universe ($T$), we mean it as the temperature ($T_\gamma$) of the photon bath. }, only  $e^\pm, \nu_i$ are particles coupled to thermal (photon) bath as the energy densities of other heavier SM particles are already Boltzmann suppressed.
 As the universe cools, once the weak interactions involving $e^{\pm}$ and $\nu_i$ drop below the Hubble expansion rate $H(T)$, neutrinos decouple from the photon bath.
Considering only the SM weak interactions, the neutrino decoupling\footnote{by $\nu_L$  we refer the whole SM neutrinos ($\nu_i$)} temperature turns out to be $2$ MeV \cite{Hannestad:2001iy,Dodelson:2003ft}. 
After that, there exist two separate baths of photon ($+e^\pm$) and $\nu_i$, each with different temperatures; $T_\gamma$ and $T_\nu$ respectively.  
Approximately, at a temperature below $T_{\gamma}\lesssim 0.5$ MeV, $e^{\pm}$ annihilate, and the entropy is transferred entirely to photon bath leading
to a increment in  $T_{\gamma}$ compared to neutrino bath.
This difference in temperature is parameterised in terms of $N_{\rm eff}$ which is given by\cite{Dodelson:2003ft},

\begin{eqnarray}
    N_{\rm eff}= \frac{8}{7} \left(\frac{11}{4} \right)^{4/3} \left(\frac{\rho_{\nu_L}(T_\nu)}{\rho_\gamma(T_\gamma)} \right)=3\times\frac{8}{7} \left(\frac{11}{4} \right)^{4/3} \left(\frac{ T_\nu^4}{T_\gamma^4} \right)
    \label{eq:neff}
\end{eqnarray}
At the time of CMB formation, the predicted value of $N_{\rm eff}$ within the SM particle content is $N_{\rm eff}^{\rm CMB}=3.046$ \cite{Mangano:2005cc}, whereas the recent Planck 2018 data \cite{Planck:2018vyg} estimates it to be $N_{\rm eff}^{\rm CMB}=2.99^{+0.33}_{-0.34}$ at $95\%$ confidence level (C.L.). 

\begin{figure}
    \centering
    \subfigure[\label{p1}]{
    \includegraphics[scale=0.35]{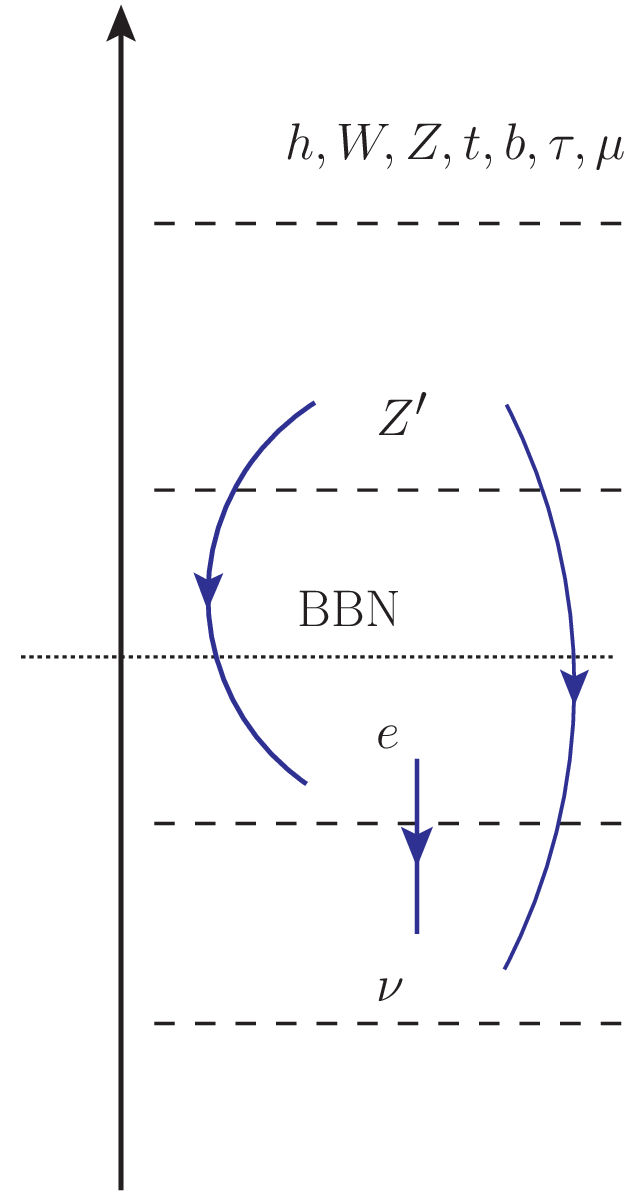}}~~~~~~~~~~~~~~
    \subfigure[\label{p2}]{
    \includegraphics[scale=0.37]{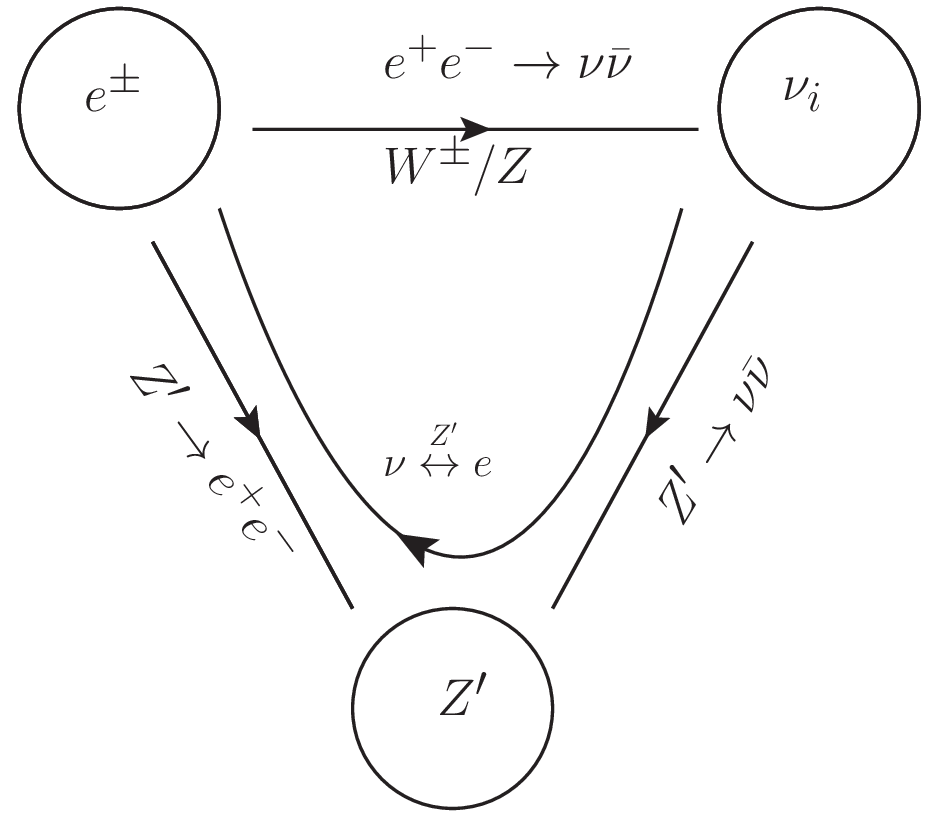}}
    \caption{Cartoon diagram of (a) particle spectrum and (b) interactions between three sectors. The vertical axis in (a) denotes the mass scale (or, temperature) and the dotted line indicates the temperature where BBN started. The blue lines in (a) signify the interactions between particles that affect $N_{\rm eff}$. Here $\nu$ in the figure signifies SM neutrinos ($\nu_i$) only.}
    \label{fig:cartoon1}
\end{figure}

 In this scenario, the  $Z'$ arising from the aforementioned $U(1)_{X}$ models introduces new interactions with both $\nu_i$ and $e^\pm$, which can potentially impact the neutrino decoupling consequently altering $N_{\rm eff}^{\rm CMB}$.
As pointed out previously, the SM particles relevant for $\nu_L$ decoupling are only $e^{\pm}$ and $\nu_i$,
whereas the energy densities of heavy SM particles are negligible due to Boltzmann suppression.
The light quarks also do not take part due to the QCD confinement at a much higher temperature around $\sim 150$ MeV.
{Hence, neutrino decoupling in presence of light $Z'$ is independent of the generation of baryons ($B$ number) gauged under the new $U(1)$ symmetry.}
Following the same argument used above, in BSM $U(1)_{X}$ scenario $Z'$ must be light enough ($M_{Z'}\lesssim \mathcal{O}(1)$ MeV) to affect $\nu_L$ decoupling which will be shown in the later part of this section. 
There are two other BSM particles present in our model: the BSM scalar $\sigma$ and the RHN $\nu_{R}$.
As we elaborated in sec.\ref{sec:model},
the BSM scalar $\sigma$ can be taken as heavy enough so that they are irrelevant for phenomenology at the MeV scale temperature. Hence, we integrate out $\sigma$ to perceive the sole effect of light $Z'$ in late time cosmology. 
In the Majorana type mass models, $\nu_{R}$ are also too heavy to affect $\nu_L$ decoupling \cite{Mohapatra:1979ia,Schechter:1980gr} and we can neglect them also.
However, in Dirac-type mass models, $\nu_{R}$ are relativistic at MeV temperature and can significantly alter $N_{\rm eff}$ \cite{Abazajian:2019oqj}.
In this work, we only consider heavy Majorana RHN and ignore their contribution to temperature evaluation.
In Fig.\ref{p1} we present the relevant mass scales by a schematic diagram.

So, in our proposed scenario, we have to trace the interactions between only three baths i.e. $e,\nu_L$ and $Z'$ to evaluate $T_\nu$ or $N_{\rm eff}$ (eq.\eqref{eq:neff}).
We describe the scenario using a cartoon diagram in Fig.\ref{p2}.
It is essential to take care of various energy transfers among these $ 3$ particles as they will play a key role in computing temperature evolution equations (see Appendix \ref{sec:apxA}). 
The energy transfer rates are dictated by the various collision processes and the distribution functions of respective particles \cite{Escudero:2018mvt,Fradette:2018hhl,Kreisch:2019yzn}.
Here we enlist the relevant processes to consider for the successful evaluation of $N_{\rm eff}$.
\begin{enumerate}
    \item {\bf SM contributions:} The SM weak interactions are active at temperature $T \sim$ MeV. At this 
    point, active neutrino annihilations ($\nu_i\Bar{\nu}_{i}\leftrightarrow e^{+}e^{-}$) as well as elastic scatterings ($\nu_i e^\pm\leftrightarrow \nu_i e^\pm$) mediated by SM $Z$ or $ W^{\pm}$ take place to maintain the required thermal 
    equilibrium of the early universe.

    \item {\bf BSM contributions to $\gamma$ bath:}  For a light $Z'$ with mass $M_{Z'} \sim \mathcal{O}$(MeV) sufficient energy density can be pumped into the thermal bath via 
    the decay and inverse decay between $Z'$ and electrons ($Z'\leftrightarrow e^{+}e^{-}$) at temperature around 
    ${\cal O}({\rm MeV})$. Additional
    contributions to the thermal bath may in principle come from scattering processes like $Z'Z'\leftrightarrow e^{+}e^{-}$, $Z'\gamma\leftrightarrow e^{+}e^{-}$. However, it turns out that 
    for a very light $Z^\prime$, with mass $M_{Z^\prime} \sim {\cal O}$ MeV, its coupling $g_X$ with SM 
    fermions is highly constrained from various experimental data, $g_X \lesssim (10^{-3} -10^{-5})$. For such a small
    coupling, the aforementioned decay process of $Z^\prime$ significantly 
    dominates over the scattering processes. Thus we ignore the scattering contributions in our numerical calculations.

    \item {\bf BSM contributions to $\nu$ bath:} Similarly, $Z'$ can transfer energy to $\nu_i$ bath through decay and inverse decay ($Z'\leftrightarrow \nu_i\Bar{\nu}_{i})$ processes. Moreover, ($\nu_i\Bar{\nu}_{i} \leftrightarrow e^{+}e^{-}$) scattering process can play an important role through one loop coupling of $Z^\prime $ with electron. 
    This one-loop coupling of $Z^\prime$ with electrons is responsible for connecting two separate thermal baths containing
    $\nu_i$ and electrons. This feature can be seen in certain scenarios of the $U(1)_X$ models and we will elaborate
    on this issue in great detail in a later section. 

    \item {\bf Within $\nu$ bath:} In this case, if we assume that different $\nu_i$ flavours have different temperatures then $\nu_i\Bar{\nu}_{i}\leftrightarrow \nu_{j}\Bar{\nu}_{j}, ~(i \ne j) $ mediated by both $Z $ and $Z'$ will
    have significant impact on the overall $\nu$ thermal bath. We will address this point in detail in the last part of this section.
\end{enumerate}

To construct the temperature equations and compute the energy transfer rates we adopt the formalism already developed in ref.\cite{Escudero:2018mvt,EscuderoAbenza:2020cmq,Escudero:2019gzq,Luo:2020sho}.
It is worth highlighting the approximations made in the formalism prescribed in ref.\cite{Escudero:2018mvt} before the description of temperature equations.
Firstly, Maxwell Boltzmann distributions were considered to characterize the phase space distribution of all particles in equilibrium, aiding in simplifying the collision term integral. The use of the Fermi Dirac distribution in the collision term 
does not alter the energy transfer rates substantially \cite{Dolgov:1992wf,Biswas:2022fga}.
Additionally, the electron's mass was ignored to simplify the collision terms as non-zero electron mass would have 
resulted in a minimal modification of the energy transfer rate, typically less than a few percent \cite{Kawasaki:2000en}.
It is to be noted that the $\nu_L$ masses can be easily neglected as the 
relevant temperatures for neutrino decoupling is sufficiently higher than $\nu_L$-mass. 

After successfully demonstrating all relevant processes and stating the assumptions, we are now set to construct the temperature evolution equations.
The temperature equations are derived from the Liouville equation for phase space distribution of particles in a thermal bath (eq.\eqref{eq:liou}) and the collision terms take care of the energy transfers among involved particles through the processes discussed before \cite{Kolb:1990vq}. Following the detailed calculations of the temperature 
evolution for the SM and BSM scenarios as displayed in appendices \ref{sec:apxA} and \ref{sec:apxB}, here we quote the final 
results of aforementioned temperature evolution equations \cite{Escudero:2018mvt,Escudero:2019gzq}:
\begin{eqnarray}
\frac{dT_{\nu_L}}{dt} &=& -  \left( 4 H  \rho_{\nu_L}   -  \frac{ \delta \rho_{\nu_{L} \to e^{\pm}}}{\delta t}+  \frac{ \delta \rho_{Z'\to\nu_{L}}}{\delta t}    \right) \left(   \frac{\partial \rho_{\nu_{L}} }{ \partial T_{\nu_{L}} }    \right)^{-1} \label{eq:nux} \\
\frac{dT_{Z'}}{dt} &=& -  \left( 3 H \left( \rho_{Z'} + P_{Z'}\right) -  \frac{ \delta \rho_{Z'\to\nu_{L}}}{\delta t} -  \frac{ \delta \rho_{Z'\to e^{\pm}}}{\delta t}    \right) \left( \frac{\partial \rho_{Z'}}{ \partial T_{Z'}  }  \right)^{-1} 
\label{eq:Tnu_mu-tau} \\
\frac{dT_{\gamma}}{dt} & =&- \left(  4 H \rho_{\gamma} + 3 H \left( \rho_{e} + p_{e}\right) + \frac{ \delta \rho_{\nu_{L}\to e^{\pm}}}{\delta t} +  \frac{ \delta \rho_{Z'\to e^{\pm}}}{\delta t}  \right)\left(    \frac{\partial \rho_{\gamma}}{\partial T_\gamma} + \frac{\partial \rho_e}{\partial T_\gamma}       \right)^{-1}
   \label{eq:Tgamma},
\end{eqnarray}
where, $\rho_{r}, P_r$ and $T_r$ signify the energy density, pressure density, and temperature of species $r$. Terms like $\frac{\delta\rho_{a \to b}}{\delta t}$ indicate the energy transfer rate from bath $a$ to $b$ and is determined by integrating the collision terms (see eq.\eqref{eq:cf}).
The energy transfer rates are discussed in great detail in Appendix \ref{sec:apxB} \footnote{{ The BSM collision terms in sec.\ref{sec:apxB} have been simplified in the limit $\sqrt{s}\sim T\ll M_{Z'}$ which is a good approximation for $\nu_L$ decoupling. However, at high temperature ($T\gg M_{Z'}$) this simplification can not be used.}}. 
Here we assume all three $\nu_i$ generations share the same temperature $T_{\nu_L}$ \footnote{as for our analysis RHNs are not relevant we will often denote $T_{\nu_L}$ as $T_{\nu}$} and $\rho_{\nu_L}$ is the summation over the energy densities of three generations of $\nu_i$. 

Note that, the equations in eq.(\ref{eq:nux}-\ref{eq:Tgamma}) are dependent on the thermal history of heavy $Z'$ when it remains in equilibrium 
with thermal bath in the early universe ($T_{Z'}\gtrsim M_{Z'}$) through its interaction with fermions ($f$). 
$Z'$ preserves its thermal equilibrium via decay, inverse decay ($Z'\leftrightarrow f \Bar{f}$) and also 
through scattering ($f \Bar{f} \leftrightarrow Z' Z'(\gamma) $) process. Thanks to processes like $f \Bar{f} \leftrightarrow Z' Z'(\gamma) $ the chemical potentials ($\mu_i(T)$) are suppressed and $Z'$ remain in chemical equilibrium with the SM bath.
The condition for the thermal equilibrium of $Z'$ incorporates a lower bound on $g_X (\gtrsim 10^{-9})$ and the lower bound may shift slightly depending on the specific choice of $U(1)_{X}$ charge and the value of $M_{Z'}$.
Alternatively, it is highly plausible that $Z'$ was initially not in thermal equilibrium in the early universe, but produced 
from other SM particles via freeze in process. In these scenarios, it is necessary to solve the coupled equations for $\mu_i(T)$ to compute $N_{\rm eff}$ \cite{Esseili:2023ldf}. 
This alternate scenario requires a detailed complementary study which will be reported elsewhere.

As mentioned above we affix to the simplest BSM scenario assuming $Z'$ in thermal equilibrium and set the initial condition for the set of equations eq.(\ref{eq:nux}-\ref{eq:Tgamma}) as $T_{\nu_L}=T_{Z'}=T_{\gamma}$ at $T_{\gamma}\gtrsim M_{Z'}$ for the reasons already discussed before.
In such case, one can further simplify the scenario assuming $T_{Z'}=T_{\nu_L}$ as $Z'$ remains coupled to $\nu$ bath for a longer time than with $\gamma$ bath \cite{Escudero:2019gzq}.
This makes the term ${ \delta \rho_{Z'\to\nu_{L}}}/{\delta t}=0$ and reduces the three equations in eq.\eqref{eq:nux}-\eqref{eq:Tgamma} into two.
However, the equations described in the above format are useful for generic scenarios.

So far, we have assumed that all three $\nu_L$ share common temperature $T_\nu$ as mentioned earlier. 
The approximation is valid as neutrino oscillations are active around MeV temperature leading to all three generations of neutrinos equilibriating with each other\cite{Akhmedov:1999uz,Hannestad:2001iy,Dolgov:2002wy,Dolgov:2002ab}. 
However one can also evaluate the temperatures assuming different temperatures $T_{\nu_i},(i=e,\mu,\tau)$ for all three generations, and the relevant equation reads as  (see eq.\eqref{eq:ultrapro1}),
\begin{eqnarray}
     \frac{dT_{\nu_i}}{dt}&=& -\left(  4 H \rho_{\nu_i} -\left(\frac{\delta \rho_{\nu_i\to e}}{\delta t}\right)_{\rm tot}-\sum_{j\neq i}\left(\frac{\delta \rho_{\nu_i \to \nu_j}}{\delta t}\right)_{\rm tot}+\frac{ \delta \rho_{Z'\to \nu_i}}{\delta t} \right)\left(\frac{\partial\rho_{\nu_i}}{\partial T_{\nu_i}}\right)^{-1} ~.
     \label{eq:nuy}
\end{eqnarray}     
Note that earlier eq.\eqref{eq:nux} differs from the one described in eq.\eqref{eq:nuy} as the later one also contains the energy transfer rate between different generations of $\nu_i$. 
However the value of $N_{\rm eff}$ does not change significantly ($\lesssim 10\%$) if one solves the temperature equation without considering different $T_{\nu_i}$ (see Appendix \ref{sec:apxC}) \cite{Escudero:2019gzq}.
For the same reason we stick to the simpler scenario assuming a common temperature of $\nu_L$ bath and solve eq.\eqref{eq:nux}-\eqref{eq:Tgamma} for numerical estimation of $N_{\rm eff}$ throughout this paper.

\begin{figure}[!tbh]
    \centering
    \subfigure[\label{a1}]{
    \includegraphics[scale=0.4]{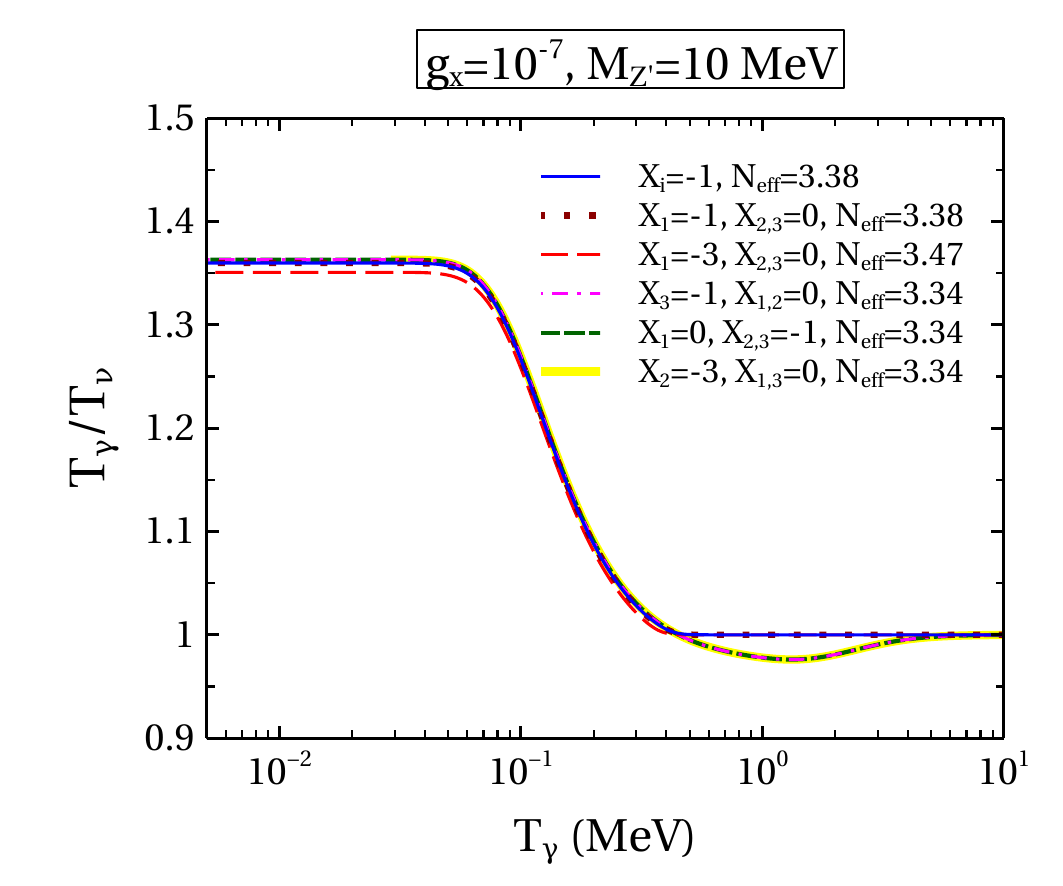}}
    \subfigure[\label{a2}]{
    \includegraphics[scale=0.4]{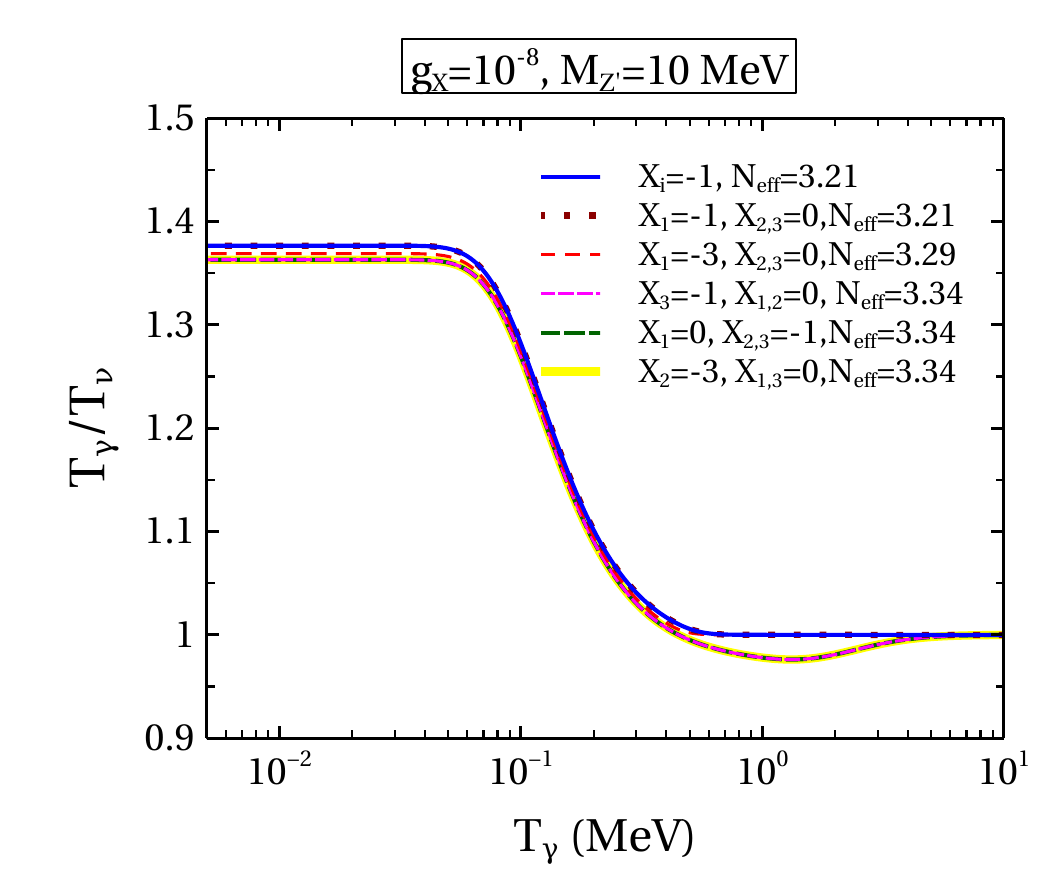}}
    \subfigure[\label{a3}]{
    \includegraphics[scale=0.4]{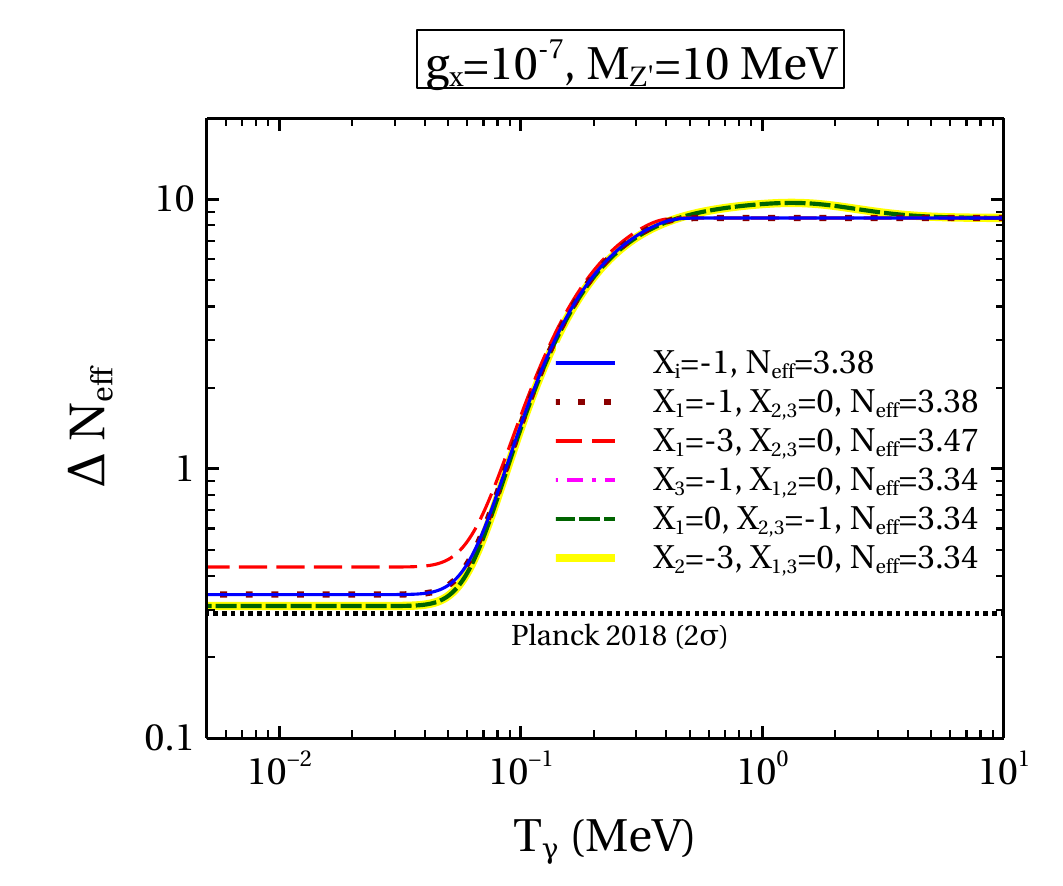}}
    \subfigure[\label{a4}]{
    \includegraphics[scale=0.4]{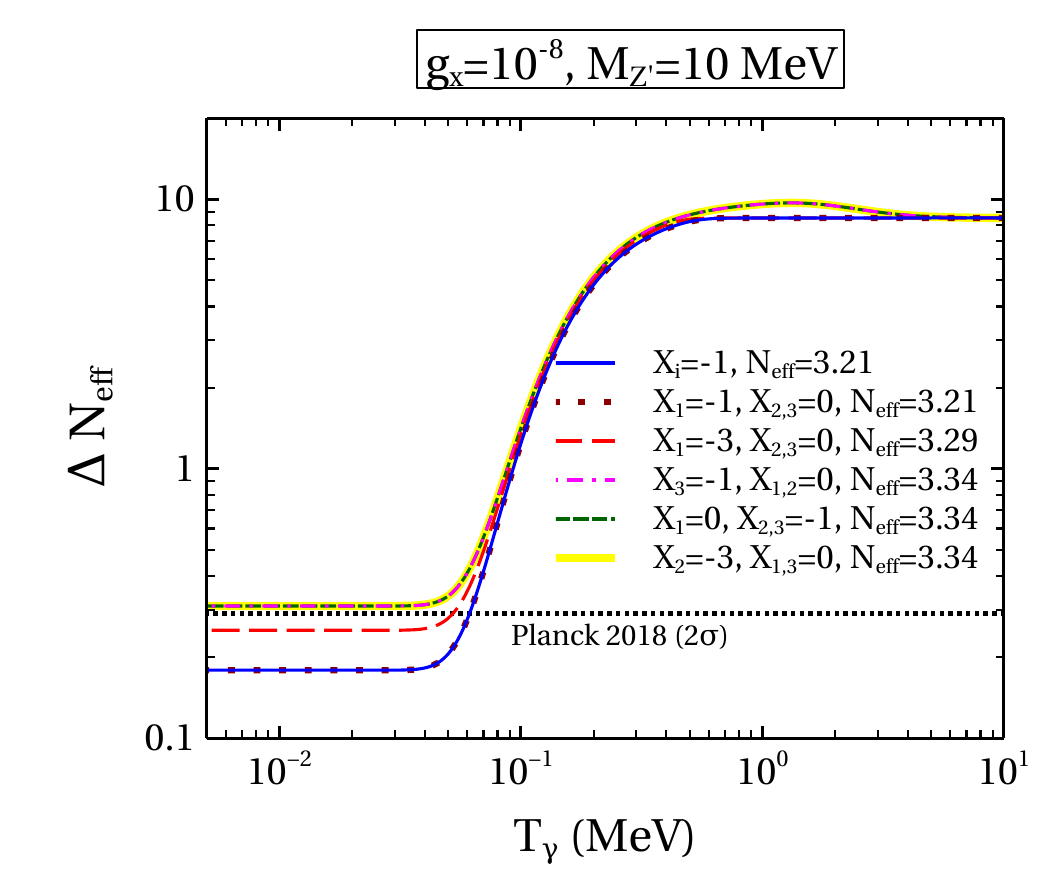}}
    \caption{Evolution of $T_{\gamma}/T_{\nu}$ with photon bath temperature $T_{\gamma}$  assuming all 3 $\nu_{L}$ share same temperature for different $U(1)_X$ charge combinations. We chose a benchmark parameter value $M_{Z'}=10$ MeV for both the plots. The couplings are taken $g_X=10^{-7}$ for (a)  and $g_X=10^{-8}$ for (b). The lines corresponding to different charge combinations are indicated by the plot legends in the figure. The legends named $N_{\rm eff}$ in the plots refer to $N_{\rm eff}^{\rm CMB}$. For this plot, we treat the $U(1)_X$ charges of leptons ($X_i$) as free parameters and for some specific values of such charges, it will lead to the popular $U(1)_X$ models.}
    \label{fig:1}
\end{figure}
After a detailed discussion of the basic framework, we are now set to perform an exhaustive numerical analysis.
In Fig.\ref{fig:1} we show case the evolution of temperature ratio ($T_{\gamma}/T_{\nu}$) as well as $\Delta N_{\rm eff}~(\equiv N_{\rm eff}^{\rm CMB}-3.046)$ with $T_{\gamma}$ for different $U(1)_{X}$ charge combinations.
Here we denote $T_\nu\equiv T_{\nu_i}$ with the assumption all three neutrinos share a common temperature.
Following the discussion in the previous paragraphs, we stress that the relevant charges for $\nu_L$ decoupling are $X_{L_i}= X_{\ell_i}\equiv X_i$ or more precisely their modulus values as their squared value will enter in the collision terms (see Appendix \ref{sec:apxB}). 
In the aforementioned plot, we present the simplest scenario where all 3 $\nu_{L}$ share the same temperature.
We consider  benchmark parameter (BP) values $M_{Z'}=10$ MeV with $g_X=10^{-7}$ (Fig.\ref{a1} \& \ref{a3}) and $g_X=10^{-8}$ (Fig.\ref{a2} \&\ref{a4}).
We will justify the importance of such a light $Z'$ in this process at the end of this section.
For such a light $M_{Z^\prime}$ and $g_X =10^{-7}$, we show the variation of  $T_{\gamma}/T_{\nu}$ with $T_\gamma$ 
and $\Delta N_{\rm eff}$ with $T_\gamma$ in Fig.\ref{a1} and Fig.\ref{a3} respectively.
From the above figure, it is very clear that around $T_\gamma\gtrsim 10$MeV,
$T_{\gamma}/T_{\nu}=1$ as both $\nu_L$ and $Z'$ was coupled to photon bath at that time. 
However, the ratio starts to increase after $T_\gamma \sim 0.5$ MeV and then saturates at low temperature ($T_\gamma\sim 10^{-2}$ MeV) as almost all the processes mentioned earlier gradually become inefficient at low $T_\gamma$.
In Fig.\ref{a3} we portray the corresponding variation in $\Delta N_{\rm eff}$ for all the $U(1)_{X}$ charge combinations in Fig.\ref{a1}. 
As around $T_\gamma \sim 10^{-2}$ MeV the values of $\Delta N_{\rm eff}$ become saturated we can surmise that it will remain unchanged till recombination epoch ($T_\gamma\sim 0.1$ eV) and say $\Delta N_{\rm eff} 
  (T_\gamma  \sim 10^{-2}{\rm MeV})\equiv\Delta N_{\rm eff}^{\rm CMB}$.    
In a similar way, for $g_X=10^{-8}$ also we display the evolution of $T_{\gamma}/T_{\nu}$ and $\Delta N_{\rm eff}$ in Fig.\ref{a2} and Fig.\ref{a4} respectively.
In Fig.\ref{a3} and Fig.\ref{a4} we also showcase the $2\sigma$ exclusion limit $\Delta N_{\rm eff}^{\rm CMB}=0.28$ from Planck 2018 \cite{Planck:2018vyg} and shown in black dotted line.
Note that this limit is only valid at CMB and it excludes any values of $\Delta N_{\rm eff}$ at that time above the black dotted line.
It is easy to infer that in the presence of the light $Z'$, the values differ from SM prediction.
Before spelling out the physical implications of the $U(1)_{X}$ scenario we tabulate our findings from Fig.\ref{fig:1} for the ease of understanding in Table\ref{tab:neff1}.
\begin{table}[!tbh]
  \centering
  \renewcommand{\arraystretch}{1.0}
  \begin{tabular}{|p{3cm}|c|c|c|p{4cm}|c|}
    \hline
    \multirow{2}{3cm}{$U(1)_\mathbb{X}$ coupling} & \multicolumn{3}{c|}{$U(1)_\mathbb{X}$ charge} & ~Notation &$N_{\rm eff}^{\rm CMB}$\\
    \cline{2-4}
                     & $|X_1|$ & $|X_2|$ & $|X_3|$&    & \\
    \hline
    \hline
                    & 1      & 1       & 1      & blue solid & 3.38 \\ 
                     \cline{2-2}\cline{3-3}\cline{4-4} \cline{5-5} \cline{6-6}
                       & 0      & 1       & 1      & green dashed & 3.34 \\
                      \cline{2-2}\cline{3-3}\cline{4-4} \cline{5-5} \cline{6-6}
    $g_{X}=10^{-7}$    & 1      & 0       & 0      & brown dotted & 3.38 \\
                     \cline{2-2}\cline{3-3}\cline{4-4} \cline{5-5} \cline{6-6}
                      & 0      & 1       & 0      & magenta dashed dot & 3.34 \\
                 \cline{2-2}\cline{3-3}\cline{4-4} \cline{5-5} \cline{6-6}
        Fig.2(a,c)        & 0      & 3       & 0      & thick yellow& 3.34 \\
                      \cline{2-2}\cline{3-3}\cline{4-4} \cline{5-5} \cline{6-6}
                      & 3      & 0       & 0      & red dashed& 3.47 \\
                      \cline{2-2}\cline{3-3}\cline{4-4} \cline{5-5} \cline{6-6}
    \hline
    \hline
                      & 1      & 1       & 1      & blue solid& 3.21 \\ 
                     \cline{2-2}\cline{3-3}\cline{4-4} \cline{5-5} \cline{6-6}
                       & 0      & 1       & 1      & green dashed& 3.34 \\
                      \cline{2-2}\cline{3-3}\cline{4-4} \cline{5-5} \cline{6-6}
     $g_{X}=10^{-8}$    & 1      & 0       & 0      &brown dotted& 3.21 \\
                    \cline{2-2}\cline{3-3}\cline{4-4} \cline{5-5} \cline{6-6}
                      & 0      & 1       & 0      & magenta dashed dot& 3.34 \\
                    \cline{2-2}\cline{3-3}\cline{4-4} \cline{5-5} \cline{6-6}
        Fig.2(b,d)   & 0      & 3       & 0      &thick yellow& 3.34 \\
                      \cline{2-2}\cline{3-3}\cline{4-4} \cline{5-5} \cline{6-6}
                      & 3      & 0       & 0      & red dashed& 3.29 \\
    \hline                   
  \end{tabular}
  \caption{Values of $ N^{\rm CMB}_{\rm eff}$ for different $U(1)_{X}$ charges inferred from Fig.\ref{fig:1}. As mentioned earlier, here we treat the $U(1)_X$ charges of leptons ($X_i$) as free parameters and for some specific values of such charges it will lead to the popular $U(1)_X$ models.}
   \label{tab:neff1}
\end{table}

Both from Fig.\ref{fig:1} and Table\ref{tab:neff1} it is evident that in our proposed $U(1)_{X}$ scenario the value of temperature ratio or $N^{\rm CMB}_{\rm eff}$ differs 
from the values predicted by SM only. 
One can interpret this feature from the new processes involved in $\nu_L$ decoupling apart from the SM weak interactions (see Fig.\ref{fig:cartoon1}).
Thus the light $Z'$ acts as the bridge between photon and $\nu_L$ bath and tries to balance their energy densities (through decays and scatterings) and hence reduces the temperature ratio from the value predicted by the SM only.
From both Fig.\ref{a1} and Fig.\ref{a2} we notice that for a fixed value of $g_X$ and $X_{L_1}\neq0$ the ratio $T_{\gamma}/T_{\nu}$ (at very late time) and $\Delta N^{\rm CMB}_{\rm eff}$ grow with an increase in $|X_{1}|$.
For $X_{1}\neq 0$ the following BSM processes affect $\nu_L$ decoupling:
(i) $Z'$ decaying to both $e^{+}e^{-}$ and $ \nu_i \Bar{\nu_i}$ and 
(ii) scattering process $\nu_{i}\Bar{\nu_{i}} \to e^+e^-$ mediated by $Z'$. 
Thus with an increase in $|X_{1}|$, the effective coupling ($X_1 g_X$) governing these BSM processes 
increases and hence boosts the BSM contribution (see eq.\eqref{eq:intx} and eq.\eqref{eq:BSMx}).  
As a result, picking higher values of  $X_{1}$ leads to a higher interaction rate between $\nu$ bath and photon bath leading to an enhancement in $\Delta N^{\rm CMB}_{\rm eff}$ or, more precisely a diminution in $T_\gamma/T_\nu$. 
%

On the contrary when $X_{1}=0$ the only BSM process relevant for   
$\nu_L$ decoupling is $Z'\leftrightarrow \nu_{\mu,\tau} \Bar{\nu_{\mu,\tau}}$ as there is no tree level coupling of $Z'$ with electrons.
At $T_\nu < M_{Z'}$ eventually all $Z'$ decay to $\nu_L$ transferring all their energy density to $\nu$ bath only.
As all the equilibrium number density of $Z'$ finally gets diluted to $\nu_L$ bath (with $100\%$ branching ratio) it does not depend on the coupling strength ($X_{2/3}g_X$).
There is no change in $N_{\rm eff}$ with the change in charge assignments ($X_2,X_3$) for $X_{1}=0$.
For the same reason described above, we infer that for $X_{1}\neq0$, $N_{\rm eff}$ increases with an increase in $g_X$ whereas with
$X_{1}=0$ it does not change at all with change in $g_X$ (comparing Fig.\ref{a3} and Fig.\ref{a4} ).
Due to the fact that for $X_{1}=0$, $Z'$ has only decay mode to $\nu_{\mu,\tau}$,
$T_\nu$ starts to increase before $e^{\pm}$ decouples ($T_\gamma\sim 0.5$ MeV).
This causes a slight dip in the $T_{\gamma}/T_{\nu}$ evolution line 
at higher temperature for $X_{1}=0$ in Fig.\ref{a1} and Fig.\ref{a2}.

{The approximate behaviour of $N_{\rm eff}$ for the case with $X_1=0$ can also be understood by analytic estimates. As mentioned above, to compute the BSM effect in $\nu_L$ decoupling, we have to consider only the $Z'\to \nu \Bar{\nu}$ decay. Using the co-moving energy conservation rule in two different epochs. The first epoch (``1") is just after $\nu_L$ decoupling from SM bath and the second (``2") is at CMB formation when $Z'$ has completely decayed to $\nu_L$. For these two epochs, one can write,
\begin{equation}
    (\rho_{\nu_L}(a_1)+\rho_{Z'}(a_1))a_1^4=(\rho_{\nu_L}(a_2)) a_2^4
    \label{eq:cons1}
\end{equation}
where $a_1 (T_1)$ and $a_2 (T_2)$ signify the scale factors at the two epochs. We are using instantaneous $\nu_L$ decoupling at $T_1\approx2$ MeV which leads to slightly less value, similar to what happens the SM case as well (discrepancy of 0.046 \cite{Escudero:2018mvt}). At $T_2=2$ MeV $\nu_L$ decouples from photon bath. However, for simplicity we assume $Z'$ to follow its initial equilibrium distribution. Thus we can use their respective equilibrium energy densities\footnote{{The equilibrium energy densities are given as $\rho_{\nu_L}(T_i)=3\times2\times\frac{7}{8} \frac{\pi^2}{30}T_i^4$ and $\rho_{Z'}=\frac{3}{2\pi^2}\int_m^\infty dE \sqrt{E^2-m^2}/(e^{E}/T_i-1)$. At $T_i=2$ MeV $\rho_{\nu_L}=2.76\times 10^{-11}$ GeV$^4$ and $\rho_{Z'}=2.20\times 10^{-12}$ GeV$^4$ for $M_{Z'}=10$ MeV.}}. Now in SM scenario (absence of $Z'$) $\rho_{\nu_L}^{\rm SM}$ only red-shifts after decoupling and hence one can write,
\begin{equation}
   \rho_{\nu_L}^{\rm SM}(a_1) a_1^4= \rho_{\nu_L}^{\rm SM}(a_2) a_2^4 
   \label{eq:cons2}
\end{equation} 
Combining eq.\eqref{eq:cons1} and eq.\eqref{eq:cons2} one can write 
\begin{eqnarray}
    \left(\dfrac{\rho_{\nu_L}}{\rho_{\nu_L}^{\rm SM}}\right)_{\rm CMB}= \dfrac{\rho_{\nu_L}(a_1)+\rho_{Z'}(a_1)}{\rho_{\nu_L}(a_1)} 
    \label{eq:cons3}
\end{eqnarray} 
Using this eq.\eqref{eq:cons3} we can rewrite the earlier definition of $N_{\rm eff}$ in eq.\eqref{eq:neff} as \cite{Ghosh:2022fws},
\begin{eqnarray}
    N_{\rm eff}^{\rm BSM}= \left(\dfrac{\rho_{\nu_L}}{\rho_{\nu_L}^{\rm SM}}\right)_{\rm CMB} N_{\rm eff}^{\rm SM}= \dfrac{\rho_{\nu_L}(a_1)+\rho_{Z'}(a_1)}{\rho_{\nu_L}(a_1)} N_{\rm eff}^{\rm SM}
    \label{eq:ana}
\end{eqnarray}
From the above eq.\eqref{eq:ana} on notices that the reason behind the enhanced $N_{\rm eff}$ in $U(1)_X$ models is the additional $Z'$ energy density at an earlier epoch.
As hinted before, for $X_1=0$ scenario, only one decay mode of $Z'$ exists and thus eventually all the $Z'$ density gets transferred to $\nu_L$ bath irrespective of $g_X X_{2,3}$. Also, it is easier to understand that for $X_1=0$ scenario $M_{Z'}$ plays the most crucial role to decide its energy density at the decoupling epoch and hence also $N_{\rm eff}$.
Now for $M_{Z'}=10 $ MeV, we get from eq.\eqref{eq:ana}, $N_{\rm eff}=3.28$  in presence of $Z'$ which is close to the one obtained numerically in Table \ref{tab:neff1}. Even though the $Z'$ density is suppressed ($\exp(-M_{Z'}/T_1)$) at $T_1=2$ MeV in the vicinity of $\nu_L$ decoupling, it injects all the energy (via decays) to $\nu_L$ sector enhancing $N_{\rm eff}$. Following the same argument, decreasing $M_{Z'}$ will lead to more contribution to $N_{\rm eff}$, as will be explored in Sec.\ref{sec:class} \footnote{{However, it is worth highlighting that these analytic estimates are drawn based on two assumptions: (1) $Z'$ decays to $\nu_L$ only and not to $e^\pm$; and (2) $Z'$ is in equilibrium. Thus, these estimates do not work for the case when $X_1\neq0$.  They also fail if  the coupling $(g_X \ll 10^{-9})$ is very small and mass $(M_{Z'} \ll 1)$ MeV where equilibrium formalism cannot be used \cite{Escudero:2019gzq,Li:2023puz}.}}. }

Before concluding the discussion about Fig.\ref{fig:1}, we point out the fact that
we compute $N_{\rm eff}$ with a basic assumption that all $\nu_L$ share same temperature
and hence all $\nu_L$ equilibrate with each other even if $Z'$ decays (transfers energy) to any one of them.
For this reason we observe similar behaviour of $T_{\gamma}/T_{\nu}$ for all $U(1)_{X}$ charge combinations with same value of $X_{1}$ and 
the same phenomenology elaborated in the earlier paragraph accounts for that. 
One should note that the $U(1)_{X}$ charge combination used in Fig.\ref{fig:1} is not an exhaustive one, yet one can easily deduce the outcome of other combinations from the reasoning we made above. For example all charge combinations with $|X_1|=1$ and $|X_1|=0$ will lead to $ N_{\rm eff}^{\rm CMB}=3.38$ and $3.34$ respectively for $g_{X}=10^{-7}$,$M_{Z'}=10$ MeV.
This is the most pivotal point of our analysis and we will revisit this in the next section. At this juncture, it is
worth pointing out that the dependence of $N^{\rm CMB}_{\rm eff}$ on $g_X$ for $\mid X_i \mid = 0$, indicating the absence of tree-level $Z^\prime e^+e^- $ coupling. However, some effective coupling between $Z'$ and $e^\pm$ can be generated depending on the specific $U(1)_X$ model \cite{Escudero:2019gzq}.
In the subsequent section, we will extensively discuss such a scenario where the induced $Z^\prime e^+e^-$ coupling plays a crucial role in deciding $N^{\rm CMB}_{\rm eff}$. 
For the ease of our notation from now on whenever we say $N_{\rm eff}$, we refer to $N_{\rm eff}^{\rm CMB}$ only.  

\begin{figure}[h]
    \centering
    \includegraphics[scale=0.5]{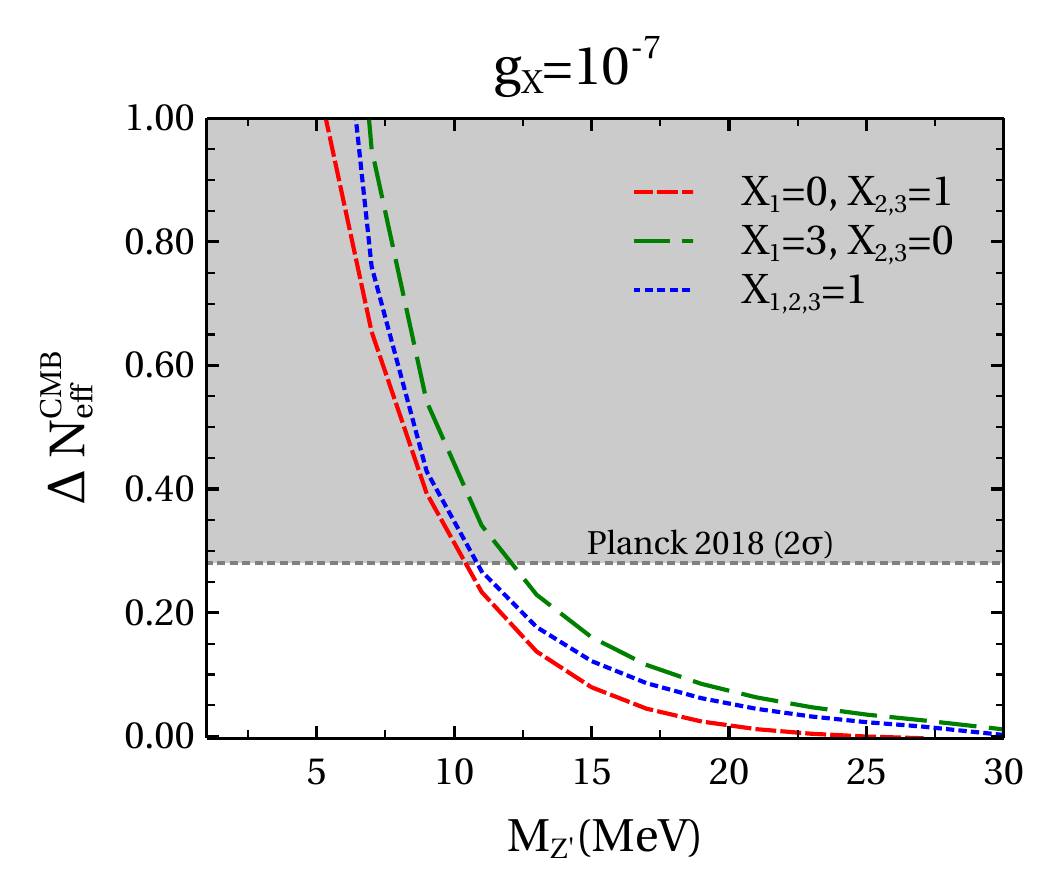}
    \caption{Variation of $\Delta N^{\rm CMB}_{\rm eff}$ with $M_{Z'}$  assuming all three $\nu_{L}$ share same temperature for different $U(1)_X$ charge combinations. We choose a fixed value of $ g_X=10^{-7}$.}
    \label{fig:mscan_P}
\end{figure}

Having discussed the dependence of $N_{\rm eff}$ on the two most important parameters of the BSM model: the $Z^\prime$ universal gauge coupling $g_X$ and $U(1)_{X}$ charge combinations, now we turn our attention 
to investigate its dependence on the light $M_{Z'}$.  
In Fig.\ref{fig:mscan_P} we show variation of $\Delta N_{\rm eff}^{\rm CMB}$ \footnote{To amplify the change in $N_{\rm eff}$  
with $M_{Z'}$ and portray more lucidly, for this particular plot we switch to 
$\Delta N_{\rm eff}^{\rm CMB}\equiv N_{\rm eff}-3.046$. The variation in $N_{\rm eff}$ will immediately follow from it.
}
with $M_{Z'}$ for a fixed coupling $g_X=10^{-7}$ with different $U(1)_{X}$ charge combinations. 
Rather than showing all the $U(1)_{X}$ charge combinations used in Fig.\ref{fig:1}, we just portray only three distinct combinations of them in Fig.\ref{fig:mscan_P}.
We indicate the combinations ($|X_{1,2,3}|=1$), ($|X_1|=0,|X_{2,3}|=1$), and ($|X_1|=3,|X_{2,3}|=0$) by blue, green and red lines respectively.
From the aforementioned figure, we observe that $\Delta N_{\rm eff}^{\rm CMB}$ decreases as $M_{Z^\prime}$ increases 
and eventually it becomes almost zero, reproducing the SM value of $N_{\rm eff}$ when $M_{Z^\prime} \gtrsim 30$ MeV.
This feature can be interpreted from our previous discussion in the context of Fig.\ref{fig:cartoon1}. 
The energy density of heavier $Z'$ gets Boltzmann suppressed at $\nu_L$ decoupling temperature ($T_\gamma\sim 2$ MeV) making the BSM contribution less significant in deciding  $\nu_L$ decoupling.
On the other hand $Z'$ mediated scattering processes between $\nu$ and $e$, also propagator suppressed for higher $M_{Z'}$.
At very high $M_{Z'}$ ($\gtrsim 30$ MeV), the BSM contribution hardly plays any role in deciding 
$\nu_L$ decoupling resulting $\Delta N_{\rm eff}^{\rm CMB}=0$.
Following the same argument we infer that a lower value of $M_{Z'}$ will enhance the BSM contribution and hence $\Delta N_{\rm eff}^{\rm CMB}$.
For a fixed value of $M_{Z'}$, the difference in $\Delta N_{\rm eff}^{\rm CMB}$ for different 
$U(1)_{X}$ charge combinations can be easily apprehended from the discussion in the context of Fig.\ref{fig:1}. 
So far we explored the dependence of $\Delta N_{\rm eff}^{\rm CMB}$ on various model parameters, and the dependence of $N_{\rm eff}$ also can be easily understood from that.
Keeping in mind the key findings from $U(1)_{X}$ $Z'$ models in the context of $N_{\rm eff}$, in the following section we will explore their contribution to alleviating the Hubble tension.

\section{Numerical results}
\label{sec:class}

\begin{figure}[!tbh]
    \centering
    \includegraphics[scale=0.4]{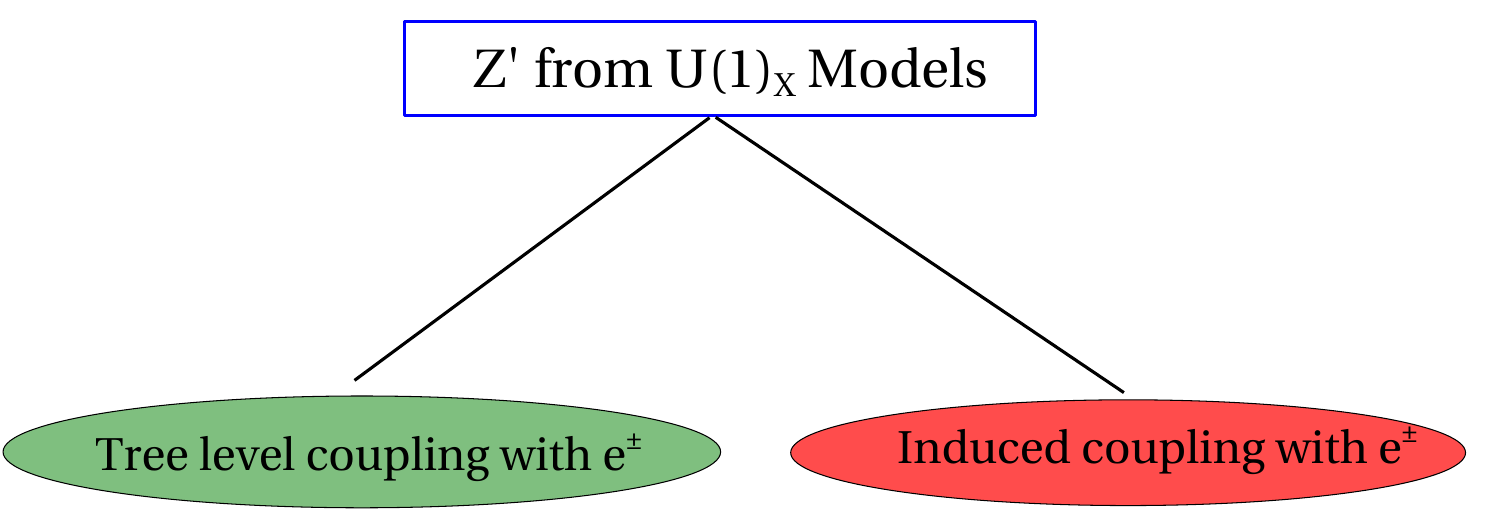}
    \caption{Classification of $Z'$ models depending on $X_1=0$ and $X_1\neq0$}
    \label{fig:class}
\end{figure}

In the previous section, we pinned down the key aspects of light $Z'$ from generic $U(1)_{X}$ extension
and showed that the $U(1)_{X}$ charge assignments play a key role 
in deciding $N_{\rm eff}$ as well as its dependence on coupling $g_X$.
In this section, we will take a closer look at the model parameters and explore their cosmological implication through exhaustive numerical scans.
Though one can have numerous $U(1)_{X}$ charge assignments as suggested in the Table \ref{tab:u1x}.,
in the context of $N_{\rm eff}$ the arbitrary charge assignments can be categorised into only two classes when we assume all $\nu_L$ share the same temperature (through oscillation).
Needless to say, it is the coupling of $Z'$ with electron (apart from $\nu_i$) that affects the $N_{\rm eff}$ and not the couplings with $\tau,\mu$ or quarks, for the reasons already discussed earlier.
Thus the light $Z'$ models can be broadly classified into two pictures: ${X}_{1}\neq0$ and ${X}_{1}= 0$, 
 more precisely, whether $Z'$ has coupling with electron or not (see Fig.\ref{fig:class}). 
 
In our discussions on the light $Z'$ phenomenology so far we have mainly focused on its tree level couplings with fermions. Yet, it is important to highlight that even in the absence of tree level $Z' e^+e^-$ interaction ($X_1=0$), $Z'$ can develop induced coupling\footnote{The source of the induced coupling is model dependent. It can originate from kinetic mixing, can be loop induced or can originate from flavor violating couplings. In this model independent section we will remain agnostic about its origin.}  with $e^\pm$ \cite{Amaral:2020tga}.
As a result, for $\mid X_1\mid = 0$, while computing $N_{\rm eff}$, we need to consider the following effective $Z^\prime e^+e^-$ interaction Lagrangian:
 \begin{equation}
     \mathcal{L}_{int}= (\epsilon e) \Bar{e} \gamma^\mu e Z'_\mu,
     \label{eq:mixL}
 \end{equation}
where $\epsilon$ is the induced effective coupling. If, for instance, the induced coupling is generated from the $\gamma -Z^\prime $ kinetic mixing at one loop level, its expression is given as  (see Appendix A of ref. \cite{Amaral:2020tga}).
\begin{equation}
    \epsilon = \frac{ g_X}{2\pi^2} \sum_{\ell} X_\ell \int_0^{1} dx~ x(1-x) \log\left( \frac{\Lambda^2}{\Delta_\ell} \right),
    \label{eq:mix}
 \end{equation}
where, $\Delta_\ell= m_\ell^2 -x(1-x)q^2$, $\Lambda$ denotes an arbitrary mass scale and the summation includes all $U(1)_{{X}}$ charged fermions with mass $m_\ell$. 
This induced coupling will have a significant impact in scenarios where $X_1=0$, as we'll shortly discover.
While we aim to maintain a model-independent discussion, the effective coupling described in eq.\eqref{eq:mixL} relies on the characteristics of particular $U(1)_{{X}}$ models. Therefore, to compare our findings with existing literature, we opt for the benchmark value of the effective coupling, setting $\epsilon=-\frac{g_X}{70}$, which is very commonly used for 
 $L_\mu - L_\tau$ models (with $X_1=0$)\cite{Banerjee:2021laz,Escudero:2019gzq}.
Henceforth, in all our numerical results, we will use this particular value of $\epsilon$.

\begin{figure}[!tbh]
    \centering
    \subfigure[\label{s1}]{
    \includegraphics[scale=0.4]{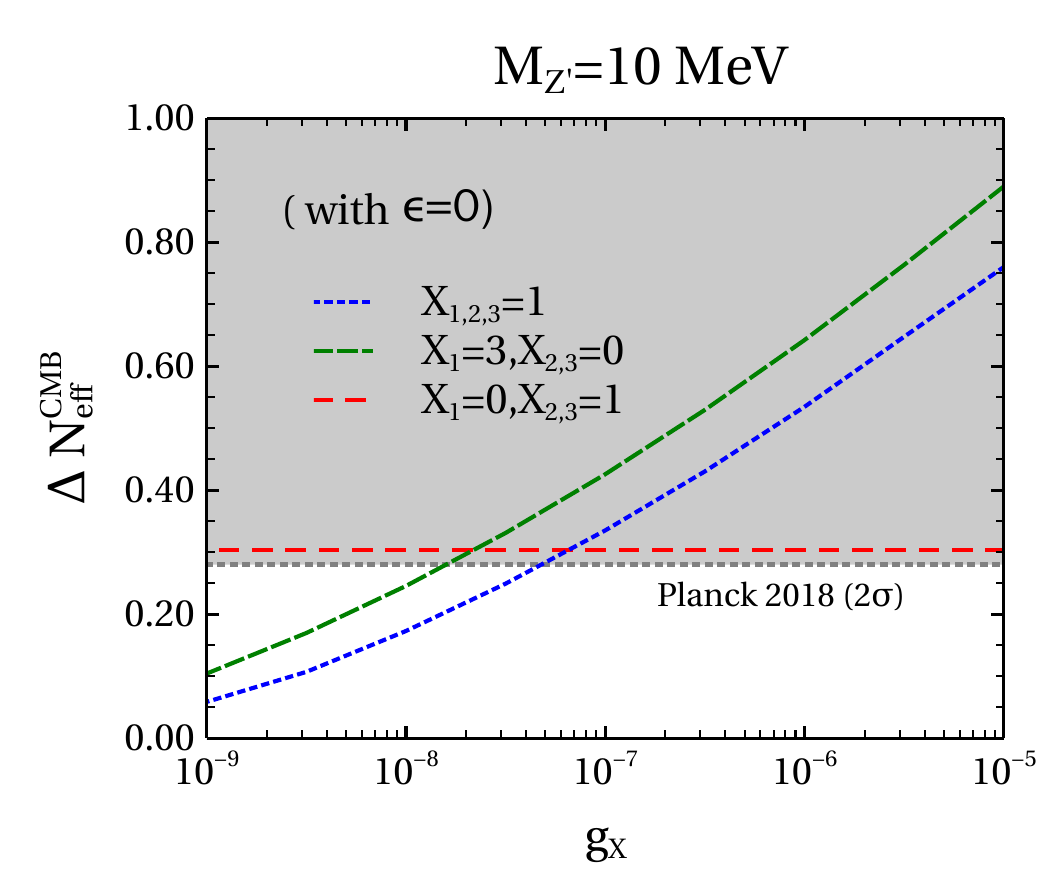}}
    \subfigure[\label{s2}]{
    \includegraphics[scale=0.4]{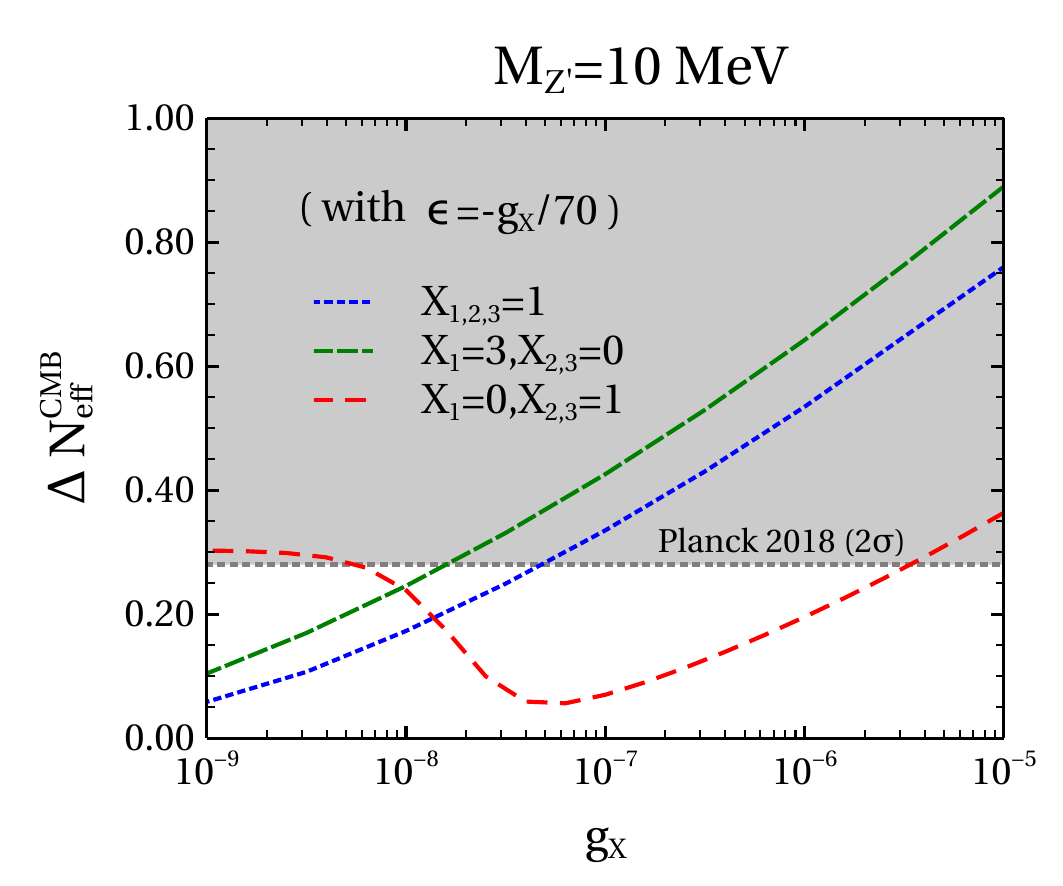}}
    \caption{Variation of $\Delta N_{\rm eff}^{\rm CMB}$ with $g_X$ in (a) with $\epsilon=0$ and in (b) with $\epsilon=-\frac{g_X}{70}$ assuming all 3 $\nu_{L}$ share same temperature for different $U(1)_X$ charge combinations. We choose a fixed value of $M_{Z'}=10$ MeV for both (a) and (b).}
    \label{fig:gscan}
\end{figure}
In Fig.\ref{fig:gscan}, we aim to investigate the role of $g_{X}$ on $\Delta N_{\rm eff}^{\rm CMB}$ while maintaining a constant $M_{Z'}=10$ MeV and as stated earlier the variation of $N_{\rm eff}$
also follows from it.
This scrutiny involves two distinct scenarios for the $Z' e^+e^-$ coupling: (a) solely with tree-level couplings ($\epsilon=0$) depicted in Fig.\ref{s1}, and (b) considering induced coupling as well ($\epsilon\neq 0$) shown in Fig.\ref{s2}.
We display the variation of $\Delta N_{\rm eff}^{\rm CMB}$ with $g_X$ for three $U(1)_{\mathbb{X}}$ charge  
combinations ($|X_{1,2,3}|=1$), ($|X_1|=0,|X_{2,3}|=1$) and ($|X_1|=3,|X_{2,3}|=0$) by blue, green and red  lines respectively.
In both Fig.\ref{s1} and Fig.\ref{s2}, the grey band corresponds to the value of $\Delta N^{\rm CMB}_{\rm eff}$
that is excluded by the $2\sigma $ upper limit obtained from the Planck 2018 measurement \cite{Planck:2018vyg}.
Note that the dependence of $\Delta N_{\rm eff}^{\rm CMB}$ (and hence $N_{\rm eff}$) on $g_X$ for the cases with $X_1\neq0$ remains the same even after including mixing.
This feature is easy to realize as the $Z' e^+e^-$ induced coupling ($\epsilon $) is suppressed by 
an order of magnitude compared to the corresponding $Z e^+e^-$ tree level interaction.
Hence, in the presence of the tree-level $Z^\prime e^+e^-$ interaction $(X_1 \ne 0)$, one can easily ignore
the contribution of $Z^\prime$ arising due to the induced coupling $\epsilon $.
For the reasons already discussed in sec.\ref{sec:neff}, the BSM contribution increases as the values of $g_{X}$ rise, leading to an increase in $\Delta N_{\rm eff}^{\rm CMB}$ (also $N_{\rm eff}$) also, as shown in the aforementioned figure.

However, comparing Fig.\ref{s1} and Fig.\ref{s2} one notices that the dependence $\Delta N_{\rm eff}^{\rm CMB}$ on $g_X$  for the case with $X_1=0$ changes drastically when the induced coupling $(\epsilon \ne 0)$ is present.
In Fig.\ref{s1}, we discern that $\Delta N_{\rm eff}^{\rm CMB}$ remains unchanged with variations in $g_X$ 
for $X_1=0$ in the absence of induced coupling ($\epsilon=0$). In this scenario, regardless of the
associated coupling, the $Z'$ decay is limited to $\nu_L$ exclusively, transferring its entire energy density, as reasoned in sec.\ref{sec:neff} \footnote{{See the discussion in context of eq.\eqref{eq:ana}.}}.
{The numerically obtained value of $N_{\rm eff}$ is also close to the analytical estimates presented in the previous section.}
This observation upholds our earlier analysis discussed in the preceding section. 

On the other hand, when $|X_1|=0$, in the presence of the induced coupling $(\epsilon \ne 0)$, 
$Z'$ can couple to $e^\pm$. Thus for a nonzero $\epsilon$ the decay $Z'\to e^+e^-$  and  $\nu_i\Bar{\nu_i} \to e^+e^-$ scattering processes mediated by $Z'$ continue during $\nu_L$ decoupling temperature ($T_\gamma\sim 1$ MeV). Among these two processes, the scattering process 
tries to balance $e$ and $\nu_L$ bath by increasing $T_\nu$ or increasing $N_{\rm eff}$.
Therefore, as $g_{X}$ increases, the contribution of BSM scenarios also increases, resulting in 
an overall rise in $\Delta N_{\rm eff}^{\rm CMB}$. This feature is reflected in Fig.\ref{s2} (red line) 
for $g_{X}\gtrsim 4 \times 10^{-8}$. On the other extreme, for $g_{X}\lesssim 4 \times 10^{-8}$, the scattering process ($\propto \epsilon^2 g_X^2$) is unable to compete with the tree level decay $Z'\to \nu_L\Bar{\nu_L}$ ($\propto g_X^2$) process. Hence, for lower values of $g_{X} (\lesssim 4 \times 10^{-8})$, $Z'$ promptly decays to $\nu$ bath transferring all its energy to $\nu$ sector and the scattering processes become inefficient to dilute this extra energy density to $e$ bath. So, when $g_{X} (\lesssim 4 \times 10^{-8})$ we see a distinct rise in $\Delta N_{\rm eff}^{\rm CMB}$ for $X_1=0$ in Fig.\ref{s2}.
When $g_{X} (\lesssim 7 \times 10^{-8})$ is very small, the BSM contribution that affects the 
evolution of the energy density of $\nu_L$ is primarily dominated by $Z'$ decay processes.
Irrespective of the coupling, at this level, the $Z'$ particle transfers all its energy density to $\nu_{L}$.
It is essential to note that at this point, the value of $\Delta N_{\rm eff}^{\rm CMB}$ for $X_1=0$ turns out to be 
identical for both scenarios with and without induced coupling when comparing Fig.\ref{s1} and Fig.\ref{s2}.
This outcome serves as validation for our previous argument. 
At this point, it is worth pointing out another aspect of this tree-level vs induced $Z'$ coupling in generating non-zero contributions of BSM physics to $\Delta N_{\rm eff}^{\rm CMB}$. When $g_X$ exceeds $10^{-8}$ and $X_1 = 0$, the other two 
 $U(1)_X$ charge combinations with non-zero $\mid X_1\mid$ result in the tree-level $Z' e^+e^-$ interaction yielding greater 
contributions to $\Delta N_{\rm eff}^{\rm CMB}$ compared to the aforementioned induced $Z'$ coupling. This distinction is particularly noticeable in Fig. \ref{s2}.

Now, we put forward the main thrust of this paper and split the analysis of light $Z'$ realized in different $U(1)_{{X}}$ models in the context of  $N_{\rm eff}$ into two categories as shown in Fig.\ref{fig:class}.
The consequences of all other $U(1)_{{X}}$ charge combinations can be easily anticipated from the broad classifications. 
Following this line of thought we will now perform numerical scans to explore the imprints in $N_{\rm eff}$ in the presence of light $Z'$ emerging from two broad classes of generic $U(1)_X$ models in the following two subsections.

\subsection{$Z'$ having tree level coupling with $e^{\pm}$}
\label{sub:x1}
\begin{figure}[!tbh]
    \centering
    \includegraphics[scale=0.4]{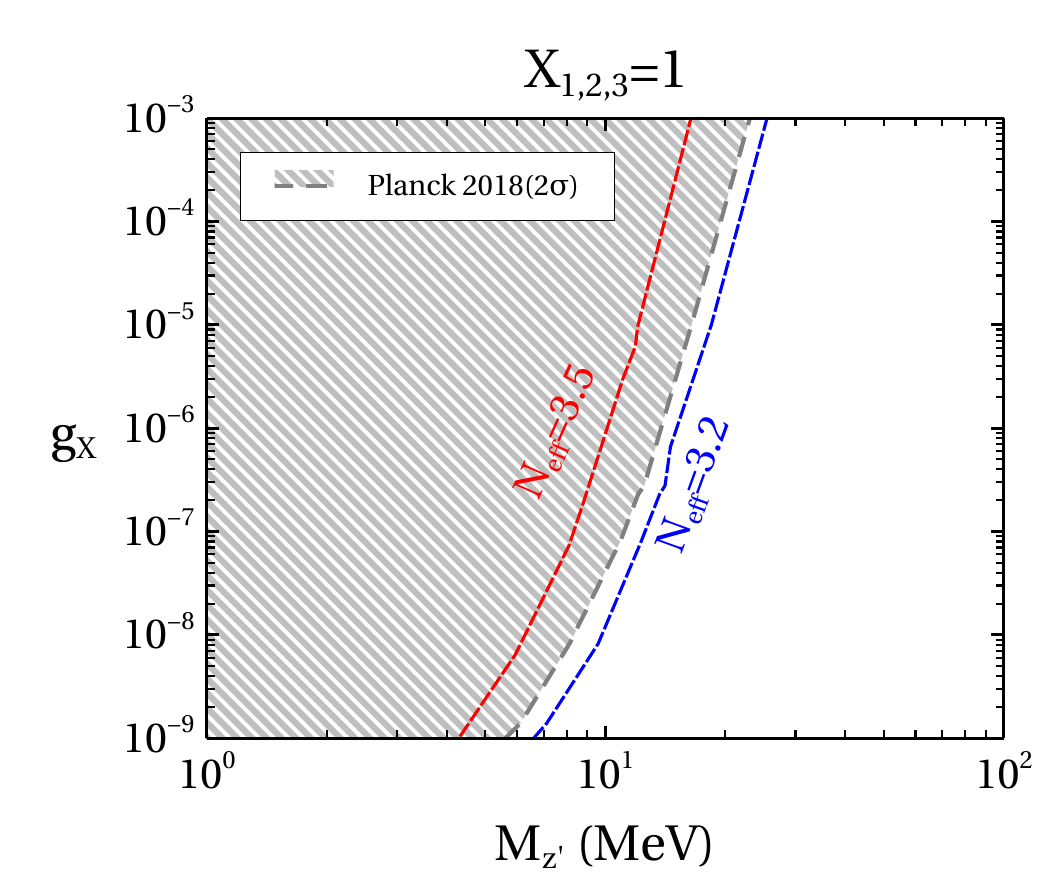}
    \caption{The parameter space for $N_{\rm eff}$ is depicted in the $M_{Z'}$ vs. $g_{X}$ plane for a generic $U(1)_{{X}}$ gauge extension with the charge assignment $|X_{1,2,3}|=1$. The 2$\sigma$ upper bound from Planck 2018 \cite{Planck:2018vyg} is depicted by the grey dashed line, excluding the parameter space to the left of that line shown by the grey region. The blue and red dashed lines indicate the values of $N_{\rm eff}^{\rm CMB}=3.2$ and $3.5$, respectively, where the $H_0$ tension can be relaxed as pointed out in ref.\cite{DiValentino:2021izs}.}
    \label{fig:case1}
\end{figure}

Here we consider the charge assignment  $|X_{1,2,3}|=1$ and show the contours of constant $N_{\rm eff}$\footnote{As stated earlier by $N_{\rm eff}$ we refer $N_{\rm eff}$ at the time of CMB.}  in $M_{Z'}$ vs. $g_{X}$ plane in Fig.\ref{fig:case1}.
In the same plot, we also showcase the $2\sigma$ bound from Planck 2018 data \cite{Planck:2018vyg} shown by the grey dashed line. The parameter space to the left of the grey dashed line, as shown by the grey region, is excluded by Planck 2028 data at $2\sigma$ \cite{Planck:2018vyg}.
From the figure we observe that for a fixed $g_{X}$, $N_{\rm eff}$ decreases with an increase in $M_{Z'}$   
as we explained in the context of Fig.\ref{fig:mscan_P}. 
For very high $M_{Z'}$ the contribution to $N_{\rm eff}$ becomes negligible.
We consider the lowest value of $g_{X}=10^{-9}$ as below that, $Z'$ fails to thermalize in the early universe.  
Following the discussion made earlier in this section, it is evident that increasing the value of $|X_1|$ will lead to a gradual shift of the contour for Planck 2018 upper limit (grey dashed line) towards right i.e. towards higher values of $M_{Z'}$.
Thus this bound applies to all $U(1)_X$ models with $|X_1|=1$ irrespective of other $U(1)_X$  charges. Note that for $|X_1|=n~(n\neq 1)$ these bounds do not apply, rather the bounds for such models should be evaluated explicitly as shown in sec.\ref{sec:examples}.

\subsection{$Z'$ having induced coupling with $e^{\pm}$}
\label{sub:x2}
\begin{figure}[!tbh]
    \centering
    \subfigure[\label{w1}]{
    \includegraphics[scale=0.4]{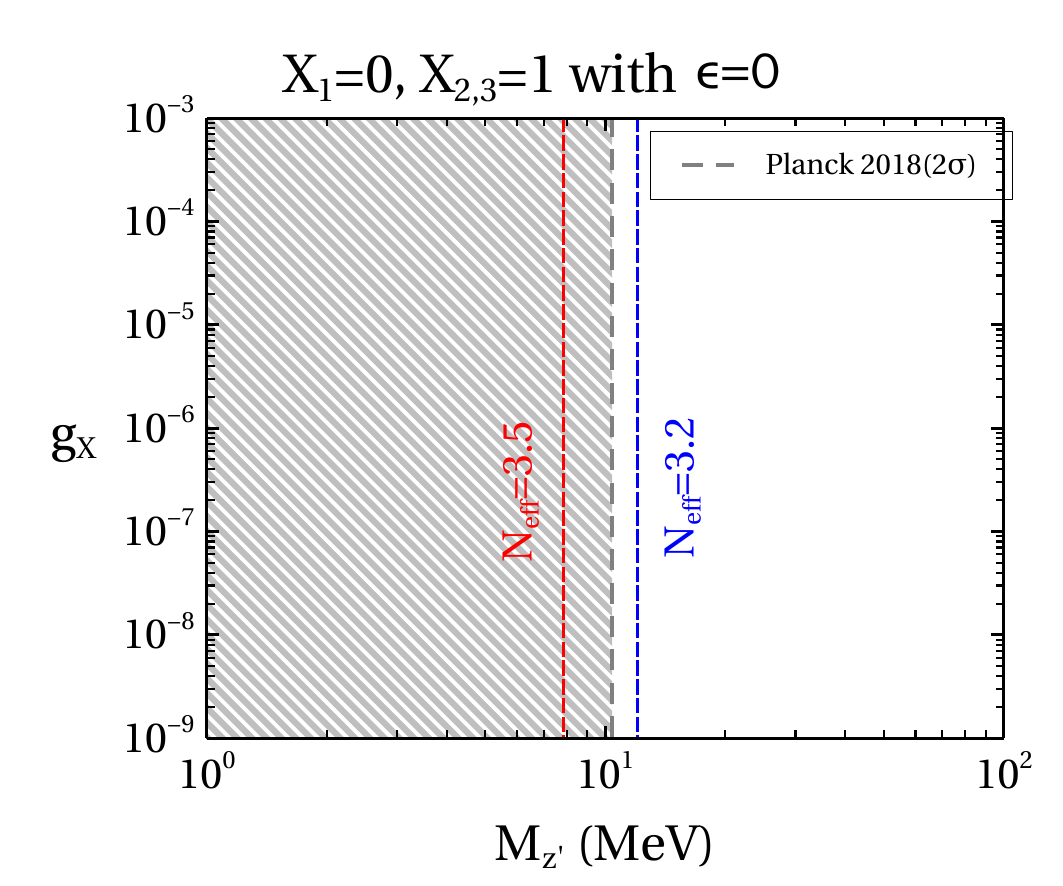}}
    \subfigure[\label{w2}]{
    \includegraphics[scale=0.4]{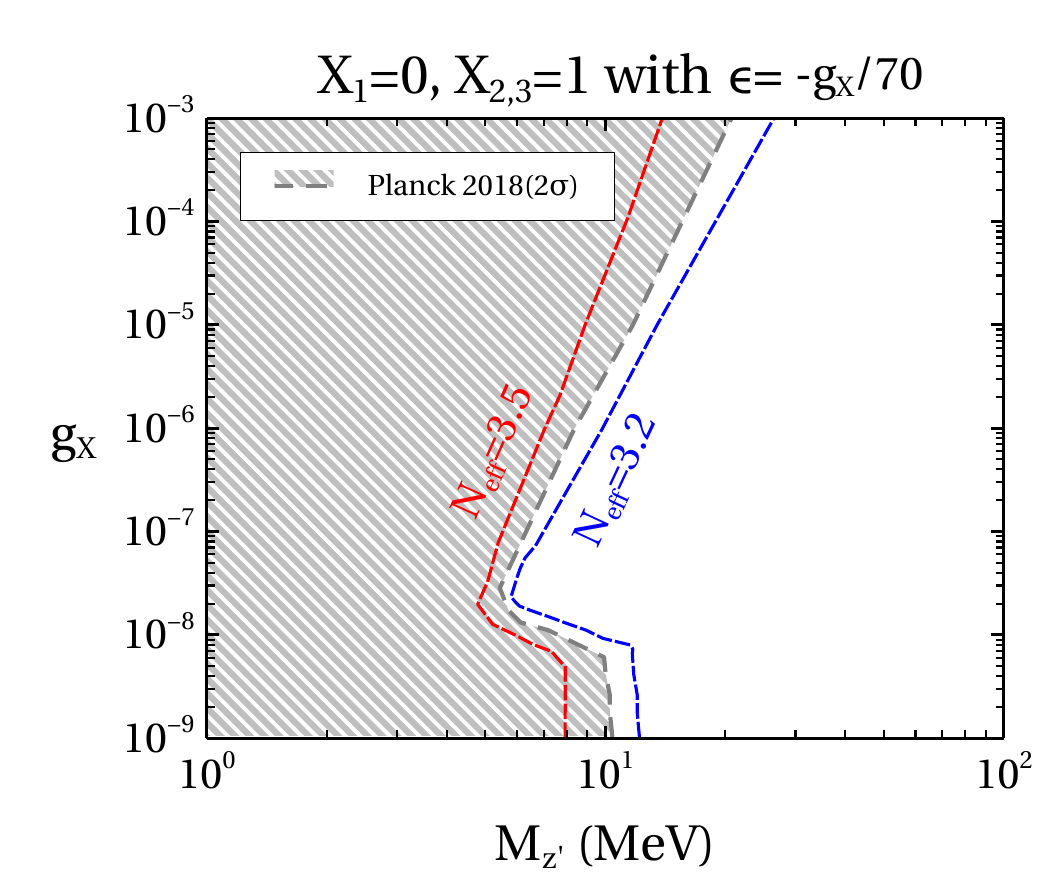}}
    \caption{Parameter space space for $N_{\rm eff}$ in $M_{Z'}$ vs. $g_{X}$ plane for a generic $U(1)_{{X}}$ gauge extension with the charge assignment $|X_1|=0,|X_{2,3}|=1$. We show the results without kinetic mixing ($\epsilon=0$) in (a) and with kinetic mixing ($\epsilon \neq 0$) in (b). The grey dashed line in each figure represents the 2$\sigma$ upper bound from Planck 2018 \cite{Planck:2018vyg}, excluding the parameter space to its left, as depicted by the grey region. The blue and red dashed lines in each figure denote $N_{\rm eff}^{\rm CMB}$ values at $3.2$ and $3.5$, respectively, indicating the intermediate regions where the $H_0$ tension can be relaxed \cite{DiValentino:2021izs}.}
    \label{fig:case2}
\end{figure}
Here we consider the following charge assignment  $|X_1|=0,|X_{2,3}|=1$ and show the contours of constant $N_{\rm eff}$ in $M_{Z'}$ vs. $g_{X}$ plane in Fig.\ref{fig:case2}. 
Similar to Fig.\ref{fig:case1} we portray the contour lines for
$N_{\rm eff}=3.2$, $N_{\rm eff}=3.5$ and $2\sigma$ upper limit from Planck 2018  \cite{Planck:2018vyg} depicted by blue, red and grey dashed lines. The grey region to the left of the grey dashed line in each figure is excluded by the 2$\sigma$ upper bound from Planck 2018 data \cite{Planck:2018vyg}.

\noindent We show the numerical results for the case $|X_1|=0$ without induced coupling ($\epsilon=0$) in Fig.\ref{w1}.
The dependence of $N_{\rm eff}$ on the model parameter $g_{X}$ is pretty straight forward as we noticed in earlier Fig.\ref{s1}.
For $|X_1|=0$, in the absence of induced coupling, $N_{\rm eff}$ stays  unchanged despite the variation in $g_{X}$,
resulting in distinct vertical lines of $N_{\rm eff}$ contours in $M_{Z'}-g_X$ plane in Fig.\ref{w1}.
Also, from the same figure we note a decrease in $N_{\rm eff}$ with an increase in $M_{Z'}$ for a fixed $g_{X}$, as elaborated earlier.
Note that, in the absence of induced coupling ($\epsilon=0 $)the contour for Planck 2018 upper limit (grey dashed line) will not change with increasing $X_{2,3}$ for such kind ($X_1=0$) of $U(1)_X$ models as argued before in the context of Fig.\ref{s1}. Following the discussions in context of eq.\eqref{eq:ana} we can analytically estimate that around $M_{Z'}=8$ and $M_{Z'}=12$ MeV the predicted values will be $N_{\rm eff}=3.48$ and $N_{\rm eff}=3.17$ respectively, which are very close to the numbers obtained numerically in Fig.\ref{w1} and are independent of $g_X$. This justifies our argument that for $X_1=0$, $N_{\rm eff}$ is only dependent on $M_{Z'}$ and not on $g_X$ for our chosen parameter space {\footnote{{Note that this argument does not hold for a very small coupling and mass i.e. $g_X\ll 10^{-9},\&~M_{Z'}\ll1$ MeV \cite{Li:2023puz}, where the equilibrium formulation can no longer be used.}}}.

 However, the results are quite different after including the induced coupling ($\epsilon\neq 0$) of $Z'$ with electrons as shown in Fig.\ref{w2}.
Due to the induced coupling of $Z'$ with electrons, the $N_{\rm eff}$ contours replicate the feature of the  $|X_{1}|=1$ case 
(as shown in Fig.\ref{fig:case1}) for higher values of $g_{X}~ (\gtrsim 10^{-8})$ with a knee like pattern around $g_{X}\sim 10^{-8}$.
Such non-trivial dependence is the consequence of the interplay between tree-level decay ( $Z'\to \nu_L \Bar{\nu_L}$) and the induced $Z'$ mediated scattering as explained earlier in detail.
We notice the bend in the contour lines in Fig.\ref{w2}
since the collision term accounting for the BSM contribution in computing $N_{\rm eff}$ is decay dominated in lower coupling region ($g_X\lesssim 4\times 10^{-8}$).
It is worth mentioning that for $\epsilon \neq 0$ the contour for Planck 2018 upper limit  (grey dashed line) will shift towards the right with increasing $X_{2,3}$ for $g_X\gtrsim 4\times 10^{-8}$ and will remain unchanged for $g_X\lesssim 4\times 10^{-8}$.
Again note that for all the models with $|X_1|=0$ and $|X_2|,|X_3|\neq 0$ this bounds apply  when one considers $\epsilon=0$. However, for $\epsilon \neq 0$ the bounds for such models with different $|X_2|,|X_3|$  should also be evaluated explicitly as shown in sec.\ref{sec:examples}.

Thus from this section, we propound that the cosmological imprints of light $Z'$ models due to different charge assignments leading to different 
$U(1)_{X}$ models can be put under the same roof following our prescription.
However, in the model independent analysis, we ignored the experimental constraints which are inevitably relevant for our parameter space.
For completeness, we will show the numerical results for some specific $U(1)_{{X}}$ models along with experimental constraints in the next section.

{We conclude this section with a comment on futuristic limits on $N_{\rm eff}$ by several proposed experiments.
We list a few such experiments and their expected limits on $N_{\rm eff}$ in Table \ref{tab:cons}. 
There also exists BBN bounds on $N_{\rm eff}$ which rule out $N_{\rm eff}>3.16$ at $1\sigma$ \cite{Fields:2019pfx}. 
However, applying BBN constraints requires dedicated analysis which is beyond the scope of this work. For our chosen parameter space the $Z'$ remains nonrelativistic at the onset of BBN and we show only CMB constraints.}
\renewcommand{\arraystretch}{1.4}
\begin{table}[tbh!]
\begin{center}
\begin{tabular}{ |c|c| } 
 \hline
 ~~Experiments~~ & ~Upper bound on $N_{\rm eff} (2\sigma)$~~ \\
 \hline
\hline
 ~~Planck 2018 \cite{Planck:2018vyg}~~ &~~ $N_{\rm eff}<3.33$~~ \\
 \hline
 ~~Simon Observatory \cite{SimonsObservatory:2018koc}~~  & $N_{\rm eff}<3.14$\\
 \hline
 CMB-S4 \cite{CMB-S4:2022ght}  & $N_{\rm eff}<3.10$\\
 \hline
 CMB-HD \cite{CMB-HD:2022bsz} &  $N_{\rm eff}<3.07$ \\
 \hline
\end{tabular}
\end{center}
\caption{{$2\sigma$ upper limit on $N_{\rm eff}$ at CMB from different experiments. Note that, all the bounds in the table indicate $N_{\rm eff}$ at CMB.}}
\label{tab:cons}
\end{table}
\renewcommand{\arraystretch}{1}



\section{Specific $U(1)_X$ Symmetries:$U(1)_{B_3-3L_j} $}
\label{sec:examples}


Let's now look at the type of $U(1)_X$ symmetry where the symmetry is flavour dependent in both quarks as well as the lepton sector namely the $U(1)_{B_3-3L_j}$ gauge symmetries~\cite{Bonilla:2017lsq}. In this case, we will consider the following three gauged $U(1)$ symmetries: $U(1)_{B_3-3L_e}$, $U(1)_{B_3-3L_\mu}$ and $U(1)_{B_3-3L_\tau}$.
The charges of three of such symmetries are listed in Table~\ref{tab:Bi-3Lj}. The charges of the other $U(1)_{B_3-3L_j}$ symmetries are analogous and can be similarly written without difficulty.

 \begin{table}[h]
\begin{center}
\begin{tabular}{| c || c | c | c | c |}
  \hline
Fields  &$SU(3)_c \times SU(2)_L \times U(1)_Y$& $U(1)_{B_3-3L_1}$  & $U(1)_{B_3-3L_2}$
& $U(1)_{B_3-3L_3}$    \\
\hline \hline
$Q_i$   & $(3, 2, \frac{1}{3})$                & $(0,0,1)$       & $(0,0,1)$      & $(0,0,1)$  \\   
$u_i$   & $(3, 1, \frac{4}{3})$                & $(0,0,1)$       & $(0,0,1)$      & $(0,0,1)$  \\ 
$d_i$   & $(3, 1, -\frac{2}{3})$               & $(0,0,1)$       & $(0,0,1)$      & $(0,0,1)$  \\
$L_i$   & $(1, 2, -1)$                         & $(-3,0,0)$      & $(0,-3,0)$     & $(0,0,-3)$  \\
$l_i$   & $(1, 1, -2)$                         & $(-3,0,0)$      & $(0,-3,0)$     & $(0,0,-3)$   \\
$\nu_{R_i}$ & $(1,1,0)$                        & $(x_1,x_2,x_3)$ & $(x_1,x_2,x_3)$& $(x_1,x_2,x_3)$\\
\hline \hline
$\Phi$   & $(1, 2, 1)$                         & $0$             & $0$            & $0$   \\
$\sigma$ & $(1, 1, 0)$                         &  $y$            &  $y$           &  $y$ \\         
\hline 
  \end{tabular}
\end{center}
\caption{Particle content and $U(1)_{B_3-3L_j}$ gauge charge assignments of Standard Model and new particles. The charges of $\nu_{R_i}$ can be fixed by anomaly cancellation condition while the charge of $\sigma$ depends on the details of the nature and model for neutrino mass generation. }
  \label{tab:Bi-3Lj}
\end{table}

For all three cases, the charges of $\nu_{R_i} $ can be fixed by the anomaly cancellation conditions. A convenient anomaly free charge assignment for  $\nu_{R_i}$ is $\nu_{R_i} = -3, \nu_{R_j} = 0$; $j \neq i$ for the  $U(1)_{B_i-3L_j}$ symmetry e.g. for say $U(1)_{B_3-3L_2}$ gauge symmetry the charges can be $\nu_{R} \sim (0 -3,0)$~\cite{Bonilla:2017lsq}. The charges of the $\sigma$ field depend on the details of the model and we will not go into details of the model building unless needed \footnote{see for example ref.\cite{Bonilla:2017lsq}}. And as we mentioned earlier in sec.\ref{sec:neff}, we assume this $ \nu_R$ and $\sigma$ to be heavy enough that they are irrelevant for the analysis of $N_{\rm eff}$.
As elaborated in the previous sections, the value of $N_{\rm eff}$ in the presence of a  light $Z'$ depends only on its leptonic couplings ($L$ number) and not on the couplings with quarks. Thus $U(1)_{B_2-3 L_1}$ and $U(1)_{B_3-3 L_1}$ models will exhibit similar imprints on $N_{\rm eff}$. However, the other experimental constraints indeed depend on the $Z'$ quark couplings and change significantly with the $B$ number.

Before exploring the phenomenology of the specific $U(1)_{X}$, we would like to outline an overview of the exclusion bounds on the mass of the light-gauge boson ($M_{Z'}$) and corresponding gauge coupling ($g_X$) within the mass range $M_{Z'}\sim \mathcal{O}(10)$ MeV, which is our point of interest. Here, we will briefly review various types of low-energy experimental observations that can be used to constrain the scenarios involving any $U(1)_X$. In the following subsections, we will present the exclusion bound identified by each low-energy experiment for a specific $U(1)_{B_3-3L_i}$ scenario ($i=e, \mu, \tau$), along with our cosmological findings.\\  

\noindent {\bf Elastic electron-neutrino scattering (E$\nu$ES):} 
The elastic scatterings of neutrinos with electrons ($\nu_\alpha~ e \to \nu_\alpha ~e,~\alpha=e,\mu,\tau$) in laboratory experiments serve as one of the probes for non-standard interaction of neutrino and electron with the light-gauged boson ($Z^\prime$). 
In SM, the $e-\nu_e$ scattering involves both charged-current (CC) and neutral-current (NC) weak interactions, whereas $e-\nu_{\mu,\tau}$ scattering is solely governed CC interaction \cite{Coloma:2022umy}. The elastic $e-\nu_{e,\mu,\tau}$ scattering can be altered in the presence of an additional NC interaction mediated by the $U(1)_X$ gauge boson $Z^\prime$, which can be probed in the low-energy scattering experiments. Note that the interference terms in the matrix amplitude between the SM (W and Z-mediated ) and BSM ($Z^\prime$-mediated), play a critical role in altering the scattering rate. 
The interference term in differential cross section is given by \cite{Amaral:2020tga},
\begin{eqnarray}\nonumber
    \left[\frac{d\sigma_{i-e}}{dE_R}\right]_{\rm Interference}&=& \frac{\sqrt{2}G_F m_e}{\pi} \left(\frac{(g_X X_1)(g_X X_i)}{(2 E_R m_e+M_{Z'}^2)} \right)\\
    &&\times \left[(g_L^i+g_R^i)\left(1-\frac{m_e E_R}{2 E_\nu^2}\right) -g_R^i\frac{E_R}{E_\nu}\left(2-\frac{E_R}{E_\nu}\right)\right],
\label{eq:recoil}    
\end{eqnarray}
where $E_R$ and $E_\nu$ denote electron recoil energy and incoming neutrino energy respectively. $g_{L,R}$ is defined in appendix \ref{sec:apxA}. For the cases where $X_1=0$ the term $(g_X X_1)$ will be replaced by $\epsilon e$ in eq.\eqref{eq:recoil}.
Thus observing the electron recoil rate imposes constraints on the $M_{Z^\prime} - g_X$ plane for a model-specific scenario.  The coupling strength of the light gauge boson with leptons varies over gauge extensions. As a result, the constraints will vary from model to model.

Borexino \cite{Shutt:2002rg} and dark matter experiments like XENON 1T \cite{XENON:2017lvq}  dedicated to measuring the electron recoil rate, can be relevant in the context of $e-\nu_\alpha$ elastic scattering \cite{Majumdar:2021vdw}. The Borexino experiment is designed to study the low-energy solar neutrinos ($\nu_e$)\footnote{E$\nu$ES event rates for atmospheric neutrinos are negligible compared to solar ones \cite{Majumdar:2021vdw}.} produced via decay of $^7$Be by observing the electron recoil rate through the neutrino-electron elastic scattering process \cite{Majumdar:2021vdw}. 
These solar neutrinos undergo flavour change as they travel from the sun to the detector. This flavor changing phenomena $\nu_e \to \nu_\alpha (\alpha=e,\mu,\tau)$ can be accounted using the transition probability $P_{e\alpha}$ \cite{Chakraborty:2021apc} . Therefore the number of events for solar neutrinos ($\nu_e$) interacting with the electrons in the Borexino will be $N_{\nu_e} \propto \sum_{\alpha=e,\mu,\tau} P_{e \alpha }~ \sigma_{e- \nu_{\alpha}}$ \cite{Chakraborty:2021apc,Coloma:2022umy}. 
Note that in Solar neutrino flux $P_{e e}(\sim 50\%)>P_{e \mu}=P_{e \tau}$ and the dominant contribution in interference term is due to $\nu_e$ as it has both CC and NC interaction with electron\cite{Coloma:2022umy}. Hence for $X_1\neq 0$ one can approximate the differential recoil rate \cite{Coloma:2022umy}, 
\begin{equation}
    \left[\frac{dR}{dE_R}\right]_{\rm BSM}\propto P_{ee} \left[\frac{d\sigma_{ee}}{dE_R}\right]_{\rm Interference}
\end{equation}
For this reason the event rate for E$\nu$ES in the presence of $Z'$ depends only on the $Z'e^+e^-$ and $Z'\nu_e\Bar{\nu}_e$ coupling ($X_1 g_X$). Thus the constraint on the parameter space of $Z'$ in $B_3-3L_e$ model will be analogous to $B-L$ model \cite{Majumdar:2021vdw}, except an overall scaling of $g_X$ due to the charge $X_1=3$
in the former one.
Note that when $X_1=0$ i.e. in the absence of tree-level coupling of $Z'$ with electron one has to consider the elastic scattering (via loop induced coupling with $Z'$) of an electron with $\nu_{\mu/\tau}$ with specific transition probabilities.
Thus the E$\nu$ES constraints for $B_3-3L_{\mu}$ model can be drawn from $L_\mu-L_\tau$ model \cite{Majumdar:2021vdw}, with the overall scaling as mentioned before.

{\bf Coherent elastic neutrino-nucleus scattering (CE$\nu$NS):} The COHERENT experiment investigates coherent elastic scattering between neutrinos and nucleus in CsI material ($\nu~N \to \nu~N$) \cite{Cadeddu:2020nbr,COHERENT:2021xmm,Banerjee:2021laz}.  This mode of interaction opens up new opportunities for studying neutrino interactions, including the introduction of a new light gauge boson in this case. 
In SM, the elastic neutrino-nucleus scattering ($\nu-N$) takes place via the NC interactions between neutrinos ($\nu_{e,\mu,\tau}$) and nucleons or more precisely with the first generation of quarks ($q=\{u,d\}$). The additional NC interaction introduced by the light gauge boson $Z^\prime$ of $U(1)_X$ can contribute to the CE$\nu$NS process. The modification to SM CE$\nu$NS, due to the $Z^\prime$ can be utilized to constrain the parameter space of $M_{Z^\prime}$ vs $g_X$ plane. Similar to E$\nu$ES, the exclusion limit is also dependent on the specific gauged scenarios. This is because the coupling strength of quarks and neutrinos with the light-gauged boson is influenced by the specific gauge choice. However, in contrast to E$\nu$ES 
experiment, CE$\nu$NS is lepton flavor independent and one has to consider interaction with all 3 $\nu$. For a detailed discussion see ref.\cite{Majumdar:2021vdw}.
For the $B_3-3L_{j}$ model there is no tree level coupling of $Z'$ with $u,d$, and hence the coupling with the nucleus will be an induced coupling (as discussed later).
The CE$\nu$NS constraints for $B_3-3L_{e}$ and $B_3-3L_{\mu}$ model can be deduced from $B-L$ and $L_\mu-L_\tau$ model \cite{Majumdar:2021vdw} respectively, with an overall scaling to take into account the induced coupling of $Z'$ with first generation quarks.

{\bf Supernova 1987A (SN1987A):} Non-standard neutrino interactions with a light gauge boson can also affect the cooling of core-collapse supernovae (SN) which is a powerful source of neutrinos ($\nu_e$) \cite{Kolb:1987qy,Akita:2022etk}. 
The non-standard neutrino interaction gives rise to the initial production of light-gauged bosons in the core of a supernova (SN), $\nu_\alpha ~\overline{\nu_\alpha} \to Z^\prime$, resulting in energy loss in the SN core \cite{Escudero:2019gzq}. After production the late decay of these $Z^\prime$ into neutrinos ($Z^\prime \to \nu_i ~\overline{\nu_i} $) has the potential to modify the neutrino flux emitted by the SN \cite{Akita:2022etk}. 
For a detailed study on supernova constraints on such light mediators see ref.\cite{Fiorillo:2022cdq,Fiorillo:2023ytr,Janka:2017vlw}.
As there is no nucleon coupling of $Z'$ in $B_3-3 L_j$ model, for simple order estimation the supernova cooling bound can be adopted from $L_\mu-L_\tau$ model \cite{Escudero:2019gzq}.
However, there can be additional processes relevant for SN 1987A cooling in the presence of light $Z^\prime $ models \cite{Croon:2020lrf} and explicit derivation of that is beyond the scope of this work. 
\\~\\
{\bf Neutrino trident:} The neutrino trident production mode, in which neutrinos interact with a target nucleus to produce a pair of charged leptons without changing the neutrino flavor ($\nu~N \to \nu~N~\mu^+~\mu^-$), is a powerful tool for probing new physics \cite{Altmannshofer:2014pba}. The gauge extension introduces new interactions between leptons and light gauge bosons, increasing the rate of neutrino trident production, as predicted by the SM. Therefore any gauge extended scenarios associated with muon ($X_2 \neq 0$) face constraints from existing neutrino trident production experimental results \cite{Bonilla:2017lsq}.   

After having a generic discussion on the constraints relevant to our analysis, in the following two subsections we present our numerical results for specific $U(1)_X$ gauge choice.
We portray the bound from $N_{\rm eff}$ on these models along with the constraints discussed above. In the same plane, we also indicate the parameter space that can relax the $H_0$ tension.

\subsection{Gauged $U(1)_{B_3-3 L_e}$ symmetry}
\label{sec:Le}
\begin{figure}[!tbh]
    \centering
    \includegraphics[scale=0.4]{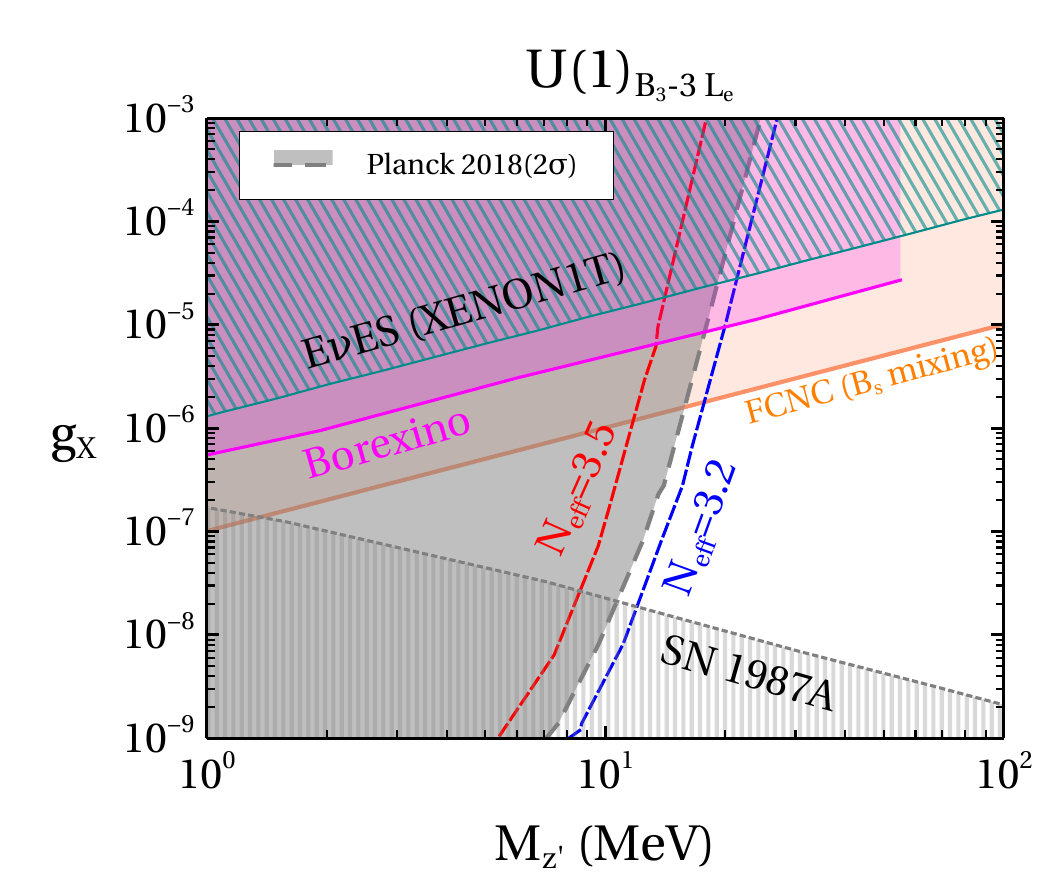}
    \caption{Constraints from cosmology and low energy experiments in $M_{Z'}$ vs. $g_{X}$ plane for a generic $U(1)_{B_3-3 L_e}$ gauge extension. The $2\sigma$ upper bound from Planck 2018 \cite{Planck:2018vyg} is shown by the grey dashed line and it excludes the parameter space to the left side of that line as shown by the grey shaded region. The region between the $N_{\rm eff}=3.2$ (blue dashed) and $N_{\rm eff}=3.5$ (red dashed) lines helps to relax the $H_0$ tension, as mentioned in the ref. \cite{DiValentino:2021izs}.}
    \label{fig:model1}
\end{figure}

In this subsection we show the values of $N_{\rm eff}$ for light $Z'$ in $U(1)_{B_3-3 L_e}$ gauge extension. We show our numerical results in $M_{Z'}$ vs. $g_{X}$ plane in Fig.\ref{fig:model1}. In the same plane, we also portray other relevant astrophysical and experimental constraints.
The grey dashed line in the same plot signify the $2\sigma$ upper limit on $N_{\rm eff}$
from Planck 2018 data \cite{Planck:2018vyg}. It excludes the parameter space to the left of that line as shown by the grey shaded region.
One can easily infer from the figure that for a fixed $g_X$ with increasing $M_{Z'}$ the value of $N_{\rm eff}$
decreases which is explained in detail in the previous section.
Note that for $U(1)_{B_3-3 L_e}$, $Z'$ has tree level coupling with $e^{\pm}$ ($X_1=-3$) and hence it belongs to the group of $U(1)_X$ models with $|X_1|\neq0$ following our discussion in previous section \ref{sub:x1}. 
For this, we notice that the behavior of $N_{\rm eff}$ with $g_X$ is similar to what we observed in the context of Fig.\ref{fig:case1}. 
However, for $U(1)_{B_3-3 L_e}$ model $|X_1|=3$ and hence for fixed  $g_X$ and $M_{Z'}$ the value of $N_{\rm eff}$ will be higher compared to the case considered in Fig.\ref{fig:case1}
for the reason already elaborated before. 
As a result we notice the $N_{\rm eff}$ lines in 
Fig. \ref{fig:model1} has moved rightwards compared to  Fig.\ref{fig:case1}.

In the same plane in Fig.\ref{fig:model1} we showcase the relevant phenomenological constraints that we mentioned earlier.
Since $Z'$ has tree level coupling with electron in $U(1)_{B_3-3 L_e}$ model,  it attracts strong constraint from E$\nu$ES \cite{Majumdar:2021vdw}.
The exclusion region from E$\nu$ES with XENON1T experiment is shown by the region shaded by dark cyan diagonal lines.
Similarly, Borexino also constrains the parameter space shown by the magenta shaded region \cite{Majumdar:2021vdw}. 
In $U(1)_{B_3-3 L_e}$ model the light $Z'$ has no tree-level coupling with the first two generations of quarks $(u,d),(c,s)$ which is important for CE$\nu$NS. However, $Z'$ can have induced coupling with first-generation quarks via CKM mixing \cite{Bonilla:2017lsq} and can contribute to CE$\nu$NS.  
In SM the CKM matrix $V_{\rm CKM}$ is constructed from charge current interaction betweeen quarks \cite{Schwartz:2014sze},
\begin{eqnarray}
    J^{\mu}_{W}&=&(\Bar{q_u})_L ~\mathcal{U}_L ~ \gamma^\mu ~\mathbb{I}_{3\times 3}~ \mathcal{D}_L^{\dagger} ~(q_d)_L
\end{eqnarray}
where, $q_u\equiv (u~ c~ t)^{\dagger}$ and $q_d\equiv (d~ s~ b)^{\dagger}$. 
$\mathcal{U}_L $ and $\mathcal{D}_L$ are rotation matrices  that diagonalize mass matrices for 
$q_u$ and $q_d$ respectively.
The CKM matrix can be constructed as $V_{\rm CKM}= \mathcal{U}_L \mathcal{D}_L^{\dagger} $ \cite{Schwartz:2014sze}.
For our estimation we take $\mathcal{U}_L =\mathbb{I}$ and hence $\mathcal{D}_L=V_{\rm CKM}^{\dagger}$\cite{Bonilla:2017lsq}.
In our scenario as $Z'$ couples to only third generation of quarks (with $\mathcal{L}_{int}\supset g_X J^{\mu}_{Z'}Z'_\mu$) it may induce flavor changing
neutral current (FCNC) and the Noether current ($ J^{\mu}_{Z'}$) is given by,
\begin{eqnarray}
     J^{\mu}_{Z'}&=&(\Bar{q_d})_L ~\mathcal{D}_L ~ \gamma^\mu ~
     \begin{pmatrix}
     0 & 0&0\\ 
     0 & 0&0\\
     0 &0&1
     \end{pmatrix}
     ~ \mathcal{D}_L^{\dagger} ~(q_d)_L.
     \label{eq:ckm}
\end{eqnarray}
After inserting the values of the elements of $\mathcal{D}_L$ matrix \cite{ParticleDataGroup:2022pth},  in eq.(\ref{eq:ckm}) the $Z'$ effective coupling with $d$ quark takes the following form:
\begin{equation}
    \mathcal{L}\supset (0.006~g_X)~Z'_{\mu} \Bar{d}_L \gamma^{\mu} d_L
\end{equation}
Using this we translate the bound from  CE$\nu$NS experiment in our scenario and obtain that it excludes the parameter space for $g_X\gtrsim 10^{-2}$. This limit is weaker than E$\nu$ES and hence not shown in Fig.\ref{fig:model1}.
As $Z'$ has coupling with $b$ quark it generates $B_s-\overline{B_s}$ mixing and hence it attracts constraint indicated by orange shaded region \cite{Bonilla:2017lsq}. 
Finally, the bound from supernova cooling is shown by the region shaded with grey vertical lines  \cite{Escudero:2019gzq}.

\subsection{Gauged $U(1)_{B_3-3 L_{\mu}}$ symmetry}
\label{sec:lmu}
\begin{figure}[!tbh]
    \centering
    \subfigure[\label{b3}]{
    \includegraphics[scale=0.4]{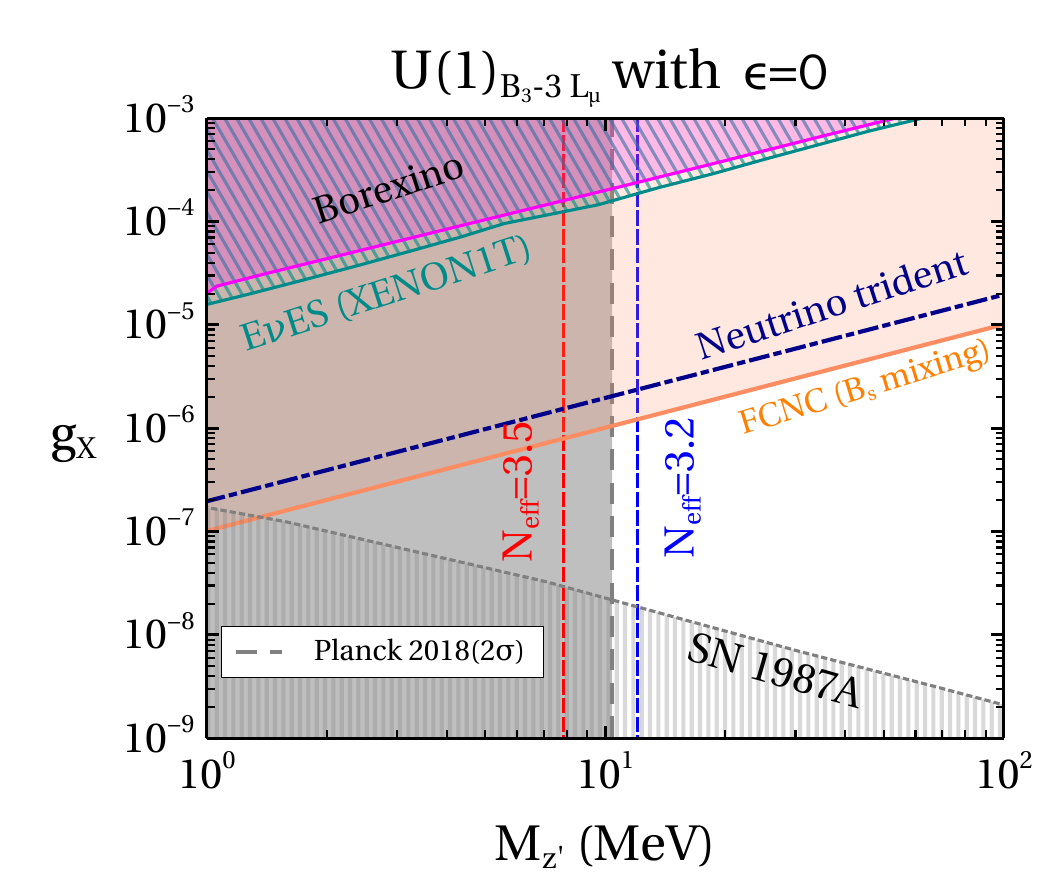}}
    \subfigure[\label{b3p}]{
    \includegraphics[scale=0.4]{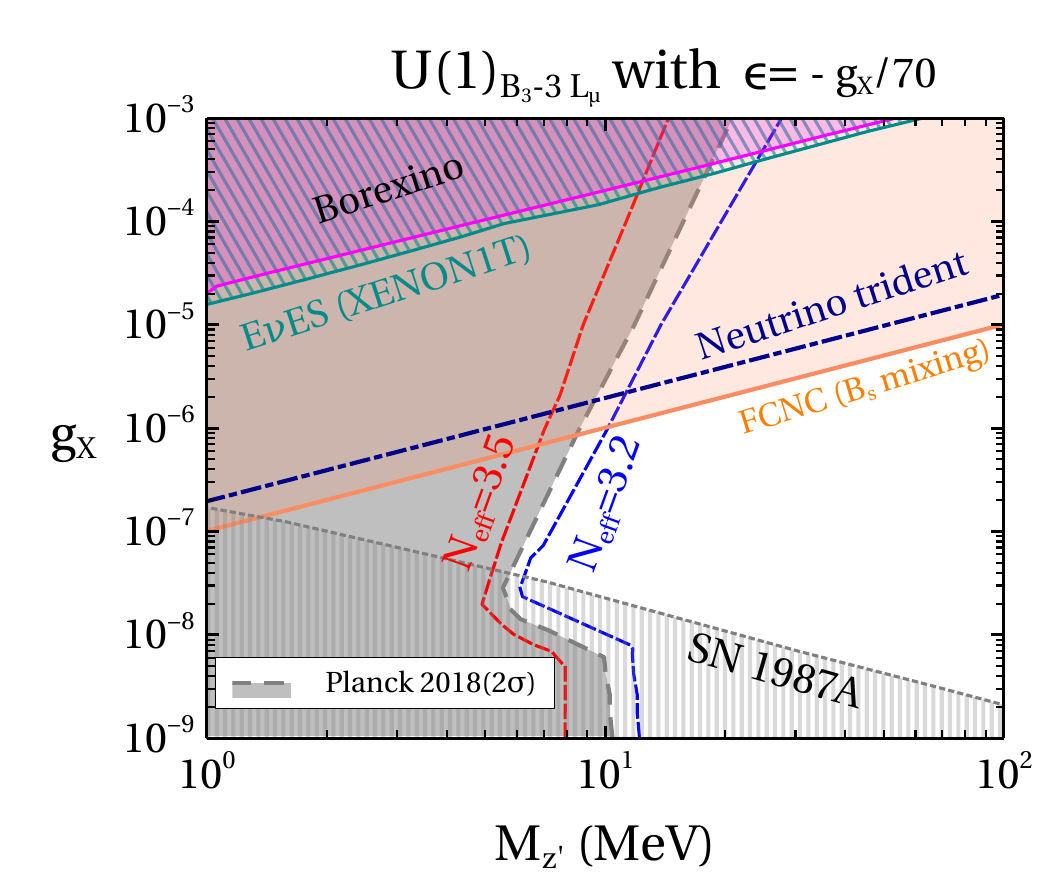}}
    \caption{Constraints from cosmology and low energy experiments in $M_{Z'}$ vs. $g_{X}$ plane for a generic $U(1)_{B_3-3 L_\mu}$ gauge extension. We present the results without loop induced mixing ($\epsilon=0$) in (a) and with loop induced mixing ($\epsilon \neq 0$) in (b). The grey dashed line represents the $2\sigma$ upper bound from Planck 2018 \cite{Planck:2018vyg}, excluding the parameter space on the left side of the line as shown by the grey shaded region. The region between the blue dashed line ($N_{\rm eff}=3.2$) and the red dashed line ($N_{\rm eff}=3.5$) indicates the parameter space where the $H_0$ tension can be relaxed as mentioned in ref.\cite{DiValentino:2021izs}.}
    \label{fig:model2}
\end{figure}

In this subsection, we present our analysis with $N_{\rm eff}$ for a light $Z'$ in the gauged extension of $U(1)_{B_3-3 L_\mu}$.
Analogous to the previous subsection we show our numerical results in $M_{Z'}$ vs. $g_{X}$ plane in Fig.\ref{fig:model2} along with all relevant astrophysical and experimental constraints.
The grey dashed line in the same plot signify the $2\sigma$ upper limit on $N_{\rm eff}$
from Planck 2018 data \cite{Planck:2018vyg} and likewise before. The grey shaded region depicts the exclusion of the parameter space to the left of the grey dashed line.

However, unlike previous scenario, for $U(1)_{B_3-3 L_\mu}$, $Z'$ has no tree level coupling with $e^{\pm}$ ($X_1=0$) and hence it belongs to other class of $U(1)_X$ models with $|X_1|=0$ as discussed in previous sub-section \ref{sub:x2}. 
We present our numerical results with the two values of effective coupling, $\epsilon=0$ and 
$\epsilon=-g_{X}/70$
in Fig.\ref{b3} and Fig.\ref{b3p} respectively.
For $\epsilon=0$ we observe no change in $N_{\rm eff}$ with $g_X$ in Fig.\ref{b3} as argued in the previous sections in the context of Fig.\ref{w1}.
On the other hand,  for $\epsilon=-g_{X}/70$ we observe the non-trivial dependence of $N_{\rm eff}$ with $g_X$ in Fig.\ref{b3} due to the interplay between scattering and decays as elaborated in detail 
in the context of Fig.\ref{w2}.

In the same plane for both the plots Fig.\ref{b3} and Fig.\ref{b3p} we portray the constraints arising from E$\nu$ES experiment with XENON1T\cite{Majumdar:2021vdw} shown by the regions shaded by dark cyan diagonal lines. 
Similarly, we showcase the exclusion region from Borexino depicted by magenta shaded region \cite{Majumdar:2021vdw}.
Note that these bounds are weaker for $U(1)_{B_3-3 L_\mu}$ compared to Fig.\ref{fig:model1}
as in this case $Z'$ has no tree level coupling with $e^{\pm}$ and the recoil rate 
gets smaller here (see eq.\eqref{eq:recoil}). 
Because, in this scenario $Z'$ couples directly with $\mu$, the Neutrino trident experiment significantly constrains the parameter space shown by the dark blue dashed dot line \cite{Bonilla:2017lsq}.
The parameter space above that line is excluded.
In this model also $Z'$ can generate $B_s-\overline{B_s}$ mixing and attracts constraint depicted by orange shaded region \cite{Bonilla:2017lsq}. 
Similar to the previous subsection the bound from  CE$\nu$NS experiment is weaker than E$\nu$ES and hence not shown in Fig.\ref{fig:model2}.
The bound from supernova cooling is shown by the region shaded with grey vertical lines \cite{Escudero:2019gzq}. 
We do not show the results for $U(1)_{B_3-3 L\tau}$ explicitly, though the imprint in $N_{\rm eff}$ for this model will be similar like the one in $U(1)_{B_3-3 L\tau}$ as both these models have $|X_1|=0$.

{For both of the aforementioned models, one can observe that for $M_{Z'}\lesssim 1$ MeV the parameter space is almost ruled out from astrophysical and laboratory searches. However, for $M_{Z'}\gtrsim 10 $ MeV parameter space remains unconstrained. Thus the MeV scale $Z'$ with $M_{Z'}\gtrsim 1$ MeV should be explored through alternate search strategies which is the main goal of this work. Since the $\nu_L$ decoupling generally happens at $T\sim $ MeV, the bound from $N_{\rm eff}$ lies in the mass region $M_{Z'}\sim$ MeV.}
\section{The Neutrino Portal Solution to the Hubble Tension}
\label{sec:h0}
The change in $N_{\rm eff}$ due to the presence of light $Z'$ gauge bosons can also lead to an interesting consequence namely a simple and elegant resolution to the cosmological Hubble constant problem. Recall that the Hubble constant problem is the discrepancy between the obtained value of the Hubble constant using early and late time probes \cite{DiValentino:2021izs, Riess:2021jrx}. As mentioned in the introduction the discrepancy in $H_0$ value can be ameliorated with increasing $N_{\rm eff}$ \cite{DiValentino:2021izs}. Several different resolutions have been proposed to address this discrepancy. One of the simplest possible resolution is the presence of light degrees of freedom changing the $N_{\rm eff}$ \cite{Escudero:2019gzq,Abazajian:2019oqj,Berbig:2020wve,Fabbrichesi:2020wbt}. There are already a few model specific works successfully addressing this problem for particular $U(1)_X$ extensions such as $U(1)_{\mu - \tau}$ \cite{Escudero:2019gzq} and $U(1)_{B- L}$ \cite{EscuderoAbenza:2020cmq}. 

The general formalism developed in this work also can be used to show that the $U(1)_X$ symmetries with light $Z'$ gauge bosons, in general, can address the Hubble constant problem reducing the tension. According to the Planck 2015 TT data \cite{Bernal:2016gxb,DiValentino:2021izs}, the $H_0$ tension issue can be relaxed upto $1.8 \sigma$ with the $N_{\rm eff}$ value between $3.2$ to $3.5$. However, the Planck 2018 polarization measurements provide a more stringent bound on $N_{\rm eff}$ \cite{Planck:2018vyg} and it is difficult to reach $H_0>70 ~{\rm Km~s}^{-1}\rm{Mpc}^{-1}$ \cite{deJesus:2022pux} within the $\Lambda$CDM. Hence there still exists a disagreement of $3.6 \sigma$ in the $H_0$ value predicted from CMB and local measurement. 
In our discussion, we only highlight the parameter space where $N_{\rm eff}$ lies between $3.2$ to $3.5$ in 
$M_{Z^\prime}-g_X$ plane, that may relax $H_0$ tension followed by the analysis in ref. \cite{Bernal:2016gxb,DiValentino:2021izs}. 
In sec.\ref{sec:class} we portray the parameter space for generic $U(1)_X$ models in Fig.\ref{fig:case1} (for $|X_1|\neq 0$)and Fig.\ref{fig:case2} ($|X_1|=0$) where the blue and red dashed lines indicate the values of $N_{\rm eff}=3.2$ and $3.5$ respectively. 
Similarly, in sec.\ref{sec:examples} we also showcase the same lines with constant $N_{\rm eff}=3.2$ (blue dashed line) and $3.5$ (red dashed line) contours for the specific models: $U_{B_3-3L_e}$ (in Fig.\ref{fig:model1}) and $U_{B_3-3L_\mu}$ (in Fig. \ref{fig:model2}). 
The region between these two contour lines may ameliorate $H_0$ tension.
However, the quoted values of $N_{\rm eff}$ can be excluded by future experiment CMB-S4\cite{CMB-S4:2022ght}.
Also, recent studies show that even with increasing $N_{\rm eff}$ can not resolve the $H_0$ tension completely \cite{Khalife:2023qbu,Schoneberg:2021qvd,deJesus:2022pux}.  
Hence, claiming any possible resolution of $H_0$ problem with quantitative effectiveness, itself requires a dedicated analysis which is beyond the scope of this work \cite{DiValentino:2021izs}.

 \section{Summary and conclusion}
\label{sec:conclusions}
The Planck experiment has a very accurate measurement of the CMB which has established stringent limits on $\Delta N_{\rm eff}$, representing the number of effective relativistic degrees of freedom during 
the early epoch of the universe \cite{Planck:2018vyg}.
This sensitivity makes $\Delta N_{\rm eff}$ a useful probe for investigating various BSM scenarios that affect the 
neutrino decoupling at the time of the CMB. 
In this work, we have studied the impact of light $Z'$ particle, arising from generic $U(1)_X$ model, on $N_{\rm eff}$. In the presence of this light $Z'$, $N_{\rm eff}$ receives two-fold contributions: the decay of 
$Z^\prime \to \nu_L {\bar \nu_L},~e^+e^- $ and  scattering of light SM leptons ($\nu_L,e$) via this $Z'$.   
At first, we considered a generic $U(1)_X$ model with arbitrary charge assignments. 
Apart from the light $Z'$ gauge boson, the model also contains a BSM scalar $\sigma$ and RHN $\nu_R$ which are necessary for the model construction.   
To understand the sole effect of the light $Z'$ in early universe temperature evolution, we assume the other BSM particles are sufficiently heavy ($\gtrsim 100$ MeV) that they decouple at the time of neutrino decoupling. 
In this work, we only consider the scenario where $Z'$ was initially ($T\gtrsim M_{Z'}$) in a thermal bath. 
We adopt the formalism developed in ref.\cite{Escudero:2019gzq} and solve the temperature equations to evaluate $N_{\rm eff}$ in sec.\ref{sec:neff}. We enumerate our findings below.
\begin{itemize}
    \item Firstly, we noticed that the $U(1)_X$ charges for $e,\nu_{e,\mu,\tau}$ play a crucial role in $N_{\rm eff}$ for our proposed generic $U(1)_X$ model in sec.\ref{sec:model} as these are the only SM particles relevant for $\nu$ decoupling. In a similar vein, we found that the mass of the BSM gauge boson $Z'$ should be around $\lesssim30$ MeV to affect neutrino decoupling.

    \item After a careful analysis in sec.\ref{sec:neff} we observed that the value of $N_{\rm eff}$ depends 
    differently on the coupling $g_X$ for two distinct scenarios those are $|X_1|=0$ and $|X_1|\neq0$. 
    Based on this fact, in sec.\ref{sec:class} we categorise $U(1)_X$ models into two classes: (a) electrons having tree level coupling with $Z'$ ($|X_1|=0$) and (b) without tree level coupling of an electron with $Z'$ ($|X_1|\neq0$). However, in the second scenario electron may have some induced coupling with $Z'$ even if $|X_1|=0$ and we explored that scenario too. For both the two classes we present the contours from the upper limit on $N_{\rm eff}$ from Planck 2018 \cite{Planck:2018vyg} in $M_{Z'}$ vs. $g_X$ plane as shown in Fig.\ref{fig:case1} and Fig.\ref{fig:case2}.
    In the same plane, we also highlight the parameter space favoured to relax $H_0$ tension shown in earlier analyses \cite{Bernal:2016gxb,Escudero:2019gzq}.

    \item For comparison with existing constraints from ground based experiments, in sec.\ref{sec:examples} we considered specific $U(1)_X$ models as examples and discuss the cosmological implications in the context of $N_{\rm eff}$. 
    We present the numerical results for $U(1)_{B_3-3 L_e}~(|X_1|=3)$ and $U(1)_{B_3-3 L_\mu}~(|X_1|=0)$ models in  sub-sec.\ref{sec:Le} and sub-sec.\ref{sec:lmu} respectively which have not been explored in the existing literature. 
    Depending on the coupling of an electron with $Z'$ these two models also lead to distinguishable $N_{\rm eff}$ contours in $M_{Z'}$ vs. $g_X$ plane as portrayed in Fig.\ref{fig:model1} and  Fig.\ref{fig:model2}.
    Our detailed analysis shows that the bounds from $N_{\rm eff}$ on the $M_{Z'}$ vs. $g_X$ plane for some of these models are stronger than $B-L$ or $L_\mu-L_\tau$ model due to higher $U(1)_X$ charge of an electron. A priori these bounds can not be scaled from the existing $N_{\rm eff}$ bound on other models with different $U(1)_X$ lepton charges from the existing literature  ($B-L$ or $L_\mu-L_\tau$).

    \item For both $U(1)_{B_3-3 L_e}$ and $U(1)_{B_3-3 L_\mu}$ models we also showcase the relevant astrophysical and laboratory constraints in Fig.\ref{fig:model1} and  Fig.\ref{fig:model2} respectively.
    We displayed the constraints from E$\nu$ES with XENON1T \cite{Majumdar:2021vdw} and Borexino \cite{Chakraborty:2021apc}, neutrino trident \cite{Bonilla:2017lsq}, $B_s-\overline{B_s}$ mixing \cite{Bonilla:2017lsq}  as well as from SN1987A \cite{Escudero:2019gzq}. We checked the constraints from CE$\nu$NS \cite{Majumdar:2021vdw} is weaker than the other constraints as our chosen $U(1)_X$ models do not have $Z'$ coupling with first generation quarks. For $U(1)_{B_3-3 L_\mu}$ model electron has no BSM gauge coupling and hence, the bounds from E$\nu$ES for this model are weaker than the other one. In the same plane, we also indicate the parameter space that can relax the $H_0$ tension \cite{Bernal:2016gxb,Escudero:2019gzq}.   
\end{itemize}

For both $U(1)_{B_3-3 L_e}$ and $U(1)_{B_3-3 L_\mu}$ models we have shown that the bounds on $N_{\rm eff}$ from Planck 2018 data \cite{Planck:2018vyg} can provide more stringent bound on the parameter space than the laboratory searches for $M_{Z'}\lesssim \mathcal{O}(1)$ MeV.
The future generation experiments like CMB-S4 ( $\Delta N_{\rm eff} =0.06$ at $2 \sigma$)
\cite{CMB-S4:2022ght} can even probe more parameter space for such models.
We also show that there is a certain parameter space still left to relax $H_0$ tension \cite{Bernal:2016gxb,Escudero:2019gzq} allowed from all kinds of constraints.
On the other hand, as $U(1)_X$ models are well motivated BSM scenarios and widely studied in several aspects, this analysis may enhance insights to explore their connection with the $H_0$ problem too.
Besides the explicit results of $B_3-3L_i$, our discussion for the generic $U(1)_X$ models shows how the the $L$ number ($X_i$)
plays a crucial role in deciding the constraint on the parameter space from $N_{\rm eff}$.
From the classification of generic $U(1)_X$ models depending on the $L$ number, one can anticipate the consequences in $N_{\rm eff}$ for various other exotic $U(1)_X$ models that have not been explored so far \cite{Coloma:2022umy,AtzoriCorona:2022moj}.
Thus our generalized prescription for $N_{\rm eff}$ is extremely helpful to put stringent constraints on the light $Z'$ parameter space from cosmology complementary to the bounds obtained from ground based experiments.


\begin{acknowledgments}
SJ thanks Sougata Ganguly for the helpful discussion and Miguel Escudero for the email conversations. The work of SJ is funded by CSIR, Government of India, under the NET JRF fellowship scheme with Award file No. 09/080(1172)/2020-EMR-I. PG
would like to acknowledge the Indian Association for the Cultivation of Science, Kolkata for the financial support.
  
\end{acknowledgments}

\appendix
\section{Neutrino decoupling in SM scenario}
\label{sec:apxA}
In this section, we recapitulate the formalism for $\nu_L$ decoupling in the SM scenario which is already developed in ref.\cite{Escudero:2018mvt}. However, for better understanding and smooth transition to the extra $U(1)_X$ scenario we just note down the key points for the SM case here.
As mentioned in sec.\ref{sec:neff} the only particles relevant at the temperature scale of 
$\nu_L$ decoupling are $\nu_i~(i=e,\mu,\tau)$,$e^{\pm}$ and $\gamma$. 
So in the SM scenario, the temperature evaluation of only these $3$ particles is needed to evaluate $T_\nu/T_\gamma$.
Initially all $\nu_i~(i=e,\mu,\tau)$ were coupled to $\gamma$ and $e^{\pm}$ bath at high temperature via weak interaction processes. 
To track the exact temperature of photon bath ($T_{\gamma}$) and neutrino bath ($T_\nu$) we start from the Liouville equation \cite{Kolb:1990vq}:
\begin{eqnarray}
    \frac{\partial f(p,t)}{\partial t}- H p \frac{\partial f(p,t)}{\partial p} = \mathcal{C}[f],
    \label{eq:liou}
\label{eq:b1}
\end{eqnarray}
where,$f(p,t)$ is the distribution function in a homogeneous and isotropic universe and $p,t$ signifies the momentum and time respectively. $\mathcal{C}[f]$ is the collision term and it describes the total change in $f(p,t)$ with time.

For SM $\nu_L$ decoupling the involved processes are $2\to2$ scattering between $\nu,e$ as shown in Table \ref{tab:amp}.
So, we write the collision terms for a particle $\chi$ undergoing $2\to2$ process say, $\chi+i\to j+k$ for which $\mathcal{C}[f]$ can be written as (assuming CP invariance),
\begin{eqnarray}\nonumber
    \mathcal{C}[f_\chi] &\equiv& - \frac{1}{2 E_\chi} \int   d\Pi_{i} d\Pi_{j} d\Pi_{k} (2\pi)^4 \delta^4 (p_\chi + p_i - p_j -p_k)\times  \\
&&|\mathcal{M}|^2_{\chi+i\to j+k} \left[f_\chi  f_{i} \left[1 \pm f_j \right]\left[1 \pm f_k \right] -  f_{j} f_k   \left[ 1 \pm f_\chi \right] \left[1 \pm f_i \right] \right]\,
\label{eq:cf}
\end{eqnarray}
where, $f_\ell$,$E_\ell$, and $p_\ell$ denote the distribution function, energy, and 4-momentum respectively for particle $\ell$. $|\mathcal{M}|^2_{\chi+i\to j+k}$ is the amplitude square for the above mentioned process.

Multiplying eq.\eqref{eq:b1} with $g_{\chi}\, E\,d^3p_{\chi}/(2\pi)^3$ and integrating over momenta will give the following energy density equation:
\begin{eqnarray}
    \frac{d\rho_\chi}{dt} + 3 H (\rho_\chi + P_\chi) = \int g_{\chi}\,  E \, \frac{d^3p_{\chi}}{(2\pi)^3} \, \mathcal{C}[f_\chi] =\frac{\delta \rho_{\chi\to j}}{\delta t}
    \label{eq:Etrans}
\end{eqnarray}
$\frac{\delta \rho_{\chi\to j}}{\delta t}$ is the energy density transfer rate from $\chi \to j({\rm or}~ k)$.
$P_\chi$ is the pressure density. 
So, we can rewrite energy density equation eq.\eqref{eq:Etrans} for $\chi$,
\begin{eqnarray}
    \frac{dT_{\chi}}{dt}&=& \left( - 3 H (\rho_{\chi}+P_\chi) +\frac{\delta \rho_{\chi}}{\delta t}\right)\left({\frac{\partial\rho_{\chi}}{\partial T_{\chi}}}\right)^{-1}
 \label{eq:tnui}
\end{eqnarray}

Now for SM $\nu_L$ decoupling we consider the energy transfers between $\nu_i~(i=e,\mu,\tau)$ to $e^\pm,\gamma$ photon bath.
For photon bath containing $\gamma,e^{\pm}$  we consider a common temperature $T_{\gamma}$ and add their energy density equations and rearrange accordingly,
\begin{eqnarray}
 \frac{d\rho_{\gamma}}{dt}+\frac{d\rho_{e}}{dt}&=&  - 4 H \rho_{\gamma} - 3 H (\rho_{e}+P_e) +\sum_{i=e,\mu,\tau}\frac{\delta \rho_{e\to \nu_i}}{\delta t}\\
  \frac{dT_{\gamma}}{dt}&=& \left(- 4 H \rho_{\gamma} - 3 H (\rho_{e}+P_e) -\sum_{i=e,\mu,\tau}\frac{\delta \rho_{\nu_i\to e}}{\delta t}\right)\left({\frac{\partial \rho_{\gamma}}{\partial  T_{\gamma}}+\frac{\partial \rho_{e}}{\partial  T_{\gamma}}}\right)^{-1},
   \label{eq:tgamma}
\end{eqnarray}
where, in the last step we have used $\frac{\delta \rho_{\nu_\chi\to j}}{\delta t}=-\frac{\delta \rho_{\nu_j\to \chi}}{\delta t}$.
Similarly, following eq.\eqref{eq:Etrans} we can write the respective temperature ($T_{\nu_i}$) equations for each type of $\nu$,
\begin{eqnarray}
    \frac{dT_{\nu_i}}{dt}&=& \left(- 4 H \rho_{\nu_i} +\frac{\delta \rho_{\nu_i}}{\delta t}\right) \left( \frac{\partial\rho_{\nu_i}}{\partial T_{\nu_i}}~\right)^{-1}
    \label{eq:tnusm}
\end{eqnarray}
Once we have the temperature equations the remaining step is to evaluate the energy transfer rates i.e. the collision terms.
The relevant processes in the SM scenario are given in Table~\ref{tab:amp}. 

\renewcommand{\arraystretch}{1.4}
\begin{table}[tbh!]
\begin{center}
\begin{tabular}{ |c|c| } 
 \hline
     Process & $2^{-5} G_F^2 \overline{|M|^2 }$\\
   \hline
 ~~$\nu_i +\Bar{\nu_i}\longleftrightarrow e^+ +e^-$ ~~ & ~~~$4[g_{L_i}^2(p_1.p_4)(p_2.p_3)+g_{R_i}^2(p_1.p_3)(p_2.p_4)]$~~~ \\ 
  \hline
 $\nu_i +e^-  \longleftrightarrow \nu_i +e^-  $ & $4[g_{L_i}^2(p_1.p_2)(p_3.p_4)+g_{R_i}^2(p_1.p_4)(p_2.p_3)]$ \\ 
 \hline
$\nu_i +e^+  \longleftrightarrow \nu_i +e^+  $ & $4[g_{L_i}^2(p_1.p_4)(p_2.p_3)+g_{R_i}^2(p_1.p_2)(p_3.p_4)]$\\ 
 \hline
 $\nu_i +\nu_{j}  \longleftrightarrow \nu_i +\nu_{j}   $ & $(p_1.p_2)(p_3.p_4)$\\
 \hline
 $\nu_i +\Bar{\nu_j}  \longleftrightarrow \nu_i +\Bar{\nu_{j}}    $ & $(p_1.p_2)(p_3.p_4)$\\
 \hline
\end{tabular}
\end{center}
\caption{Squared amplitudes for interactions relevant for  $\nu$ decoupling in SM. The $i$ in subscripts stand for different generations of $\nu_L$ \cite{Dolgov:2002wy}. }
\label{tab:amp}
\end{table}
\renewcommand{\arraystretch}{1}
All the processes are either elastic scatterings between $\nu_i$ and $e$ or annihilations and all are mediated by
$W^{\pm},Z$. As we are focusing on the MeV scale we can integrate out the mediators and write in terms of Fermi constant $G_F$. In Table \ref{tab:amp},
$p_{1,2}$ and $p_{3,4}$ denote the  momentum of incoming and outgoing particles. Here, $g_{L_e} = \frac{1}{2} + s_W^2$, $g_{R_e} = s_W^2$, $g_{L_{\mu,\,\tau}} = -\frac{1}{2} + s_W^2$, $g_{R_{\mu,\,\tau}} = s_W^2$ and $s_W= \sin \theta_W$ where $\theta_W$ is  Weinberg angle. The reason behind the difference of $1$ between $g_{L_e}$ and $g_{R_{\mu,\,\tau}}$ is the additional charged current process available for $\nu_e$.

{\bf Collision terms for $\nu,e^\pm$ scattering}

\renewcommand{\arraystretch}{1.5}
\begin{table}[tbh!]
\begin{center}
\begin{tabular}{ |c|c| } 
 \hline
     Process & $\pi^5 G_F^{-2} \frac{\delta \rho}{\delta t}$\\
   \hline
 ~~$\nu_i +\bar{\nu}_i\longleftrightarrow e^+ +e^-$ ~~ & ~~~$64  (g_{L_i}^2+g_{R_i}^2) \left[ T_{\gamma}^9-  T_{\nu_i}^9 \right]$~~~ \\ 
  \hline
 $\nu_i +e^\pm  \longleftrightarrow \nu_i +e^\pm  $ & $112(g_{L_i}^2+g_{R_i}^2) ~T_{\gamma}^4
    T_{\nu_i}^4 \left[ T_{\gamma}-  T_{\nu_i} \right]$ \\  
 \hline
 $\nu_i +\bar{\nu}_i  \longleftrightarrow \nu_j +\bar{\nu}_{j} $ & $32 \left[ T_{\nu_j}^9-  T_{\nu_i}^9 \right]$\\
 \hline
 $\nu_i +\nu_{j}  \longleftrightarrow \nu_i +\nu_{j}   $ & $48 T_{\nu_j}^4
    T_{\nu_i}^4 \left[ T_{\nu_j}-  T_{\nu_i} \right]$\\
 \hline
 $\nu_i +\bar{\nu}_j  \longleftrightarrow \nu_i +\bar{\nu}_{j}    $ & $8 T_{\nu_j}^4
    T_{\nu_i}^4 \left[ T_{\nu_j}-  T_{\nu_i} \right]$\\
 \hline
\end{tabular}
\end{center}
\caption{Collision terms for $\nu,e^\pm$ scattering assuming MB distribution function \cite{Luo:2020sho,Escudero:2018mvt}. The $i$ in subscripts stand for different generations of $\nu_L$.}
\label{tab:col}
\end{table}
\renewcommand{\arraystretch}{1}
The first 2 processes are responsible for energy transfer from $\nu_e$ to $e$ sector (photon bath) and the rest are responsible for energy transfer from $\nu_e$ to $\nu_{\mu/\tau}$ sector.
Finally, we enumerate below all the energy transfer rates for the SM scenario following the formalism in ref.\cite{Luo:2020sho,Escudero:2018mvt}:
\begin{eqnarray}
   \left(\frac{\delta \rho_{\nu_e \to e}}{\delta t}\right)_{\rm SM}&=&\frac{G_F^2}{\pi^5} \left\{(1+4s_W^2+8s_W^4) F_{\rm MB}(T_{\gamma},T_{\nu_e})\right\} \label{eq:delnu1} \\
   \left(\frac{\delta \rho_{\nu_{\mu/\tau}\to e}}{\delta t}\right)_{\rm SM}&=&\frac{G_F^2}{\pi^5} \left\{(1-4s_W^2+8s_W^4) F_{\rm MB}(T_{\gamma},T_{\nu_\mu})\right\}
   \label{eq:delnu2}\\
   \left(\frac{\delta \rho_{\nu_e \to \mu/\tau}}{\delta t}\right)_{\rm SM}&=&\frac{G_F^2} {\pi^5} F_{\rm MB}(T_{\nu_\mu},T_{\nu_e}),
   \label{eq:delnui}
\end{eqnarray}
where,
\begin{eqnarray}
    F_{\rm MB}(T_1,T_2) = 32\, (T_1^9-T_2^9) + 56 \, T_1^4\,T_2^4 \, (T_1-T_2)\, .
\end{eqnarray}
Plugging these collision terms in eq.\eqref{eq:tgamma} and eq.\eqref{eq:tnusm} we can track the temperature evolution.

\section{Neutrino decoupling in presence of a $U(1)_{X}$ gauge boson}
\label{sec:apxB}
The relevant lagrangian for the calculation of $N_{\rm eff}$ in $U(1)_{X}$ model has the following generic form,
\begin{equation}
    \mathcal{L}_{\rm int}\supset Z^{\prime}_\alpha J^{\alpha}_{\mathbb{X}}, 
\end{equation}
where, $J^{\alpha}_{\mathbb{X}}$ is the Noether current associated with $U(1)_{\mathbb{X}}$ and given by,
\begin{eqnarray}\nonumber
 J^{\alpha}_{\mathbb{X}} &=& g_{X} \left( X_3\bar \tau \gamma^\alpha \tau + X_3 \bar \nu_{\tau} \gamma^\alpha P_L \nu_{\tau}  + X_2\bar \mu \gamma^\alpha \mu + X_2\bar \nu_{\mu} \gamma^\alpha P_L \nu_{\mu} \right) \\
 && +g_{X} \left(  X_1 \bar e \gamma^\alpha e + X_1\bar \nu_{e} \gamma^\alpha P_L \nu_{e} \right)
\end{eqnarray}

We have not written the corresponding interactions with quark sectors since around $\sim 2$ MeV quarks are already confined \cite{Zhitnitsky:2008ha}. 
The $\mu,\tau$ particles also do not take part as they have suppressed energy density around MeV temperature.
In the presence of these $Z'$ the following things will affect neutrino decoupling (see Fig.\ref{fig:cartoon1}).
\begin{enumerate}
    \item Depending on charge assignments ($X_1\neq 0$) $Z'$ can decay to $e^+e^-$ and the inverse decay can also happen upto $T_{\gamma}=M_{Z'}$. So there will be energy transfer from $Z'$ to $e^\pm$ sector($\gamma$ bath) and vice versa. Similarly, $Z'$ can transfer energy to $\nu$ bath with the decays and inverse decays to $\nu_{i}\Bar{\nu_{i}}$.

    \item Through $Z'$ portal $\nu_{i}+\overline{\nu_{i}} \longleftrightarrow e^+ e^-$, $\nu_{i}+e^\pm \longleftrightarrow {\nu_{i}}+ e^\pm~,i=(e,\mu,\tau) $ processes can happen (depending on $X_{1,2,3}$)
    leading to an additional energy transfer between $\nu_i$ sector  and  $e^\pm$.
\end{enumerate}
We will discuss the above points in the following subsections. As discussed in Sec.\ref{sec:neff}, in this work we will only focus on the scenarios where $Z'$ was in thermal equilibrium with $e,\nu_i$ at $T>T^\nu_{\rm dec}$. Throughout this section we will simplify the collision terms in the effective operator limit assuming $\sqrt{s} \sim T \ll M_{Z'}$, for the purpose of $\nu_L$ decoupling calculations.
\subsection{$Z'$ decay to $e^+e^-$ }
We consider the process 
\begin{equation}
    Z'(k)\leftrightarrow e^-(p_1)+e^+(p_2),
\end{equation}
where, $k,p_1,p_2$ denote the four momenta of respective particles.
Following eq.\eqref{eq:cf} and eq.\eqref{eq:Etrans} we calculate the energy transfer rate from $Z'$ to $e$ bath,
\begin{eqnarray}\nonumber
    \int g_{Z'} E_{Z'} \frac{d^3 k}{(2\pi)^3 2 E_{Z'}}  \mathcal{C}[f]&=&-\int g_{Z'}  \frac{d^3  k}{(2\pi)^3 2 E_{Z'}}    g_{1}\frac{d^3  p_1}{(2\pi)^3 2 E_1}
 g_{2} \frac{d^3  p_2}{(2\pi)^3 2 E_2}  \, E_{Z'} \frac{1}{g_{Z'} g_1 g_2}|{\cal M}_{e^+e^- \to Z'}|^2\\
 && (f^{\rm (eq)}_{z'}(k)-f^{\rm (eq)}_e( p_1) f^{\rm (eq)}_e(p_2))   (2\pi)^4 \delta^4(p_1+p_2 - k)
 \label{eq:rhoz}
\end{eqnarray}
Here the degrees of freedom and energy of the corresponding particle are denoted as $g_\ell\& E_\ell,~(\ell\equiv Z',1,2)$ respectively. 
$f^{eq}_{i},~(i\equiv Z',e)$ signify the equilibrium distribution function of $i$ particle

The decay width of $Z'$ to $e^+ e^-$ is given by, 
\begin{eqnarray}
    \Gamma_{Z'\to e^+ e^-}&=& \frac{1}{2 M_{Z'}} \int  \frac{d^3  p_1}{(2\pi)^3 2 E_1} \frac{d^3  p_2}{(2\pi)^3 2 E_2} \frac{1}{g_{Z'}}|{\cal M}_{Z'\to e^+ e^-}|^2  (2\pi)^4 \delta^4(p_1+p_2 - p_k) \\
    &=& X_1^2 \frac{g_X^2 M_{Z'}}{12 \pi} (1+\frac{2 m_e^2}{M^2_{Z'}})\sqrt{1-\frac{4 m_e^2}{M^2_{Z'}}}
    \label{eq:dec}
\end{eqnarray}
where $m_e$ is the electron mass.

Using eq.\eqref{eq:dec} we can simplify above eq.\eqref{eq:rhoz} as
\begin{eqnarray}
    \int g_{z'} E_k \frac{d^3 k}{(2\pi)^3 2 E_k}  \mathcal{C}[f]
    &=& \frac{3 M_{Z'}^3 \Gamma_{Z' \to e^+e^-} }{2 \pi^2 } \left[T_\gamma~K_2\left( \frac{M_{Z'}}{T_\gamma}\right) - T_{Z'} K_2\left(\frac{M_{Z'}}{T_{Z'}}\right) \right]
\end{eqnarray}

Therefore, from eq.\eqref{eq:cf} we write
\begin{equation}
    \frac{ \delta \rho_{Z'\to e}}{\delta t}  =  \frac{3 m_{Z'}^3}{2\pi^2} \left[ T_{\gamma} K_2\left(\frac{M_{Z'}}{T_{\gamma}}\right) - T_{Z'} K_2\left(\frac{M_{Z'}}{T_{Z'}}\right) \right]   \Gamma_{Z' \to e^+ e^-} ,~
    \label{eq:rhoZ}
\end{equation}
$T_{Z'}$ denotes temperature of $Z'$ bath.
\subsection{$Z'$ decay to $\nu_i \Bar{\nu_i}$ }
Here we consider the processes 
\begin{equation}
    Z'(k)\leftrightarrow \nu_i(p_1)+ \Bar{\nu_i}(p_2),~{\rm where~}i=e,\mu,\tau.
\end{equation}
The corresponding decay widths of $Z'$ to $i$ type $\nu$ are given as,
\begin{eqnarray}
    \Gamma_{Z'\to \nu_i \Bar{\nu_i}}&=& X_i^2 \frac{g_X^2 M_{Z'}}{24 \pi}.
    \label{eq:dec1}
\end{eqnarray}
Similar to the previous section we can write the energy density transfer rates from $Z'$ to each generation of $\nu$ as,
\begin{equation}
    \frac{ \delta \rho_{Z'\to \nu_i}}{\delta t}  =  \frac{3 m_{Z'}^3}{2\pi^2} \left[ T_{\nu_i} K_2\left(\frac{M_{Z'}}{T_{\nu_i}}\right) - T_{Z'} K_2\left(\frac{M_{Z'}}{T_{Z'}}\right) \right]   \Gamma_{Z' \to \nu_i \Bar{\nu_i}} ,~
    \label{eq:rhoZ2}
\end{equation}
\subsection{ Energy transfer from $\nu_{i}$ to $e^+e^-$ mediated by BSM $Z'$}
In the presence of light $Z'$ there will be additional energy transfer between $e^{\pm}$ and $\nu_i$ apart from the SM weak processes (mediated by SM $W\&Z$).
The two relevant processes are: 
\begin{eqnarray}
     \nu (p_1) +\bar \nu (p_2)  & \leftrightarrow &e^-(p_3) +e^+(p_4)\\
    \nu (p_1)+ e^\pm (p_2) &\leftrightarrow& \nu (p_3) +e^\pm (p_4)
\end{eqnarray}
We denote the amplitudes for SM and BSM processes as $\mathcal{M}_{\rm SM}$ and $ \mathcal{M}_{\rm BSM}$ respectively. Then the total amplitude $\mathcal{M}_{\rm tot}$ can be written as 
\begin{eqnarray}
    |\mathcal{M}_{\rm tot}|^2= |\mathcal{M}_{\rm SM}|^2+ 2 Re(\mathcal{M}_{\rm SM}\mathcal{M}^{\dagger}_{\rm BSM})+|\mathcal{M}_{\rm BSM}|^2
    \label{eq:ams}
\end{eqnarray}
Plugging this in eq.\eqref{eq:cf} we will get the total energy transfer rate ($\nu\leftrightarrow e$) 
Three sources contribute to the total energy transfer rate; pure SM, SM-BSM interference, and pure BSM respectively. 
For simplicity, we calculate the energy transfer rates for each of them separately.
The amplitudes for pure SM contributions can be found in sec.\ref{sec:apxA}.
And the energy transfer rates for pure SM contributions are given in eq.(\ref{eq:delnu1}- \ref{eq:delnui}) \cite{Escudero:2018mvt}.

The amplitude for BSM processes (with proper momentum assignment) will be ($s\ll M^2_{Z'}$)
\begin{eqnarray}
    \mathcal{M}_{BSM}=-\frac{(X_i g_{X})(X_1 g_{X})}{M_{Z'}^2}[ \bar \nu_i\gamma^\mu P_L \nu_i]~[ \bar e \gamma_\mu e]
\end{eqnarray}
The interference amplitudes in eq.\eqref{eq:ams} are tabulated in Tab.-\ref{tab:ampi}.
\renewcommand{\arraystretch}{1.4}
\begin{table}[tbh!]
\begin{center}
\begin{tabular}{ |c|c| } 
 \hline
     Process & $2^{-3} /\sqrt{2} ~G_F ~(X_i g_X)^{-1} ~(X_1 g_X)^{-1}M_{Z'}^{2}~ \overline{|M|^2 }$\\
   \hline
 $\nu_i +e^-  \longleftrightarrow \nu_i +e^-  $ & $~1/2~[g_L (p_1.p_2)(p_3.p_4)+g_R (p_1.p_4)(p_2.p_3)]$ \\ 
 \hline
$\nu_i +e^+  \longleftrightarrow \nu_i +e^+  $ & $~1/2~[g_L (p_1.p_4)(p_2.p_3)+g_R (p_1.p_2)(p_3.p_4)]$\\ 
 \hline
 ~~$\nu_i +\Bar{\nu_i}\longleftrightarrow e^+ +e^-$ ~~ & ~~~$[g_L (p_1.p_4)(p_2.p_3)+g_R (p_1.p_3)(p_2.p_4)]$~~~ \\ 
  \hline
\end{tabular}
\end{center}
\caption{Relevant SM-BSM interference terms for  $\nu$ decoupling in SM. The interactions are written with $i~(i=e,\mu,\tau)$ type $\nu$. Collision terms in the effective operator limit assuming $s\ll M_{Z'}^2$,}
\label{tab:ampi}
\end{table}
\renewcommand{\arraystretch}{1}

Following the formalism developed in ref.\cite{Luo:2020sho} we write the collision terms for the interference terms:
\begin{eqnarray}
\int  \frac{d^3p}{2\pi^3}  p ~\mathcal{C}[f]_{\nu_i + e \leftrightarrow \nu_i +e}& = & (X_i g_X)(X_1 g_X)~\frac{ 7 \sqrt{2} G_F}{\pi^5 M_{Z'}^2} (g_L +g_R) ~T_{\gamma}^4
    T_{\nu_i}^4 \left[ T_{\gamma}-  T_{\nu_i} \right] \\
\int  \frac{d^3p}{2\pi^3}  p ~\mathcal{C}[f]_{\nu_i \bar{\nu_i} \leftrightarrow e e}& = &(X_i g_X)(X_1 g_X)~\frac{ 4 \sqrt{2} G_F}{\pi^5 M_{Z'}^2} (g_L +g_R) ~ \left[ T^9_{\gamma}-  T^9_{\nu_i} \right]
\label{eq:intE}
\end{eqnarray}
The first one in the above two collision terms accounts for elastic scattering of $\nu_i$ with both $e^\pm$. Here we have multiplied with an additional factor $2$ to count the effect of $\Bar{\nu_i}$. The second collision term in eq.\eqref{eq:intE} is due to annihilations.
So, the total energy transfer rate from $\nu$ to $e$ bath due to the interference with $U(1)_X$ gauge boson, 
\begin{eqnarray}
   \left(\frac{\delta \rho_{\nu_i\to e}}{\delta t}\right)_{\rm int}&=& 
    (X_i g_X)(X_1 g_X)~ \frac{\sqrt{2}G_F}{8\pi^5 M_{Z'}^2} \left\{(g_L +g_R) F_{\rm MB}(T_{\gamma},T_{\nu_i}) \right \},
    \label{eq:intx}
\end{eqnarray}

In a similar fashion, we now move to calculate the collision term due to pure BSM amplitudes (third term in eq.\eqref{eq:ams}).
The pure BSM amplitudes in eq.\eqref{eq:ams} are tabulated in Tab.-\ref{tab:ampix}.
\renewcommand{\arraystretch}{1.4}
\begin{table}[tbh!]
\begin{center}
\begin{tabular}{ |c|c| } 
 \hline
     Process & $2^{-4} ~(X_i g_X)^{-2} ~(X_1 g_X)^{-2}~M_{Z'}^{2}~ \overline{|M|^2 }$\\
   \hline
 $\nu_i +e^-  \longleftrightarrow \nu_i +e^-  $ & $~1/2~[ (p_1.p_2)(p_3.p_4)+(p_1.p_4)(p_2.p_3)]$ \\ 
 \hline
$\nu_i +e^+  \longleftrightarrow \nu_i +e^+  $ & $~1/2~[(p_1.p_4)(p_2.p_3)+ (p_1.p_2)(p_3.p_4)]$\\ 
 \hline
 ~~$\nu_i +\Bar{\nu_i}\longleftrightarrow e^+ +e^-$ ~~ & ~~~$[ (p_1.p_4)(p_2.p_3)+ (p_1.p_3)(p_2.p_4)]$~~~ \\ 
  \hline
\end{tabular}
\end{center}
\caption{Relevant BSM  terms for  $\nu$ decoupling in the SM. The interactions are written with $i~(i=e,\mu,\tau)$ type $\nu$. Collision terms in the effective operator limit assuming $s\ll M_{Z'}^2$.}
\label{tab:ampix}
\end{table}
\renewcommand{\arraystretch}{1}

The collision terms will be:
\begin{eqnarray}
\int  \frac{d^3p}{2\pi^3}  p ~\mathcal{C}[f]_{\nu_i + e \leftrightarrow \nu_i +e}& = & (X_i g_X)^2(X_1 g_X)^2~\frac{ 14 }{\pi^5 M_{Z'}^4}  ~T_{\gamma}^4
    T_{\nu_i}^4 \left[ T_{\gamma}-  T_{\nu_i} \right] \\
\int  \frac{d^3p}{2\pi^3}  p ~\mathcal{C}[f]_{\nu_i \bar{\nu_i} \leftrightarrow e e}& = &(X_i g_X)^2(X_1 g_X)^2~\frac{ 8}{\pi^5 M_{Z'}^4}  ~ \left[ T^9_{\gamma}-  T^9_{\nu_i} \right]
\label{eq:BSME}
\end{eqnarray}
So, the total energy transfer rate from $\nu$ to $e$ bath due to the pure BSM matrix element of  $U(1)_X$ gauge boson, 
\begin{eqnarray}
   \left(\frac{\delta \rho_{\nu_i\to e}}{\delta t}\right)_{\rm BSM}&=& 
    (X_i g_X)^2(X_1 g_X)^2~ \frac{1}{4\pi^5 M_{Z'}^4}  F_{\rm MB}(T_{\gamma},T_{\nu_i}),
    \label{eq:BSMx}
\end{eqnarray}
So the total energy transfer rate from $\nu_i$ to $e$ bath will be
\begin{equation}
    \left(\frac{\delta \rho_{\nu_i\to e}}{\delta t}\right)_{\rm tot}=\left(\frac{\delta \rho_{\nu_i\to e}}{\delta t}\right)_{\rm SM}+\left(\frac{\delta \rho_{\nu_i\to e}}{\delta t}\right)_{\rm int}+\left(\frac{\delta \rho_{\nu_i\to e}}{\delta t}\right)_{\rm BSM}
    \label{eq:deltot}
\end{equation}

\uline{\bf $X_1=0$}: Here we would like to highlight the scenarios where $X_1=0$. In such case apparently, it seems that the BSM contributions in eq.\eqref{eq:rhoz},eq.\eqref{eq:BSME},eq.\eqref{eq:BSMx} vanish. However, in such cases even though $Z'$ doesn't have any tree level coupling with $e^\pm$, it can couple to $e^\pm$ through induced couplings which may generate from kinetic mixing, fermion loops or CKM mixing. If the loop factor induced coupling is denoted as $\epsilon$, then in such cases we have to just replace the term $(X_1 g_X)$ in eq.\eqref{eq:rhoz},eq.\eqref{eq:BSME},eq.\eqref{eq:BSMx} with $\epsilon~e$.
\begin{equation}
    (X_1 g_X)\longrightarrow (\epsilon e) ~~~({\rm for}~~ X_1=0)
\end{equation}
So, even if $X_1=0$, there can be energy transfer from $Z'$ to $e$ bath as well as $\nu_i$ to $e$ bath by $Z'$ mediation.

Note that eq.\eqref{eq:deltot} differs from ref.\cite{Escudero:2019gzq} where the authors ignored the SM-BSM interference term in the calculation of energy transfer rates. 
However, our BSM contribution term $\left(\frac{\delta \rho_{\nu_i\to e}}{\delta t}\right)_{\rm BSM}$ in eq.\eqref{eq:BSMx} agrees with ref.\cite{Escudero:2019gzq}  for a specific choice of $U(1)_X~ (~U(1)_{L_\mu-L_\tau})$.

\subsection{Temperature evolution}
Now with all the collision terms, we are set to formulate the energy evaluation equations.
Following the prescription in appendix-\ref{sec:apxA} we can write the following energy density evaluation equations:
\begin{eqnarray}
     \frac{d\rho_{\nu_i}}{dt}&=& -\left(  4 H \rho_{\nu_i} -\left(\frac{\delta \rho_{\nu_i\to e}}{\delta t}\right)_{\rm tot}-\sum_{j\neq i}\left(\frac{\delta \rho_{\nu_i \to \nu_j}}{\delta t}\right)_{\rm tot}+\frac{ \delta \rho_{Z'\to \nu_i}}{\delta t} \right)\\
     \frac{d\rho_{Z'}}{dt}&=&  -\left( 3 H (\rho_{Z'}+P_{Z'}) -\sum_i\frac{\delta \rho_{Z'\to \nu_i}}{\delta t}-\frac{\delta \rho_{Z'\to e}}{\delta t}\right)\\
     \frac{d\rho_{\gamma}}{dt}  &=&- \left(  4 H \rho_{\gamma} + 3 H \left( \rho_{e} + p_{e}\right) +\sum_i \left(\frac{\delta \rho_{\nu_i\to e}}{\delta t}\right)_{\rm tot} +  \frac{ \delta \rho_{Z'\to e}}{\delta t}  \right),
     \label{eq:ultra}
\end{eqnarray}
where the $\sum_i$ indicates the summation of energy density transfer rates over all $3$ generatation of $\nu~(e,\mu,\tau)$.
And $\left(\frac{\delta \rho_{\nu_i \to \nu_j}}{\delta t}\right)_{\rm tot}$ bears a similar kind of form as in the SM case. However, in the presence of the light $Z'$, it will get some additional contribution from $Z'$ mediation. So the $G_F^2$ in eq.\eqref{eq:delnui} will be simply replaced by $G_F\to (G_F+(X_i g_X)(X_j g_X)/M_{Z'}^2)$. 
Similar to the previous section we can write the temperature evaluation equations using the partial derivatives and we get, 
\begin{eqnarray}
     \frac{dT_{\nu_i}}{dt}&=& -\left(  4 H \rho_{\nu_i} -\left(\frac{\delta \rho_{\nu_i\to e}}{\delta t}\right)_{\rm tot}-\sum_{j\neq i}\left(\frac{\delta \rho_{\nu_i \to \nu_j}}{\delta t}\right)_{\rm tot}+\frac{ \delta \rho_{Z'\to \nu_i}}{\delta t} \right)\left(\frac{\partial\rho_{\nu_i}}{\partial T_{\nu_i}}\right)^{-1} \label{eq:ultrapro1}\\
     \frac{dT_{Z'}}{dt}&=&  -\left( 3 H (\rho_{Z'}+P_{Z'}) -\sum_i\frac{\delta \rho_{Z'\to \nu_i}}{\delta t}-\frac{\delta \rho_{Z'\to e}}{\delta t}\right)\left(\frac{\partial\rho_{Z'}}{\partial T_{Z'}}\right)^{-1} \label{eq:ultrapro2}\\
     \frac{dT_{\gamma}}{dt}  &=&- \left(  4 H \rho_{\gamma} + 3 H \left( \rho_{e} + p_{e}\right) + \sum_i\left(\frac{ \delta \rho_{\nu_i\to e}}{\delta t} \right)_{\rm tot}+  \frac{ \delta \rho_{Z'\to e}}{\delta t}  \right)\left(    \frac{\partial \rho_{\gamma}}{\partial T_\gamma} + \frac{\partial \rho_e}{\partial T_\gamma}       \right)^{-1}
     \label{eq:ultrapro}
\end{eqnarray}

Hence solving the aforementioned five Boltzmann equations on can track the evolution of $\nu_L$ decoupling. 
By exploiting the neutrino oscillations which are active around $\sim$ MeV temperature \cite{Akhmedov:1999uz,Hannestad:2001iy,Dolgov:2002wy,Dolgov:2002ab}, one can further simplify the scenario by assuming all 3 $\nu_i$ equilibrate with each other and acquire a common temperature i.e. $T_{\nu_e}=T_{\nu_\mu}=T_{\nu_\tau}\equiv T_\nu$. 
The benefit of such an assumption is that we no longer have to keep track of the energy transfer rates within different $\nu$ sectors i.e. $\left(\frac{\delta \rho_{\nu_i \to \nu_j}}{\delta t}\right)_{\rm tot}=0$ in eq.\eqref{eq:ultrapro1}. In such case the three $T_{\nu_i}$ equations  (eq.\eqref{eq:ultrapro1}) simply reduces to a single temperature ($T_\nu$) evaluation equation:
\begin{eqnarray}
\frac{dT_{\nu}}{dt}&=& -\left(  4 H \rho_{\nu} -\left(\frac{\delta \rho_{\nu\to e}}{\delta t}\right)_{\rm tot}+\frac{ \delta \rho_{Z'\to \nu}}{\delta t} \right)\left(\frac{\partial\rho_{\nu}}{\partial T_{\nu}}\right)^{-1}, \label{eq:ultraf}
\end{eqnarray}     
where, $\rho_\nu=\sum_{i=e,\mu,\tau}\rho_{\nu_i}$,
$\left(\frac{\delta \rho_{\nu\to e}}{\delta t}\right)_{\rm tot}=\sum_{i=e,\mu,\tau}\left(\frac{\delta \rho_{\nu_i\to e}}{\delta t}\right)_{\rm tot}$ and $\frac{ \delta \rho_{Z'\to \nu}}{\delta t}=\sum_{i=e,\mu,\tau} \frac{ \delta \rho_{Z'\to \nu_i}}{\delta t}$


\section{Evaluation of $N_{\rm eff}$ with different $T_{\nu_i}$}
\label{sec:apxC}
\begin{figure}[!tbh]
    \centering
    \subfigure[\label{t1}]{
    \includegraphics[scale=0.28]{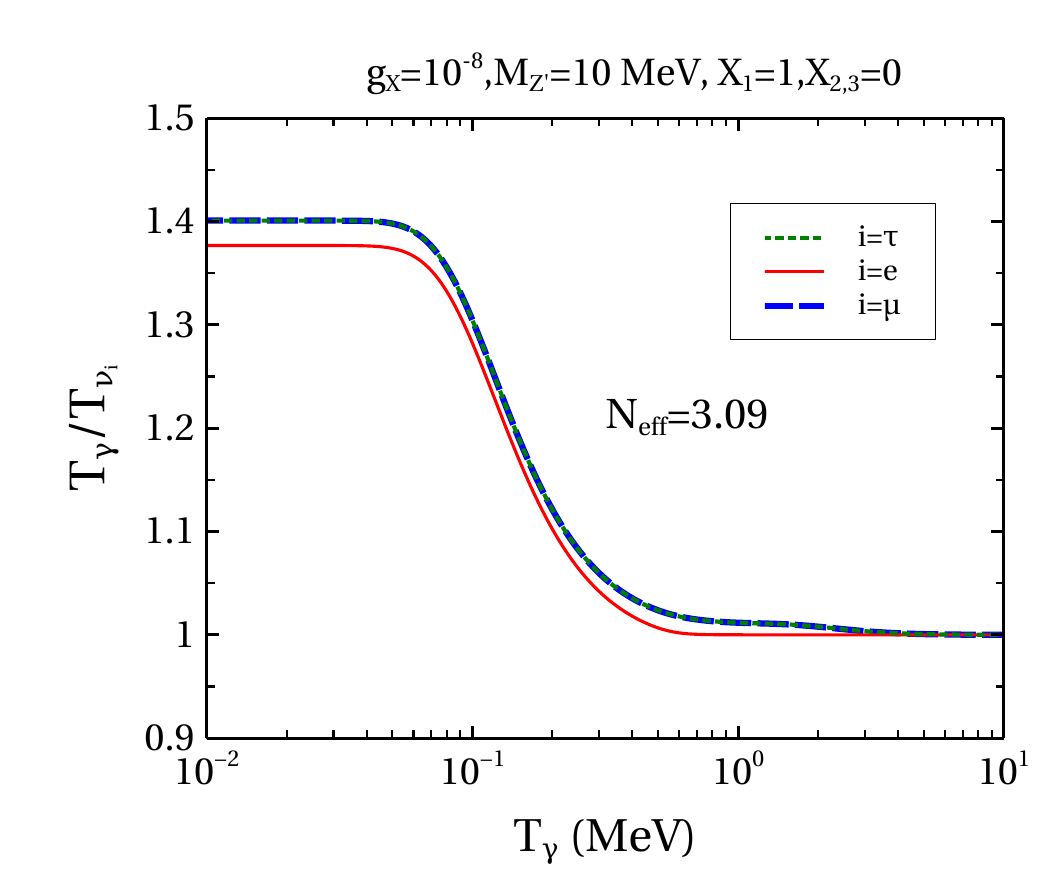}}
    \subfigure[\label{t2}]{
    \includegraphics[scale=0.28]{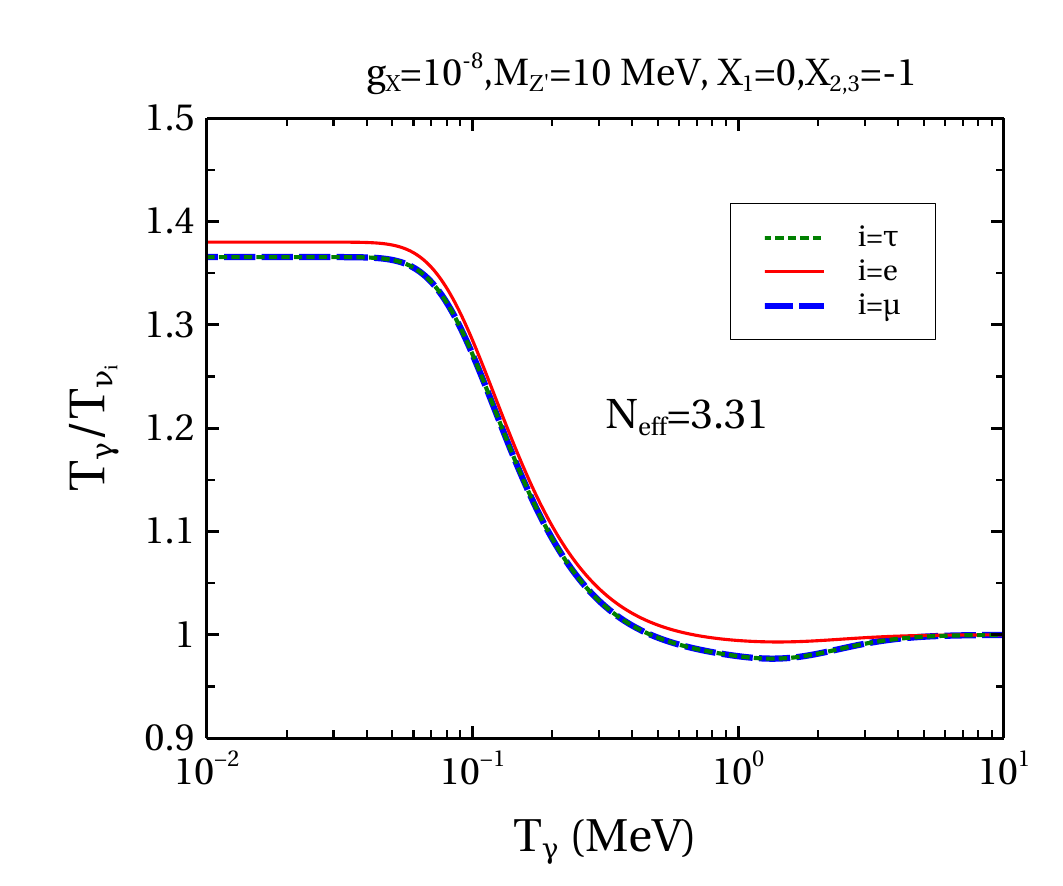}}
    \subfigure[\label{t3}]{
    \includegraphics[scale=0.28]{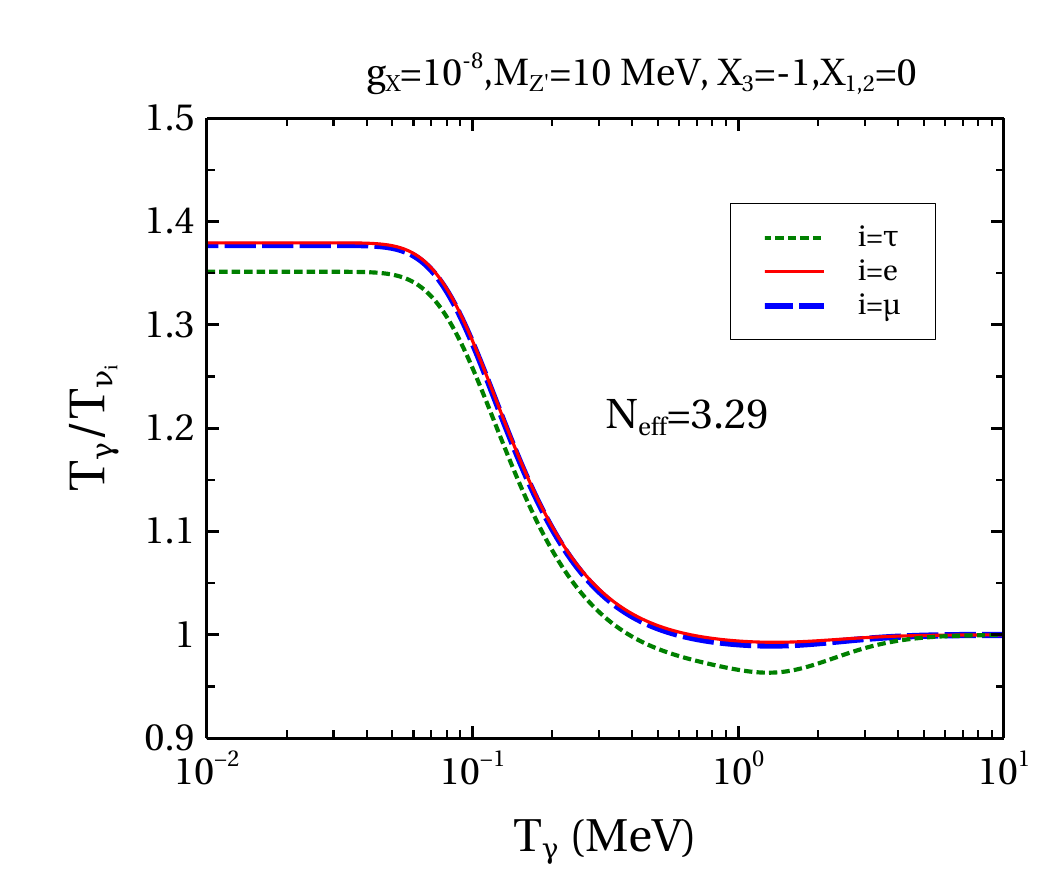}}
    \caption{Evolution of $T_{\gamma}/T_{\nu}$ with photon bath temperature $T_{\gamma}$  assuming all 3 $\nu_i$ with different temperatures for different $U(1)_X$ charge combinations. We chose a benchmark parameter value $M_{Z'}=10$ MeV and $g_X=10^{-7}$ for all 3 plots. We consider 3 different charge combination (a)$X_1=-1,X_{2,3}=0$ (b)$X_2=-1,X_{1,3}=0$ 
    and (a)$X_3=-1,X_{1,2}=0$. Here even if $X_1=0$ $N_{eff}$ changes slightly depending on $X_{2,3}$ (in contrast to the case considered in the papaer.)}
    \label{fig:Tnudiff}
\end{figure}

Throughout the paper we evaluated $N_{\rm eff} $ assuming all three $\nu_i,~(i\equiv e,\mu,\tau)$ share the same temperature.
However, the values of $N_{\rm eff} $ change if we allow all $\nu_i$ to develop different temperatures as prescribed in context of eq.(\ref{eq:ultrapro1}-\ref{eq:ultrapro}).
We show our results for $g_X=10^{-8},~M_{Z'}=10$ MeV for three different $U(1)_X$ charge combinations in Fig.\ref{fig:Tnudiff} assuming only tree level couplings (i.e. $\epsilon=0$).

We consider $|X_1|=1,|X_{2,3}|=0$, $|X_1|=0,|X_{2,3}|=1$ and $|X_3|=-1,|X_{1,2}|=0$ in Fig.\ref{t1},Fig.\ref{t2} and Fig.\ref{t3}
respectively.
The red, blue and green lines indicate the temperature ratio $T_\gamma/T_{\nu_i}$ for $\nu_e, \nu_\mu$ and $\nu_\tau$ respectively. 
Note that for $|X_1|=1,|X_{2,3}|=0$ only $\nu_e$ has BSM interactions
which lead to enhancing $T_{\nu_e}$, whereas $T_{\nu_{\mu/\tau}}$ reproduces the same value as predicted in SM scenario (where, $(T_\gamma/T_{\nu_{i}})_{\rm SM}\sim1.4$ \cite{Escudero:2018mvt}) as portrayed in Fig.\ref{t1}.
It is interesting to note that this value of $N_{\rm eff}=3.09$ is less ($\sim 8\%$) than the one predicted for the same charge configuration in sec.\ref{sec:neff}.
This is due to the fact we overestimated the neutrino energy increment in presence of $Z'$ while assuming all 3 $\nu_i$ share the same temperature.

For Fig.\ref{t2} and Fig.\ref{t3} also we notice the calculated values of $N_{\rm eff}$ is slightly less than what was predicted in sec.\ref{sec:neff}.
For $|X_1|=0,|X_{2,3}|=1$, $Z'$ promptly decays to $\nu_{\mu}$ and $\nu_\tau$ and not in electron. This causes a sudden dip (before $T_\gamma\approx 0.5$ MeV ) in the respective temperature ratio curves shown by the blue and green line in Fig.\ref{t2}.
For the same reason we notice similar dip in the temperature ratio curve (green line) for $\nu_\tau$  in Fig.\ref{t3}.
The value of $N_{\rm eff}$ is slightly higher for $|X_1|=0,|X_{2,3}|=1$ in Fig.\ref{t2} compared to the one obtained for Fig.\ref{t3} as the former one contains two type of $\nu_i$ having BSM interactions.
Also it is worth pointing that the value of $N_{\rm eff}$ obtained for $|X_1|=1,|X_{2,3}|=0$ is less compared to the other two as here $Z'$ decays to both $\nu_e$ and $e$ (i.e. photon bath) whereas in the other two cases $Z'$ decays only to $\nu_i$ bath.

\bibliographystyle{utphys}
\bibliography{bibliography}

\providecommand{\href}[2]{#2}\begingroup\raggedright\begin{thebibliography}{10}

\bibitem{Dodelson:2003ft}
S.~Dodelson, {\em {Modern Cosmology}}.
\newblock Academic Press, Amsterdam, 2003.

\bibitem{Mangano:2005cc}
G.~Mangano, G.~Miele, S.~Pastor, T.~Pinto, O.~Pisanti, and P.~D. Serpico,
  ``{Relic neutrino decoupling including flavor oscillations},''
  \href{http://dx.doi.org/10.1016/j.nuclphysb.2005.09.041}{{\em Nucl. Phys. B}
  {\bfseries 729} (2005) 221--234},
  \href{http://arxiv.org/abs/hep-ph/0506164}{{\ttfamily arXiv:hep-ph/0506164}}.

\bibitem{Grohs:2015tfy}
E.~Grohs, G.~M. Fuller, C.~T. Kishimoto, M.~W. Paris, and A.~Vlasenko,
  ``{Neutrino energy transport in weak decoupling and big bang
  nucleosynthesis},'' \href{http://dx.doi.org/10.1103/PhysRevD.93.083522}{{\em
  Phys. Rev. D} {\bfseries 93} no.~8, (2016) 083522},
  \href{http://arxiv.org/abs/1512.02205}{{\ttfamily arXiv:1512.02205
  [astro-ph.CO]}}.

\bibitem{Planck:2018vyg}
{\bfseries Planck} Collaboration, N.~Aghanim {\em et~al.}, ``{Planck 2018
  results. VI. Cosmological parameters},''
  \href{http://dx.doi.org/10.1051/0004-6361/201833910}{{\em Astron. Astrophys.}
  {\bfseries 641} (2020) A6}, \href{http://arxiv.org/abs/1807.06209}{{\ttfamily
  arXiv:1807.06209 [astro-ph.CO]}}. [Erratum: Astron.Astrophys. 652, C4
  (2021)].

\bibitem{Escudero:2018mvt}
M.~Escudero, ``{Neutrino decoupling beyond the Standard Model: CMB constraints
  on the Dark Matter mass with a fast and precise $N_{\rm eff}$ evaluation},''
  \href{http://dx.doi.org/10.1088/1475-7516/2019/02/007}{{\em JCAP} {\bfseries
  02} (2019) 007}, \href{http://arxiv.org/abs/1812.05605}{{\ttfamily
  arXiv:1812.05605 [hep-ph]}}.

\bibitem{Escudero:2019gzq}
M.~Escudero, D.~Hooper, G.~Krnjaic, and M.~Pierre, ``{Cosmology with A Very
  Light L$_{\mu}$ \ensuremath{-} L$_{\tau}$ Gauge Boson},''
  \href{http://dx.doi.org/10.1007/JHEP03(2019)071}{{\em JHEP} {\bfseries 03}
  (2019) 071}, \href{http://arxiv.org/abs/1901.02010}{{\ttfamily
  arXiv:1901.02010 [hep-ph]}}.

\bibitem{Esseili:2023ldf}
H.~Esseili and G.~D. Kribs, ``{Cosmological Implications of Gauged $U(1)_{B-L}$
  on $\Delta N_{\rm eff}$ in the CMB and BBN},''
  \href{http://arxiv.org/abs/2308.07955}{{\ttfamily arXiv:2308.07955
  [hep-ph]}}.

\bibitem{Ganguly:2022ujt}
S.~Ganguly, S.~Roy, and A.~K. Saha, ``{Imprints of MeV scale hidden dark sector
  at Planck data},''
  \href{http://dx.doi.org/10.1016/j.physletb.2022.137463}{{\em Phys. Lett. B}
  {\bfseries 834} (2022) 137463},
  \href{http://arxiv.org/abs/2201.00854}{{\ttfamily arXiv:2201.00854
  [hep-ph]}}.

\bibitem{Abazajian:2019oqj}
K.~N. Abazajian and J.~Heeck, ``{Observing Dirac neutrinos in the cosmic
  microwave background},''
  \href{http://dx.doi.org/10.1103/PhysRevD.100.075027}{{\em Phys. Rev. D}
  {\bfseries 100} (2019) 075027},
  \href{http://arxiv.org/abs/1908.03286}{{\ttfamily arXiv:1908.03286
  [hep-ph]}}.

\bibitem{Poulin:2018cxd}
V.~Poulin, T.~L. Smith, T.~Karwal, and M.~Kamionkowski, ``{Early Dark Energy
  Can Resolve The Hubble Tension},''
  \href{http://dx.doi.org/10.1103/PhysRevLett.122.221301}{{\em Phys. Rev.
  Lett.} {\bfseries 122} no.~22, (2019) 221301},
  \href{http://arxiv.org/abs/1811.04083}{{\ttfamily arXiv:1811.04083
  [astro-ph.CO]}}.

\bibitem{Ghosh:2022fws}
D.~K. Ghosh, S.~Jeesun, and D.~Nanda, ``{Long-lived inert Higgs boson in a fast
  expanding universe and its imprint on the cosmic microwave background},''
  \href{http://dx.doi.org/10.1103/PhysRevD.106.115001}{{\em Phys. Rev. D}
  {\bfseries 106} no.~11, (2022) 115001},
  \href{http://arxiv.org/abs/2206.04940}{{\ttfamily arXiv:2206.04940
  [hep-ph]}}.

\bibitem{Ghosh:2023ocl}
D.~K. Ghosh, P.~Ghosh, and S.~Jeesun, ``{CMB signature of non-thermal Dark
  Matter produced from self-interacting dark sector},''
  \href{http://dx.doi.org/10.1088/1475-7516/2023/07/012}{{\em JCAP} {\bfseries
  07} (2023) 012}, \href{http://arxiv.org/abs/2301.13754}{{\ttfamily
  arXiv:2301.13754 [hep-ph]}}.

\bibitem{Poulin:2018dzj}
V.~Poulin, T.~L. Smith, D.~Grin, T.~Karwal, and M.~Kamionkowski,
  ``{Cosmological implications of ultralight axionlike fields},''
  \href{http://dx.doi.org/10.1103/PhysRevD.98.083525}{{\em Phys. Rev. D}
  {\bfseries 98} no.~8, (2018) 083525},
  \href{http://arxiv.org/abs/1806.10608}{{\ttfamily arXiv:1806.10608
  [astro-ph.CO]}}.

\bibitem{Bringmann:2018jpr}
T.~Bringmann, F.~Kahlhoefer, K.~Schmidt-Hoberg, and P.~Walia, ``{Converting
  nonrelativistic dark matter to radiation},''
  \href{http://dx.doi.org/10.1103/PhysRevD.98.023543}{{\em Phys. Rev. D}
  {\bfseries 98} no.~2, (2018) 023543},
  \href{http://arxiv.org/abs/1803.03644}{{\ttfamily arXiv:1803.03644
  [astro-ph.CO]}}.

\bibitem{Li:2023puz}
S.-P. Li and X.-J. Xu, ``{N$_{eff}$ constraints on light mediators coupled to
  neutrinos: the dilution-resistant effect},''
  \href{http://dx.doi.org/10.1007/JHEP10(2023)012}{{\em JHEP} {\bfseries 10}
  (2023) 012}, \href{http://arxiv.org/abs/2307.13967}{{\ttfamily
  arXiv:2307.13967 [hep-ph]}}.

\bibitem{Biswas:2022fga}
A.~Biswas, D.~K. Ghosh, and D.~Nanda, ``{Concealing Dirac neutrinos from cosmic
  microwave background},''
  \href{http://dx.doi.org/10.1088/1475-7516/2022/10/006}{{\em JCAP} {\bfseries
  10} (2022) 006}, \href{http://arxiv.org/abs/2206.13710}{{\ttfamily
  arXiv:2206.13710 [hep-ph]}}.

\bibitem{Berbig:2020wve}
M.~Berbig, S.~Jana, and A.~Trautner, ``{The Hubble tension and a renormalizable
  model of gauged neutrino self-interactions},''
  \href{http://dx.doi.org/10.1103/PhysRevD.102.115008}{{\em Phys. Rev. D}
  {\bfseries 102} no.~11, (2020) 115008},
  \href{http://arxiv.org/abs/2004.13039}{{\ttfamily arXiv:2004.13039
  [hep-ph]}}.

\bibitem{Fabbrichesi:2020wbt}
M.~Fabbrichesi, E.~Gabrielli, and G.~Lanfranchi, ``{The Dark Photon},''
  \href{http://arxiv.org/abs/2005.01515}{{\ttfamily arXiv:2005.01515
  [hep-ph]}}.

\bibitem{Shutt:2002rg}
{\bfseries Borexino} Collaboration, T.~Shutt, ``{Borexino: A status report},''
  \href{http://dx.doi.org/10.1016/S0920-5632(02)01498-6}{{\em Nucl. Phys. B
  Proc. Suppl.} {\bfseries 110} (2002) 323--325}.

\bibitem{Majumdar:2021vdw}
A.~Majumdar, D.~K. Papoulias, and R.~Srivastava, ``{Dark matter detectors as a
  novel probe for light new physics},''
  \href{http://dx.doi.org/10.1103/PhysRevD.106.013001}{{\em Phys. Rev. D}
  {\bfseries 106} no.~1, (2022) 013001},
  \href{http://arxiv.org/abs/2112.03309}{{\ttfamily arXiv:2112.03309
  [hep-ph]}}.

\bibitem{Ma:2015raa}
E.~Ma and R.~Srivastava, ``{Dirac or inverse seesaw neutrino masses from gauged
  $B–L$ symmetry},'' \href{http://dx.doi.org/10.1142/S0217732315300207}{{\em
  Mod. Phys. Lett. A} {\bfseries 30} no.~26, (2015) 1530020},
  \href{http://arxiv.org/abs/1504.00111}{{\ttfamily arXiv:1504.00111
  [hep-ph]}}.

\bibitem{Bonilla:2017lsq}
C.~Bonilla, T.~Modak, R.~Srivastava, and J.~W.~F. Valle, ``{$U(1)_{B_3-3L_\mu}$
  gauge symmetry as a simple description of $b\to s$ anomalies},''
  \href{http://dx.doi.org/10.1103/PhysRevD.98.095002}{{\em Phys. Rev. D}
  {\bfseries 98} no.~9, (2018) 095002},
  \href{http://arxiv.org/abs/1705.00915}{{\ttfamily arXiv:1705.00915
  [hep-ph]}}.

\bibitem{Dutta:2020jsy}
B.~Dutta, S.~Ghosh, and J.~Kumar, ``{Contributions to $\Delta N_{eff}$ from the
  dark photon of $U(1)_{T3R}$},''
  \href{http://dx.doi.org/10.1103/PhysRevD.102.015013}{{\em Phys. Rev. D}
  {\bfseries 102} no.~1, (2020) 015013},
  \href{http://arxiv.org/abs/2002.01137}{{\ttfamily arXiv:2002.01137
  [hep-ph]}}.

\bibitem{CMS:2016cfx}
{\bfseries CMS} Collaboration, V.~Khachatryan {\em et~al.}, ``{Search for
  narrow resonances in dilepton mass spectra in proton-proton collisions at
  $\sqrt{s}$ = 13 TeV and combination with 8 TeV data},''
  \href{http://dx.doi.org/10.1016/j.physletb.2017.02.010}{{\em Phys. Lett. B}
  {\bfseries 768} (2017) 57--80},
  \href{http://arxiv.org/abs/1609.05391}{{\ttfamily arXiv:1609.05391
  [hep-ex]}}.

\bibitem{ATLAS:2019erb}
{\bfseries ATLAS} Collaboration, G.~Aad {\em et~al.}, ``{Search for high-mass
  dilepton resonances using 139 fb$^{-1}$ of $pp$ collision data collected at
  $\sqrt{s}=$13 TeV with the ATLAS detector},''
  \href{http://dx.doi.org/10.1016/j.physletb.2019.07.016}{{\em Phys. Lett. B}
  {\bfseries 796} (2019) 68--87},
  \href{http://arxiv.org/abs/1903.06248}{{\ttfamily arXiv:1903.06248
  [hep-ex]}}.

\bibitem{Das:2016zue}
A.~Das, S.~Oda, N.~Okada, and D.-s. Takahashi, ``{Classically conformal U(1)'
  extended standard model, electroweak vacuum stability, and LHC Run-2
  bounds},'' \href{http://dx.doi.org/10.1103/PhysRevD.93.115038}{{\em Phys.
  Rev. D} {\bfseries 93} no.~11, (2016) 115038},
  \href{http://arxiv.org/abs/1605.01157}{{\ttfamily arXiv:1605.01157
  [hep-ph]}}.

\bibitem{Accomando:2017qcs}
E.~Accomando, L.~Delle~Rose, S.~Moretti, E.~Olaiya, and C.~H.
  Shepherd-Themistocleous, ``{Extra Higgs boson and Z$^{\prime}$ as portals to
  signatures of heavy neutrinos at the LHC},''
  \href{http://dx.doi.org/10.1007/JHEP02(2018)109}{{\em JHEP} {\bfseries 02}
  (2018) 109}, \href{http://arxiv.org/abs/1708.03650}{{\ttfamily
  arXiv:1708.03650 [hep-ph]}}.

\bibitem{Coloma:2022umy}
P.~Coloma, P.~Coloma, M.~C. Gonzalez-Garcia, M.~C. Gonzalez-Garcia, M.~Maltoni,
  M.~Maltoni, J.~a.~P. Pinheiro, J.~a.~P. Pinheiro, S.~Urrea, and S.~Urrea,
  ``{Constraining new physics with Borexino Phase-II spectral data},''
  \href{http://dx.doi.org/10.1007/JHEP07(2022)138}{{\em JHEP} {\bfseries 07}
  (2022) 138}, \href{http://arxiv.org/abs/2204.03011}{{\ttfamily
  arXiv:2204.03011 [hep-ph]}}. [Erratum: JHEP 11, 138 (2022)].

\bibitem{COHERENT:2021xmm}
{\bfseries COHERENT} Collaboration, D.~Akimov {\em et~al.}, ``{Measurement of
  the Coherent Elastic Neutrino-Nucleus Scattering Cross Section on CsI by
  COHERENT},'' \href{http://dx.doi.org/10.1103/PhysRevLett.129.081801}{{\em
  Phys. Rev. Lett.} {\bfseries 129} no.~8, (2022) 081801},
  \href{http://arxiv.org/abs/2110.07730}{{\ttfamily arXiv:2110.07730
  [hep-ex]}}.

\bibitem{DiValentino:2021izs}
E.~Di~Valentino, O.~Mena, S.~Pan, L.~Visinelli, W.~Yang, A.~Melchiorri, D.~F.
  Mota, A.~G. Riess, and J.~Silk, ``{In the realm of the Hubble
  tension\textemdash{}a review of solutions},''
  \href{http://dx.doi.org/10.1088/1361-6382/ac086d}{{\em Class. Quant. Grav.}
  {\bfseries 38} no.~15, (2021) 153001},
  \href{http://arxiv.org/abs/2103.01183}{{\ttfamily arXiv:2103.01183
  [astro-ph.CO]}}.

\bibitem{Riess:2018byc}
A.~G. Riess {\em et~al.}, ``{Milky Way Cepheid Standards for Measuring Cosmic
  Distances and Application to Gaia DR2: Implications for the Hubble
  Constant},'' \href{http://dx.doi.org/10.3847/1538-4357/aac82e}{{\em
  Astrophys. J.} {\bfseries 861} no.~2, (2018) 126},
  \href{http://arxiv.org/abs/1804.10655}{{\ttfamily arXiv:1804.10655
  [astro-ph.CO]}}.

\bibitem{Riess:2016jrr}
A.~G. Riess {\em et~al.}, ``{A 2.4 $\%$ Determination of the Local Value of the
  Hubble Constant},'' \href{http://dx.doi.org/10.3847/0004-637X/826/1/56}{{\em
  Astrophys. J.} {\bfseries 826} no.~1, (2016) 56},
  \href{http://arxiv.org/abs/1604.01424}{{\ttfamily arXiv:1604.01424
  [astro-ph.CO]}}.

\bibitem{Riess:2021jrx}
A.~G. Riess {\em et~al.}, ``{A Comprehensive Measurement of the Local Value of
  the Hubble Constant with 1 km/s/Mpc Uncertainty from the Hubble Space
  Telescope and the SH0ES Team},''
  \href{http://dx.doi.org/10.3847/2041-8213/ac5c5b}{{\em Astrophys. J. Lett.}
  {\bfseries 934} no.~1, (2022) L7},
  \href{http://arxiv.org/abs/2112.04510}{{\ttfamily arXiv:2112.04510
  [astro-ph.CO]}}.

\bibitem{Vagnozzi:2019ezj}
S.~Vagnozzi, ``{New physics in light of the $H_0$ tension: An alternative
  view},'' \href{http://dx.doi.org/10.1103/PhysRevD.102.023518}{{\em Phys. Rev.
  D} {\bfseries 102} no.~2, (2020) 023518},
  \href{http://arxiv.org/abs/1907.07569}{{\ttfamily arXiv:1907.07569
  [astro-ph.CO]}}.

\bibitem{Vagnozzi:2023nrq}
S.~Vagnozzi, ``{Seven Hints That Early-Time New Physics Alone Is Not Sufficient
  to Solve the Hubble Tension},''
  \href{http://dx.doi.org/10.3390/universe9090393}{{\em Universe} {\bfseries 9}
  no.~9, (2023) 393}, \href{http://arxiv.org/abs/2308.16628}{{\ttfamily
  arXiv:2308.16628 [astro-ph.CO]}}.

\bibitem{Planck:2015fie}
{\bfseries Planck} Collaboration, P.~A.~R. Ade {\em et~al.}, ``{Planck 2015
  results. XIII. Cosmological parameters},''
  \href{http://dx.doi.org/10.1051/0004-6361/201525830}{{\em Astron. Astrophys.}
  {\bfseries 594} (2016) A13},
  \href{http://arxiv.org/abs/1502.01589}{{\ttfamily arXiv:1502.01589
  [astro-ph.CO]}}.

\bibitem{Efstathiou:2013via}
G.~Efstathiou, ``{H0 Revisited},''
  \href{http://dx.doi.org/10.1093/mnras/stu278}{{\em Mon. Not. Roy. Astron.
  Soc.} {\bfseries 440} no.~2, (2014) 1138--1152},
  \href{http://arxiv.org/abs/1311.3461}{{\ttfamily arXiv:1311.3461
  [astro-ph.CO]}}.

\bibitem{Freedman:2017yms}
W.~L. Freedman, ``{Cosmology at a Crossroads},''
  \href{http://dx.doi.org/10.1038/s41550-017-0121}{{\em Nature Astron.}
  {\bfseries 1} (2017) 0121}, \href{http://arxiv.org/abs/1706.02739}{{\ttfamily
  arXiv:1706.02739 [astro-ph.CO]}}.

\bibitem{Bernal:2016gxb}
J.~L. Bernal, L.~Verde, and A.~G. Riess, ``{The trouble with $H_0$},''
  \href{http://dx.doi.org/10.1088/1475-7516/2016/10/019}{{\em JCAP} {\bfseries
  10} (2016) 019}, \href{http://arxiv.org/abs/1607.05617}{{\ttfamily
  arXiv:1607.05617 [astro-ph.CO]}}.

\bibitem{Araki:2021xdk}
T.~Araki, K.~Asai, K.~Honda, R.~Kasuya, J.~Sato, T.~Shimomura, and M.~J.~S.
  Yang, ``{Resolving the Hubble tension in a U(1)$_{L_\mu-L_\tau}$ model with
  the Majoron},'' \href{http://dx.doi.org/10.1093/ptep/ptab108}{{\em PTEP}
  {\bfseries 2021} no.~10, (2021) 103B05},
  \href{http://arxiv.org/abs/2103.07167}{{\ttfamily arXiv:2103.07167
  [hep-ph]}}.

\bibitem{Hildebrandt:2018yau}
H.~Hildebrandt {\em et~al.}, ``{KiDS+VIKING-450: Cosmic shear tomography with
  optical and infrared data},''
  \href{http://dx.doi.org/10.1051/0004-6361/201834878}{{\em Astron. Astrophys.}
  {\bfseries 633} (2020) A69},
  \href{http://arxiv.org/abs/1812.06076}{{\ttfamily arXiv:1812.06076
  [astro-ph.CO]}}.

\bibitem{deJesus:2022pux}
A.~S. de~Jesus, N.~Pinto-Neto, F.~S. Queiroz, J.~Silk, and D.~R. da~Silva,
  ``{The hubble rate trouble: an effective field theory of dark matter},''
  \href{http://dx.doi.org/10.1140/epjc/s10052-023-11366-5}{{\em Eur. Phys. J.
  C} {\bfseries 83} no.~3, (2023) 203},
  \href{http://arxiv.org/abs/2212.13272}{{\ttfamily arXiv:2212.13272
  [hep-ph]}}.

\bibitem{AtzoriCorona:2022moj}
M.~Atzori~Corona, M.~Cadeddu, N.~Cargioli, F.~Dordei, C.~Giunti, Y.~F. Li,
  E.~Picciau, C.~A. Ternes, and Y.~Y. Zhang, ``{Probing light mediators and
  $(g-2)_\mu$ through detection of coherent elastic neutrino nucleus scattering
  at COHERENT},'' \href{http://dx.doi.org/10.1007/JHEP05(2022)109}{{\em JHEP}
  {\bfseries 05} (2022) 109}, \href{http://arxiv.org/abs/2202.11002}{{\ttfamily
  arXiv:2202.11002 [hep-ph]}}.

\bibitem{Sabti:2019mhn}
N.~Sabti, J.~Alvey, M.~Escudero, M.~Fairbairn, and D.~Blas, ``{Refined Bounds
  on MeV-scale Thermal Dark Sectors from BBN and the CMB},''
  \href{http://dx.doi.org/10.1088/1475-7516/2020/01/004}{{\em JCAP} {\bfseries
  01} (2020) 004}, \href{http://arxiv.org/abs/1910.01649}{{\ttfamily
  arXiv:1910.01649 [hep-ph]}}.

\bibitem{Banerjee:2021laz}
H.~Banerjee, B.~Dutta, and S.~Roy, ``{Probing
  L\ensuremath{\mu}-L\ensuremath{\tau} models with CE\ensuremath{\nu}NS: A new
  look at the combined COHERENT CsI and Ar data},''
  \href{http://dx.doi.org/10.1103/PhysRevD.104.015015}{{\em Phys. Rev. D}
  {\bfseries 104} no.~1, (2021) 015015},
  \href{http://arxiv.org/abs/2103.10196}{{\ttfamily arXiv:2103.10196
  [hep-ph]}}.

\bibitem{Hannestad:2001iy}
S.~Hannestad, ``{Oscillation effects on neutrino decoupling in the early
  universe},'' \href{http://dx.doi.org/10.1103/PhysRevD.65.083006}{{\em Phys.
  Rev. D} {\bfseries 65} (2002) 083006},
  \href{http://arxiv.org/abs/astro-ph/0111423}{{\ttfamily
  arXiv:astro-ph/0111423}}.

\bibitem{Mohapatra:1979ia}
R.~N. Mohapatra and G.~Senjanovic, ``{Neutrino Mass and Spontaneous Parity
  Nonconservation},'' \href{http://dx.doi.org/10.1103/PhysRevLett.44.912}{{\em
  Phys. Rev. Lett.} {\bfseries 44} (1980) 912}.

\bibitem{Schechter:1980gr}
J.~Schechter and J.~W.~F. Valle, ``{Neutrino Masses in SU(2) x U(1)
  Theories},'' \href{http://dx.doi.org/10.1103/PhysRevD.22.2227}{{\em Phys.
  Rev. D} {\bfseries 22} (1980) 2227}.

\bibitem{Fradette:2018hhl}
A.~Fradette, M.~Pospelov, J.~Pradler, and A.~Ritz, ``{Cosmological beam dump:
  constraints on dark scalars mixed with the Higgs boson},''
  \href{http://dx.doi.org/10.1103/PhysRevD.99.075004}{{\em Phys. Rev. D}
  {\bfseries 99} no.~7, (2019) 075004},
  \href{http://arxiv.org/abs/1812.07585}{{\ttfamily arXiv:1812.07585
  [hep-ph]}}.

\bibitem{Kreisch:2019yzn}
C.~D. Kreisch, F.-Y. Cyr-Racine, and O.~Dor\'e, ``{Neutrino puzzle: Anomalies,
  interactions, and cosmological tensions},''
  \href{http://dx.doi.org/10.1103/PhysRevD.101.123505}{{\em Phys. Rev. D}
  {\bfseries 101} no.~12, (2020) 123505},
  \href{http://arxiv.org/abs/1902.00534}{{\ttfamily arXiv:1902.00534
  [astro-ph.CO]}}.

\bibitem{EscuderoAbenza:2020cmq}
M.~Escudero~Abenza, ``{Precision early universe thermodynamics made simple:
  $N_{\rm eff}$ and neutrino decoupling in the Standard Model and beyond},''
  \href{http://dx.doi.org/10.1088/1475-7516/2020/05/048}{{\em JCAP} {\bfseries
  05} (2020) 048}, \href{http://arxiv.org/abs/2001.04466}{{\ttfamily
  arXiv:2001.04466 [hep-ph]}}.

\bibitem{Luo:2020sho}
X.~Luo, W.~Rodejohann, and X.-J. Xu, ``{Dirac neutrinos and $N_{{\rm eff}}$},''
  \href{http://dx.doi.org/10.1088/1475-7516/2020/06/058}{{\em JCAP} {\bfseries
  06} (2020) 058}, \href{http://arxiv.org/abs/2005.01629}{{\ttfamily
  arXiv:2005.01629 [hep-ph]}}.

\bibitem{Dolgov:1992wf}
A.~D. Dolgov and K.~Kainulainen, ``{Fermi-Dirac corrections to the relic
  abundances},'' \href{http://dx.doi.org/10.1016/0550-3213(93)90646-7}{{\em
  Nucl. Phys. B} {\bfseries 402} (1993) 349--359},
  \href{http://arxiv.org/abs/hep-ph/9211231}{{\ttfamily arXiv:hep-ph/9211231}}.

\bibitem{Kawasaki:2000en}
M.~Kawasaki, K.~Kohri, and N.~Sugiyama, ``{MeV scale reheating temperature and
  thermalization of neutrino background},''
  \href{http://dx.doi.org/10.1103/PhysRevD.62.023506}{{\em Phys. Rev. D}
  {\bfseries 62} (2000) 023506},
  \href{http://arxiv.org/abs/astro-ph/0002127}{{\ttfamily
  arXiv:astro-ph/0002127}}.

\bibitem{Kolb:1990vq}
E.~W. Kolb and M.~S. Turner,
  \href{http://dx.doi.org/10.1201/9780429492860}{{\em {The Early Universe}}},
  vol.~69.
\newblock 1990.

\bibitem{Akhmedov:1999uz}
E.~K. Akhmedov, ``{Neutrino physics},'' in {\em {ICTP Summer School in Particle
  Physics}}, pp.~103--164.
\newblock 6, 1999.
\newblock \href{http://arxiv.org/abs/hep-ph/0001264}{{\ttfamily
  arXiv:hep-ph/0001264}}.

\bibitem{Dolgov:2002wy}
A.~D. Dolgov, ``{Neutrinos in cosmology},''
  \href{http://dx.doi.org/10.1016/S0370-1573(02)00139-4}{{\em Phys. Rept.}
  {\bfseries 370} (2002) 333--535},
  \href{http://arxiv.org/abs/hep-ph/0202122}{{\ttfamily arXiv:hep-ph/0202122}}.

\bibitem{Dolgov:2002ab}
A.~D. Dolgov, S.~H. Hansen, S.~Pastor, S.~T. Petcov, G.~G. Raffelt, and D.~V.
  Semikoz, ``{Cosmological bounds on neutrino degeneracy improved by flavor
  oscillations},'' \href{http://dx.doi.org/10.1016/S0550-3213(02)00274-2}{{\em
  Nucl. Phys. B} {\bfseries 632} (2002) 363--382},
  \href{http://arxiv.org/abs/hep-ph/0201287}{{\ttfamily arXiv:hep-ph/0201287}}.

\bibitem{Amaral:2020tga}
D.~W. P.~d. Amaral, D.~G. Cerdeno, P.~Foldenauer, and E.~Reid, ``{Solar
  neutrino probes of the muon anomalous magnetic moment in the gauged $
  \mathrm{U}{(1)}_{L_{\mu }-{L}_{\tau }} $},''
  \href{http://dx.doi.org/10.1007/JHEP12(2020)155}{{\em JHEP} {\bfseries 12}
  (2020) 155}, \href{http://arxiv.org/abs/2006.11225}{{\ttfamily
  arXiv:2006.11225 [hep-ph]}}.

\bibitem{Fields:2019pfx}
B.~D. Fields, K.~A. Olive, T.-H. Yeh, and C.~Young, ``{Big-Bang Nucleosynthesis
  after Planck},'' \href{http://dx.doi.org/10.1088/1475-7516/2020/03/010}{{\em
  JCAP} {\bfseries 03} (2020) 010},
  \href{http://arxiv.org/abs/1912.01132}{{\ttfamily arXiv:1912.01132
  [astro-ph.CO]}}. [Erratum: JCAP 11, E02 (2020)].

\bibitem{SimonsObservatory:2018koc}
{\bfseries Simons Observatory} Collaboration, P.~Ade {\em et~al.}, ``{The
  Simons Observatory: Science goals and forecasts},''
  \href{http://dx.doi.org/10.1088/1475-7516/2019/02/056}{{\em JCAP} {\bfseries
  02} (2019) 056}, \href{http://arxiv.org/abs/1808.07445}{{\ttfamily
  arXiv:1808.07445 [astro-ph.CO]}}.

\bibitem{CMB-S4:2022ght}
{\bfseries CMB-S4} Collaboration, K.~Abazajian {\em et~al.}, ``{Snowmass 2021
  CMB-S4 White Paper},'' \href{http://arxiv.org/abs/2203.08024}{{\ttfamily
  arXiv:2203.08024 [astro-ph.CO]}}.

\bibitem{CMB-HD:2022bsz}
{\bfseries CMB-HD} Collaboration, S.~Aiola {\em et~al.}, ``{Snowmass2021 CMB-HD
  White Paper},'' \href{http://arxiv.org/abs/2203.05728}{{\ttfamily
  arXiv:2203.05728 [astro-ph.CO]}}.

\bibitem{XENON:2017lvq}
{\bfseries XENON} Collaboration, E.~Aprile {\em et~al.}, ``{The XENON1T Dark
  Matter Experiment},''
  \href{http://dx.doi.org/10.1140/epjc/s10052-017-5326-3}{{\em Eur. Phys. J. C}
  {\bfseries 77} no.~12, (2017) 881},
  \href{http://arxiv.org/abs/1708.07051}{{\ttfamily arXiv:1708.07051
  [astro-ph.IM]}}.

\bibitem{Chakraborty:2021apc}
K.~Chakraborty, A.~Das, S.~Goswami, and S.~Roy, ``{Constraining general U(1)
  interactions from neutrino-electron scattering measurements at DUNE near
  detector},'' \href{http://dx.doi.org/10.1007/JHEP04(2022)008}{{\em JHEP}
  {\bfseries 04} (2022) 008}, \href{http://arxiv.org/abs/2111.08767}{{\ttfamily
  arXiv:2111.08767 [hep-ph]}}.

\bibitem{Cadeddu:2020nbr}
M.~Cadeddu, N.~Cargioli, F.~Dordei, C.~Giunti, Y.~F. Li, E.~Picciau, and Y.~Y.
  Zhang, ``{Constraints on light vector mediators through coherent elastic
  neutrino nucleus scattering data from COHERENT},''
  \href{http://dx.doi.org/10.1007/JHEP01(2021)116}{{\em JHEP} {\bfseries 01}
  (2021) 116}, \href{http://arxiv.org/abs/2008.05022}{{\ttfamily
  arXiv:2008.05022 [hep-ph]}}.

\bibitem{Kolb:1987qy}
E.~W. Kolb and M.~S. Turner, ``{Supernova SN 1987a and the Secret Interactions
  of Neutrinos},'' \href{http://dx.doi.org/10.1103/PhysRevD.36.2895}{{\em Phys.
  Rev. D} {\bfseries 36} (1987) 2895}.

\bibitem{Akita:2022etk}
K.~Akita, S.~H. Im, and M.~Masud, ``{Probing non-standard neutrino interactions
  with a light boson from next galactic and diffuse supernova neutrinos},''
  \href{http://dx.doi.org/10.1007/JHEP12(2022)050}{{\em JHEP} {\bfseries 12}
  (2022) 050}, \href{http://arxiv.org/abs/2206.06852}{{\ttfamily
  arXiv:2206.06852 [hep-ph]}}.

\bibitem{Fiorillo:2022cdq}
D.~F.~G. Fiorillo, G.~G. Raffelt, and E.~Vitagliano, ``{Strong Supernova 1987A
  Constraints on Bosons Decaying to Neutrinos},''
  \href{http://dx.doi.org/10.1103/PhysRevLett.131.021001}{{\em Phys. Rev.
  Lett.} {\bfseries 131} no.~2, (2023) 021001},
  \href{http://arxiv.org/abs/2209.11773}{{\ttfamily arXiv:2209.11773
  [hep-ph]}}.

\bibitem{Fiorillo:2023ytr}
D.~F.~G. Fiorillo, G.~G. Raffelt, and E.~Vitagliano, ``{Large Neutrino Secret
  Interactions Have a Small Impact on Supernovae},''
  \href{http://dx.doi.org/10.1103/PhysRevLett.132.021002}{{\em Phys. Rev.
  Lett.} {\bfseries 132} no.~2, (2024) 021002},
  \href{http://arxiv.org/abs/2307.15115}{{\ttfamily arXiv:2307.15115
  [hep-ph]}}.

\bibitem{Janka:2017vlw}
H.~T. Janka, ``{Neutrino Emission from Supernovae},''
  \href{http://arxiv.org/abs/1702.08713}{{\ttfamily arXiv:1702.08713
  [astro-ph.HE]}}.

\bibitem{Croon:2020lrf}
D.~Croon, G.~Elor, R.~K. Leane, and S.~D. McDermott, ``{Supernova Muons: New
  Constraints on $Z$' Bosons, Axions and ALPs},''
  \href{http://dx.doi.org/10.1007/JHEP01(2021)107}{{\em JHEP} {\bfseries 01}
  (2021) 107}, \href{http://arxiv.org/abs/2006.13942}{{\ttfamily
  arXiv:2006.13942 [hep-ph]}}.

\bibitem{Altmannshofer:2014pba}
W.~Altmannshofer, S.~Gori, M.~Pospelov, and I.~Yavin, ``{Neutrino Trident
  Production: A Powerful Probe of New Physics with Neutrino Beams},''
  \href{http://dx.doi.org/10.1103/PhysRevLett.113.091801}{{\em Phys. Rev.
  Lett.} {\bfseries 113} (2014) 091801},
  \href{http://arxiv.org/abs/1406.2332}{{\ttfamily arXiv:1406.2332 [hep-ph]}}.

\bibitem{Schwartz:2014sze}
M.~D. Schwartz, {\em {Quantum Field Theory and the Standard Model}}.
\newblock Cambridge University Press, 3, 2014.

\bibitem{ParticleDataGroup:2022pth}
{\bfseries Particle Data Group} Collaboration, R.~L. Workman {\em et~al.},
  ``{Review of Particle Physics},''
  \href{http://dx.doi.org/10.1093/ptep/ptac097}{{\em PTEP} {\bfseries 2022}
  (2022) 083C01}.

\bibitem{Khalife:2023qbu}
A.~R. Khalife, M.~B. Zanjani, S.~Galli, S.~G\"unther, J.~Lesgourgues, and
  K.~Benabed, ``{Review of Hubble tension solutions with new SH0ES and SPT-3G
  data},'' \href{http://arxiv.org/abs/2312.09814}{{\ttfamily arXiv:2312.09814
  [astro-ph.CO]}}.

\bibitem{Schoneberg:2021qvd}
N.~Sch\"oneberg, G.~Franco~Abell\'an, A.~P\'erez~S\'anchez, S.~J. Witte,
  V.~Poulin, and J.~Lesgourgues, ``{The H0 Olympics: A fair ranking of proposed
  models},'' \href{http://dx.doi.org/10.1016/j.physrep.2022.07.001}{{\em Phys.
  Rept.} {\bfseries 984} (2022) 1--55},
  \href{http://arxiv.org/abs/2107.10291}{{\ttfamily arXiv:2107.10291
  [astro-ph.CO]}}.

\bibitem{Zhitnitsky:2008ha}
A.~R. Zhitnitsky, ``{Confinement Deconfinement Phase Transition in Hot and
  Dense QCD at Large N(c)},''
  \href{http://dx.doi.org/10.1016/j.nuclphysa.2008.09.011}{{\em Nucl. Phys. A}
  {\bfseries 813} (2008) 279--292},
  \href{http://arxiv.org/abs/0808.1447}{{\ttfamily arXiv:0808.1447 [hep-ph]}}.

\end{thebibliography}\endgroup

\end{document}